\documentclass[11pt,a4paper]{article}
\usepackage{amsmath,amssymb,theorem,epsfig,url,psfrag,eepic,amsmath,mathtools,mathrsfs,amsbsy,dsfont,upgreek,pict2e}
\usepackage{jheppub}
\usepackage{graphicx}
\usepackage{tikz,pgfplots}
\usepackage[normalem]{ulem}
\usetikzlibrary{arrows.meta,calc,chains,decorations.markings,decorations.pathmorphing,decorations.pathreplacing,fadings,matrix,positioning,shapes,shapes.geometric,shapes.symbols,trees}
\pgfplotsset{compat=1.9}
\tikzset{->-/.style={postaction=decorate,decoration={markings,mark=at position .65 with {\arrow[scale=0.55]{Straight Barb}}}}}
\tikzset{cross/.style={path picture={\draw[black]
			(path picture bounding box.south east) -- (path picture bounding box.north west) (path picture bounding box.south west) -- (path picture bounding box.north east);}}}
\expandafter\def\expandafter\bfseries\expandafter{\bfseries\ifmmode\else\boldmath\fi}
\expandafter\def\expandafter\mdseries\expandafter{\mdseries\ifmmode\else\unboldmath\fi}
\expandafter\def\expandafter\normalfont\expandafter{\normalfont\ifmmode\else\unboldmath\fi}

\newcommand{\remark}[2][]{\ignorespaces}
\newcommand{\ind}[1]{{\scriptscriptstyle{#1}}}

\newcommand{\atopfrac}{\genfrac{}{}{0pt}{}}
\makeatletter
\def\Appendix{\appendix
  \def\@seccntformat##1{Appendix~\csname the##1\endcsname.~~}}
\makeatother
\makeatletter
\@addtoreset{equation}{section}

\makeatother
\title{\textbf{Dual description of $\eta$-deformed OSP sigma models}\vspace*{.3cm}}
\date{}
\author[a,b]{Mikhail Alfimov,}
\author[a,c]{Boris Feigin,}
\author[d]{Ben Hoare}
\author[c,e]{and Alexey Litvinov}
\affiliation[a]{National Research University Higher School of Economics, 119048 Moscow, Russia}
\affiliation[b]{P.N. Lebedev Physical Institute of the Russian Academy of Sciences, 119991 Moscow, Russia}
\affiliation[c]{Landau Institute for Theoretical Physics, 142432 Chernogolovka, Russia}
\affiliation[d]{Department of Mathematical Sciences, Durham University, Durham DH1 3LE, UK}
\affiliation[e]{Center for Advanced Studies, Skolkovo Institute of Science and Technology, 143026 Moscow, Russia}
\emailAdd{malfimov@hse.ru}
\emailAdd{borfeigin@gmail.com}
\emailAdd{ben.hoare@durham.ac.uk}
\emailAdd{litvinov@itp.ac.ru}
\abstract{We study the dual description of the $\eta$-deformed $OSP(N|2m)$ sigma model in the asymptotically free regime ($N>2m+2$).
Compared to the case of classical Lie groups, for supergroups there are inequivalent $\eta$-deformations corresponding to different choices of simple roots.
For a class of such deformations we propose the system of screening charges depending on a continuous parameter $b$, which  defines the $\eta$-deformed $OSP(N|2m)$ sigma model in the limit $b\rightarrow\infty$ and a certain Toda QFT as $b\rightarrow0$.
In the sigma model regime we show that the leading UV asymptotic of the $\eta$-deformed model coincides with a perturbed Gaussian theory.
In the perturbative regime $b\rightarrow0$ we show that the tree-level two-particle scattering matrix matches the expansion of the trigonometric $OSP(N|2m)$ $S$-matrix.}
\begin{document}
\maketitle
\section{Introduction}
In recent years there has been an increased interest in integrable deformations of sigma models.
In particular, in the so called $\eta$-deformation, which was originally formulated for principle chiral model in \cite{Klimcik:2008eq} and generalized to $G/H$ symmetric space sigma models in \cite{Delduc:2013fga}.
In this paper we initiate the detailed investigation of the $\eta$-deformed $G/H$ symmetric space sigma model and their Toda QFT duals when $G$ and $H$ are supergroups.
Our primary focus will be on superspheres $OSP(N|2m)/OSP(N-1|2m)$, generalizing the corresponding construction for spheres $SO(N)/SO(N-1)$ \cite{Fateev:2018yos,Litvinov:2018bou}.

It is known from previous work that the regime of interest for the dual description is imaginary $\eta$.
Writing $\eta=i\kappa$, the Euclidean action of the classical $\eta$-deformed $G/H$ symmetric space sigma model takes the form \cite{Delduc:2013fga}
\begin{equation}\label{Coset-action-deformed}
   \mathcal{A}_{\textrm{class}}=\frac{\kappa}{4\pi\nu}\int d^2\xi \, \textrm{Tr}\big[
   \mathbf{g}^{-1}\partial\mathbf{g}\,P \frac{1}{1-i\kappa\mathcal{R}_{\mathbf{g}}P}\,
   \mathbf{g}^{-1}\bar{\partial}\mathbf{g}\big] ~,
\end{equation}
where $\mathbf{g}\in G$ and we take $G$ to be semi-simple (or basic in the case of supergroups).
The trace $\mathrm{Tr}$ is the (normalised) invariant bilinear form, which in the case of supergroups should be replaced by the supertrace $\textrm{STr}$, and $P$ is the projector onto the ``coset space.''
The linear operator $\mathcal{R}$ is a particular skew-symmetric solution of modified classical Yang-Baxter (YB) equation of Drinfel'd-Jimbo type and $\mathcal{R}_{\mathbf{g}}=\textrm{Ad}_{\mathbf{g}}^{\vphantom{-1}}\mathcal{R}\textrm{Ad}_\mathbf{g}^{-1}$.
For any such $\mathcal{R}$ the theory \eqref{Coset-action-deformed} is classically integrable and admits a Lax pair representation \cite{Delduc:2013fga}.
The undeformed model is recovered by setting $\nu = \kappa R^{-2}$ and taking $\kappa \to 0$.

The model \eqref{Coset-action-deformed} is renormalisable at one loop with only $\kappa$ running \cite{Valent:2009nv,Sfetsos:2009dj,Squellari:2014jfa,Sfetsos:2015nya}.
Together with the form of the action \eqref{Coset-action-deformed}, this suggests identifying the sigma model coupling $\nu$ with the Planck constant $\hbar$.
For a certain class of backgrounds, such as $SO(N)/SO(N-1)$, it is believed that the deformation \eqref{Coset-action-deformed} defines fully renormalizable QFT
\begin{equation}\label{SM-action}
\mathcal{A}_{\hbar}=\frac{1}{4\pi}\int d^2\xi \, \left(G_{\mu\nu}+B_{\mu\nu}\right)\partial X^{\mu}\bar{\partial} X^{\nu} = \mathcal{A}_{\textrm{class}} + O(\hbar^0)~,
\end{equation}
where the metric $G_{\mu\nu}$ and the Kalb-Ramond field $B_{\mu\nu}$ depend on both the deformation parameter $\kappa$ and $\hbar$, and admit the semiclassical expansion
\begin{equation}\label{GB-hbar-expansion}
   G_{\mu\nu}=\frac{1}{\hbar}G_{\mu\nu}^{(0)}(\kappa)+G_{\mu\nu}^{(1)}(\kappa)+O(\hbar)~,\qquad
   B_{\mu\nu}=\frac{1}{\hbar}B_{\mu\nu}^{(0)}(\kappa)+B_{\mu\nu}^{(1)}(\kappa)+O(\hbar)~,
\end{equation}
where $G_{\mu\nu}^{(0)}(\kappa)$ and $B_{\mu\nu}^{(0)}(\kappa)$ follow from the classical action \eqref{Coset-action-deformed}.
That the model \eqref{SM-action} is renormalizable in general with only one running coupling $\kappa=\kappa(t)$ is a conjecture,
\unskip\footnote{Here $t=\log\frac{\Lambda^{*}}{\Lambda}$ where $\Lambda$ is the running scale.
The UV limit corresponds to $t\rightarrow-\infty$.}
and is thought to be closely related to the quantum integrability of the model \cite{Hoare:2020fye}.
Some checks of the conjecture going beyond the one-loop order were performed in \cite{Hoare:2019ark,Hoare:2019mcc,Borsato:2019oip} for the simplest models (see also, e.g., \cite{Hassler:2020tvz,Borsato:2020wwk,Codina:2020yma}).
Nevertheless, it is likely that the expansion \eqref{GB-hbar-expansion} exists, meaning that there is a particular renormalization scheme in which $\kappa$, being the only coupling, flows according to some all-loop beta function
\begin{equation}\label{all-loop-beta-function}
   \frac{d}{dt}\kappa=\beta(\kappa,\hbar)=\hbar\beta_{1}(\kappa)+\hbar^{2}\beta_{2}(\kappa)+\dots ~.
\end{equation}
Assuming that the theory \eqref{SM-action} displays asymptotic freedom implies that the equation \eqref{all-loop-beta-function} has a UV fixed point $\kappa=\kappa_{\textrm{UV}}$ where the metric $G_{\mu\nu}$ becomes flat and the Kalb-Ramond field $B_{\mu\nu}$ vanishes.

The semiclassical expansion \eqref{GB-hbar-expansion} corresponds to small $\hbar$ expansion while keeping $\kappa$-fixed, i.e. $t\sim \hbar^{-1}$.
Physically it is also interesting to consider the small $\hbar$  expansion with  $t$ fixed, i.e. $\kappa=\hbar\kappa_{0}$ with $\kappa_{0}$ fixed.
It is then easy to see from \eqref{Coset-action-deformed}, \eqref{GB-hbar-expansion} and \eqref{all-loop-beta-function} that at leading order in $\hbar\rightarrow0$ the model reduces to the undeformed one
\begin{equation}\label{Coset-action-undeformed}
   \mathcal{A}_{0}=\frac{\kappa_{0}}{4\pi}\int d^2\xi \, \textrm{Tr}\big[
   \mathbf{g}^{-1}\partial\mathbf{g} \, P \, \mathbf{g}^{-1}\bar{\partial}\mathbf{g}\big] ~,
\end{equation}
where the coupling $\kappa_{0}$ runs with its the all-loop beta-function $\beta_{0}(\kappa_{0})$ obtained from $\beta(\kappa,\hbar)$
\begin{equation}
  \beta(\hbar\kappa_{0},\hbar)=\hbar\beta_{0}(\kappa_{0})+O(\hbar^{2}) ~.
\end{equation}
Note that in the undeformed model \eqref{Coset-action-undeformed} there is only one parameter $\kappa_{0}$ that plays the role of both the running coupling constant and the inverse Planck constant.

In the case of $SO(N)/SO(N-1)$, the Lorentzian version of \eqref{Coset-action-undeformed}, i.e. the $O(N)$ sigma model, corresponds to a massive integrable QFT.
This QFT is governed by the celebrated rational $S$-matrix of Zamolodchikov and Zamolodchikov \cite{Zamolodchikov:1978xm}, which solves the Yang-Baxter (YB) equation.
It is well known that rational solutions of the YB equation admit one-parametric and two-parametric deformations known as trigonometric and elliptic.
In our case it is natural to assume that the deformed model \eqref{SM-action} should correspond to a certain trigonometric solution of the YB equation
\begin{equation}
  S_{\hbar}(\theta)=S(\theta)+O(\hbar)\qquad \hbar\rightarrow0 ~,
\end{equation}
where $S(\theta)$ corresponds to the rational $S$-matrix.
Solutions with the desired symmetry, i.e. with the matrix structure fixed by quantum group symmetry $U_q(\widehat{\mathfrak{so}}(N))$, were constructed by Bazhanov and Jimbo \cite{Bazhanov:1984gu,Jimbo:1985ua}.
Although, one still has to find an overall unitarizing factor to construct the physical $S$-matrix.
For the $O(N)$ model such a factor was recently conjectured in \cite{Fateev:2018yos}.

Importantly, the presence of two parameters in the deformed model \eqref{SM-action}, rather than one, allows us to consider the expansion with $\hbar$ fixed and $\kappa$ approaching the UV fixed point $\kappa\rightarrow\kappa_{\textrm{UV}}$.
According to the general philosophy of RG group flow, the action \eqref{SM-action} should describe an RG trajectory corresponding a free Gaussian CFT perturbed by certain relevant operators $\mathcal{O}_{r}$ with $\Delta=\bar{\Delta}<1$
\begin{equation}\label{action-vicinity-generic}
  \mathcal{A}_{\hbar}=\frac{1}{8\pi}\int d^2\xi \, \big[ \partial\boldsymbol{X}\cdot\bar{\partial}\boldsymbol{X}+\sum_{r}\lambda_{r}\mathcal{O}^{(r)}+\dots \big] \quad
  \text{where}\quad \mathcal{O}^{(r)}=\mathcal{O}^{(r)}_{\mu\nu}(\boldsymbol{X})\partial X^{\mu}\bar{\partial}X^{\nu} ~.
\end{equation}
The coupling constants $\lambda_{r}$ develop positive scaling dimensions $\Delta=\bar{\Delta}$ and hence in the UV limit become small
\begin{equation}\label{lambda-initial}
   \lambda_{r}(t)=\lambda_{r}^{(0)}e^{2(1-\Delta)t}~,\qquad t\rightarrow-\infty~.
\end{equation}
Typically $\Delta>\frac{1}{2}$ and hence the action \eqref{action-vicinity-generic} requires renormalization, which can modify the existing couplings \eqref{lambda-initial} and generate new ones as well.

Classically the field $\mathcal{O}^{(r)}$ is exactly marginal, i.e. $\Delta=\bar{\Delta}=1$.
In order to have a quantum anomalous dimension it should involve exponents.
A natural guess would be that $\mathcal{O}^{(r)}$ is a graviton operator
\begin{equation}
 \mathcal{O}^{(r)}=\mathcal{O}^{(r)}_{\mu\nu}e^{(\boldsymbol{\beta}_{r}\cdot\boldsymbol{X})}\partial X^{\mu}\bar{\partial}X^{\nu} ~,
\end{equation}
for some constant $\mathcal{O}^{(r)}_{\mu\nu}$, and hence $\Delta=\bar{\Delta}=1-(\boldsymbol{\beta}_{r}\cdot\boldsymbol{\beta}_{r})$.
In the integrable case, this operator is further constrained to commute with an infinite tower of integrals of motion
\begin{equation}
   [\mathcal{O}^{(r)},\mathbf{I}_{k}]=[\mathcal{O}^{(r)},\bar{\mathbf{I}}_{k}]=0 ~.
\end{equation}
We conjecture that this requirement can be satisfied by taking
\unskip\footnote{We have chosen $\mathcal{O}^{(r)}$ in \eqref{Or-conjecture} such that $\boldsymbol{C}$-symmetry is preserved, but $\boldsymbol{P}$- and $\boldsymbol{T}$-symmetry may be broken, since $\boldsymbol{\alpha}_{r}$ typically involves complex coefficients. Alternatively, one could choose
\begin{equation}
   \mathcal{O}_r=(\boldsymbol{\alpha}_{r}\cdot\partial\boldsymbol{X})(\boldsymbol{\alpha}_{r}\cdot\bar{\partial}\boldsymbol{X})e^{(\boldsymbol{\beta}_{r}\cdot\boldsymbol{X})}
\end{equation}
which preserves $\boldsymbol{P}$- and $\boldsymbol{T}$-symmetry, but breaks $\boldsymbol{C}$-symmetry.}
\begin{equation}\label{Or-conjecture}
  \mathcal{O}_r=\bigl|(\boldsymbol{\alpha}_{r}\cdot\partial\boldsymbol{X})\bigr|^{2}e^{(\boldsymbol{\beta}_{r}\cdot\boldsymbol{X})} ~,
\end{equation}
for some vectors $\boldsymbol{\alpha}_{r}$. 

Each interaction term of the form \eqref{Or-conjecture} defines a screening charge
\begin{equation}\label{screening-charge}
   \mathcal{S}_{r}=\oint dz \, (\boldsymbol{\alpha}_{r}\cdot\partial\boldsymbol{X})e^{(\boldsymbol{\beta}_{r}\cdot\boldsymbol{X})} ~,
\end{equation}
where, here, $\boldsymbol{X}=\boldsymbol{X}(z)$ denotes the \emph{holomorphic} part of the total field $\boldsymbol{X}(z,\bar{z})$.
This is known as the Wakimoto screening charge \cite{Wakimoto:1986gf} (see section \ref{Screenings} for more details).
For any choice of two linearly independent vectors  $\boldsymbol{\alpha}_{r}$ and $\boldsymbol{\beta}_{r}$ the  Wakimoto screening charge  $\mathcal{S}_{r}$ defines the $W$-algebra of the coset CFT $\hat{\mathfrak{su}}(2)_{k}/\hat{\mathfrak{u}}(1)$  with $k=-2-2(\boldsymbol{\beta}_{r}\cdot\boldsymbol{\beta}_{r})^{-1}$.
By definition the holomorphic currents $W_k(z)$ of this algebra commute with the screening charge \eqref{screening-charge}
\begin{equation}
   [\mathcal{S}_{r},W_{k}(z)]=0 ~,
\end{equation}
It is well known that the same algebra can be also defined as a commutant of a pair of fermionic screening charges
\begin{equation}\label{fermionic-screenings}
   \mathcal{S}_{\pm}=\oint dz \, e^{(\boldsymbol{\alpha}^{\pm}_{r}\cdot\boldsymbol{X})} ~,
\end{equation}
where
\begin{equation}\label{Wakimoto-fermionic-map}
  (\boldsymbol{\alpha}^{\pm}_{r}\cdot\boldsymbol{\alpha}^{\pm}_{r})=-1~,\qquad
  \boldsymbol{\beta}_{r}=\frac{2}{(\boldsymbol{\alpha}^{+}_{r}+\boldsymbol{\alpha}^{-}_{r})^{2}}(\boldsymbol{\alpha}^{+}_{r}+\boldsymbol{\alpha}^{-}_{r})~,
\qquad
  \boldsymbol{\alpha}_{r}=\xi_1\boldsymbol{\alpha}^{+}_{r}+\xi_2\boldsymbol{\alpha}^{-}_{r}~,
\end{equation}
for some $\xi_1$ and $\xi_2$ whose  values are not determined.
Indeed, since the integrand in \eqref{screening-charge} is defined modulo total derivatives, there are different choices of operators \eqref{Or-conjecture} for a given system of fermionic screening charges \eqref{fermionic-screenings}.

The integrable system spanned by the integrals of motion $\mathbf{I}_{k}$ belongs to the intersection of all the $W$-algebras that can be defined as the commutant of either the Wakimoto \eqref{screening-charge} or fermionic \eqref{fermionic-screenings} screening charges.
Taking them all to be of Wakimoto type leads to the theory \eqref{action-vicinity-generic}. Alternatively, taking them all to be fermionic screening charges corresponds to a Toda QFT
\begin{equation}\label{action-vicinity-generic-Toda}
  \mathcal{A}_{\hbar}^{(\mathrm{D})}=\frac{1}{8\pi}\int d^2\xi \, \big[\partial\boldsymbol{X}\cdot\bar{\partial}\boldsymbol{X}+
  \sum_{r}\sum_{\pm}e^{(\boldsymbol{\alpha}^{\pm}_{r}\cdot\boldsymbol{X})}+\dots\big] ~.
\end{equation}
This theory also requires counterterms; however, its UV structure is much more straightforward than that of \eqref{action-vicinity-generic}.
In particular, only finitely many counterterms are needed.
The resulting action defines a renormalizable QFT and, in Minkowski space, can be used to compute the perturbative $S$-matrix.

Both theories \eqref{action-vicinity-generic} and \eqref{action-vicinity-generic-Toda} are associated to the same system of integrals of motion in the vicinity of the Gaussian point.
Under the assumption that this integrable system is protected against perturbation theory, it is plausible that both actions correspond to the same QFT in different regimes.
It is clear that $\boldsymbol{\beta}_{r}$ should scale as $\boldsymbol{\beta}_{r}\sim\sqrt{\hbar}$ and hence \eqref{Wakimoto-fermionic-map} implies that
\begin{equation}
  \boldsymbol{\alpha}^{+}_{r}+\boldsymbol{\alpha}^{-}_{r}\sim \frac{1}{\sqrt{\hbar}} ~.
\end{equation}
Therefore, the Toda QFT \eqref{action-vicinity-generic-Toda} is expected to describe the strongly quantum regime of the sigma model \eqref{action-vicinity-generic} and vice versa.
We conjecture that this is a general phenomenon: any quantum integrable $\eta$-deformed sigma model admits a dual Toda theory description.
Our check of this conjecture consists of the following steps:
\begin{enumerate}
\item Identify a system of fermionic/Wakimoto screening charges that commute with an infinite system of integrals of motion $\mathbf{I}_{k}$.
\item Check that the UV expansion of the $\eta$-deformed sigma model \eqref{action-vicinity-generic} is controlled by the Wakimoto screening charges.
\unskip\footnote{In general, an all-loop sigma model action is not known. One can only compare the leading $\hbar\rightarrow0$ asymptotic. However, for the $O(3)$ case, the proposed all-loop action in \cite{Hoare:2019ark} matches exactly with the Wakimoto screening charges \cite{Alfimovunpub}.}
\item Compute the perturbative $S$-matrix for the lightest fundamental particles of the theory  \eqref{action-vicinity-generic-Toda} and check that it coincides with the expansion of the trigonometric deformation of the rational $S$-matrix that corresponds to the undeformed sigma model.
\end{enumerate}

These checks have been carried out for the deformed $O(N)$ sigma model in \cite{Fateev:2018yos,Litvinov:2018bou}.
Our aim in this paper is to generalize these results to sigma models on the supermanifold $OSP(N|2m)/OSP(N-1|2m)$.
The undeformed $OSP(N|2m)$ model can be viewed as the fermionization of the $O(N+2m)$ sigma model
\begin{equation}
   \mathcal{A}=\frac{1}{4\pi}\int d^2\xi \, \sum_{k=1}^{N+2m}(\partial_{a}\varphi_{k})^{2}~, \qquad \sum_{k=1}^{N+2m}\varphi_{k}^{2}=R^{2} ~.
\end{equation}
We pick any $m$ pairs of fields, say $\{(\varphi_{N+1},\varphi_{N+2}),\dots,(\varphi_{N+2m-1},\varphi_{N+2m})\}$ and formally replace
\begin{equation}\label{fermionization-trick}
  \varphi_{N+2k-1}+i\varphi_{N+2k}\rightarrow \psi_{k}~,\qquad
  \varphi_{N+2k-1}-i\varphi_{N+2k}\rightarrow \bar{\psi}_{k}~,
\end{equation}
where the new fields are $\psi_{k}$ and $\bar{\psi}_{k}$ are fermionic scalars.
The resulting model is
\begin{equation}\label{OSP-SM-undeformed}
   \mathcal{A}=\frac{1}{4\pi}\int d^2\xi \, \Big[\sum_{k=1}^{N}(\partial_{a}\varphi_{k})^{2}+\sum_{j=1}^{m}\partial_{a}\psi_{j}\partial_{a}\bar{\psi}_{j}\Big]~, \qquad 
   \sum_{k=1}^{N}\varphi_{k}^{2}+\sum_{j=1}^{m}\psi_{j}\bar{\psi}_{j}=R^{2} ~.
\end{equation}
Since the path integral over the fermionic fields is the reciprocal of that over $2m$ of $N$ bosonic fields \cite{Parisi:1979ka}, the one-loop RG flow equation for the radius $R$ is
\begin{equation}
   \frac{d R^{2}}{dt}=-(N-2m-2) ~.
\end{equation}
For real $R$ the model flows to strong coupling for $N>2m+2$, becomes conformal for $N=2m+2$ and zero-charged for $N<2m+2$.
We will focus on the regime $N>2m+2$, where the model can be described by a scattering theory of $N$ bosonic and $m$ charged fermionic massive particles transforming in the fundamental representation of $OSP(N|2m)$.
The corresponding $S$-matrix has been conjectured by Saleur and Wehefritz-Kaufmann in \cite{Saleur:2001cw}, generalizing the $O(N)$ $S$-matrix \cite{Zamolodchikov:1978xm}. 

Trigonometric solutions of the YB equation with $OSP(N|2m)$  symmetry have been found by Bazhanov and Shadrikov \cite{Bazhanov:1986av} (see also \cite{Galleas:2004zz,Galleas:2006kd}).
The key difference compared to the $O(N)$ case, is that there are inequivalent solutions corresponding to different choices of simple roots.
The same phenomenon also appears in the $\eta$-deformed model \eqref{Coset-action-deformed}: inequivalent choices of simple roots lead to different operators $\mathcal{R}$ and hence different deformations (see \cite{Delduc:2014kha,Hoare:2016ibq,Hoare:2018ngg} for the discussions of same issue in the context of the $AdS$ superstring).
A similar freedom shows up at the level of screening charges as well, of which there are inequivalent systems that correspond to equivalent integrable systems.
While the precise dictionary between these choices remains to be understood, in this paper we describe the relationship for a certain class of $\eta$-deformations/system of screenings/$S$-matrices, confirming our conjecture with explicit calculations for small values of $N$ and $m$.

This paper is organized as follows. In section \ref{Screenings} we introduce and discuss different systems of screening charges corresponding to the $OSP(N|2m)$ theory for $N>2m+2$.
We define the admissible systems of screening charges that underpin the duality.
In section \ref{QFT} we describe  the weak-coupling Toda QFT ($b\rightarrow0$) and the strong-coupling sigma model ($b\rightarrow\infty$) associated to a given admissible system of screening charges.
In section \ref{SM} we investigate $\eta$-deformations of $OSP$ sigma models corresponding to inequivalent choices of simple roots and in section \ref{S-matrix} we review trigonometric $OSP(N|2m)$ solutions to the YB equation and define the corresponding $S$-matrices.
We conclude in section \ref{concl} by pulling together these three pieces of knowledge and formulating our main result, explaining how a certain class of $\eta$-deformations/system of screenings/$S$-matrices are related by duality. In the appendices we collect supplementary formulae. 
\section{Screening charges}\label{Screenings}
The large class of $W$-algebras that commute with exponential screening operators has been studied in \cite{2015arXiv151208779B,Litvinov:2016mgi}.
We start by recalling the formulation of this problem.
Let $\boldsymbol{\varphi}(z)=(\varphi_{1}(z),\dots,\varphi_{N}(z))$ be an $N$-component holomorphic bosonic field normalized as
\begin{equation}
   \varphi_{i}(z)\varphi_{j}(z')=-\delta_{ij}\log(z-z')+\dots\quad\text{at}\quad z\rightarrow z'\;,
\end{equation}
and $\vec{\boldsymbol{\alpha}}=(\boldsymbol{\alpha}_{1},\dots,\boldsymbol{\alpha}_{N})$ be a set of linearly independent vectors.
We define the associated $W_{\vec{\boldsymbol{\alpha}}}$-algebra
to be the set of currents $W_{s}(z)$ with integer spins $s$, which are differential polynomials of $\partial\boldsymbol{\varphi}(z)$ of degree $s$, such that
\begin{equation}\label{comm-cond}
  \oint_{\mathcal{C}_{z}} d\xi \, e^{(\boldsymbol{\alpha}_{r}\cdot\boldsymbol{\varphi}(\xi))}W_{s}(z)=0\quad\text{for all}\quad r=1,\dots,N~,
\end{equation}
where $\mathcal{C}_{z}$ is the contour encircling the point $z$.
This condition is highly restrictive and for generic set of vectors $\vec{\boldsymbol{\alpha}}$ the algebra $W_{\vec{\boldsymbol{\alpha}}}$ is relatively small.
It can be shown that typically the only non-trivial current has spin $2$ and is given by
\unskip\footnote{Here and below we assume Wick ordering.}
\begin{equation}\label{W2}
   W_{2}(z)=-\frac{1}{2}(\partial\boldsymbol{\varphi}(z)\cdot\partial\boldsymbol{\varphi}(z))+(\boldsymbol{\rho}\cdot\partial^{2}\boldsymbol{\varphi}(z))~,\qquad
   \boldsymbol{\rho}=\sum_{r=1}^{N}\left(1+\frac{(\boldsymbol{\alpha}_{r}\cdot\boldsymbol{\alpha}_{r})}{2}\right)\hat{\boldsymbol{\alpha}}_{r} ~,
\end{equation}
where $\vec{\hat{\boldsymbol{\alpha}}}$ is the dual set of vectors satisfying $(\boldsymbol{\alpha}_{r}\cdot\hat{\boldsymbol{\alpha}}_{s})=\delta_{r,s}$.
The modes of the current \eqref{W2} satisfy the Virasoro algebra with the central charge
\begin{equation}\label{c-Virasoro}
  c=N+12(\boldsymbol{\rho}\cdot\boldsymbol{\rho})~.
\end{equation}
The existence of independent $W$-currents of higher spins, i.e. not algebraically expressible through $W_{2}$, is possible only if special conditions on the set $\vec{\boldsymbol{\alpha}}$ hold. 

All $W$-algebras with a non-trivial current of spin $3$ have been classified in \cite{2015arXiv151208779B,Litvinov:2016mgi}. They are known to correspond to the following representations of the affine Yangian of $\mathfrak{gl}(1)$ (usually denoted as $Y\big(\widehat{\mathfrak{gl}}(1)\big)$) \cite{Tsymbaliuk:2014fvq}
\begin{equation}\label{Yangian-generic-rep}
  \mathcal{F}_{k_1}\otimes\dots\otimes\mathcal{F}_{k_{N+1}} ~,
\end{equation}
where $\mathcal{F}_{k}$ with $k=1,2,3$ are three inequivalent Fock modules of $Y\big(\widehat{\mathfrak{gl}}(1)\big)$. To each pair of neighboring factors in  \eqref{Yangian-generic-rep} one associates the screening charge
\begin{equation}\label{generic-screening}
  \mathcal{F}_{k_i}\otimes\mathcal{F}_{k_{i+1}} \quad\longrightarrow \quad \oint dz\,e^{\varkappa_{k_{i+1}k_i}\varphi_{i+1}-\varkappa_{k_ik_{i+1}}\varphi_{i}} ~,
\end{equation}
where the $3\times3$ matrix $\boldsymbol{\varkappa}=\varkappa_{ij}$ depends on a free parameter $b$ and is given by
\begin{equation}
  \varkappa=
  \begin{pmatrix}
   b&b&\frac{1}{b}\\
   i\beta&i\beta&-\frac{ib}{\beta}\\
   \frac{i\beta}{b}&-\frac{i}{\beta}&\frac{i\beta}{b}
  \end{pmatrix}
  \quad\text{with}\quad\beta=\sqrt{1+b^2}~.
\end{equation}
Changing the order of the factors in \eqref{Yangian-generic-rep} leads to isomorphic $W$-algebras intertwined by the $R$-matrix of $Y\big(\widehat{\mathfrak{gl}}(1)\big)$.
However, the choice of ordering in  \eqref{Yangian-generic-rep} is important when constructing a QFT from the associated screening charges. 

In \eqref{generic-screening} two types of roots can appear.
These are characterized by whether their norm is a function of $b$ or fixed such that $(\boldsymbol{\alpha}_{r}\cdot \boldsymbol{\alpha}_{r})=-1$.
If the two representations in \eqref{generic-screening} are of the same type then the norm $(\boldsymbol{\alpha}_{r}\cdot \boldsymbol{\alpha}_{r})$ is unfixed and we call the corresponding vector a bosonic root
\begin{equation*}
\begin{picture}(200,30)(200,115)
    \Thicklines
    \unitlength 5pt 
    \put(40,25){\circle{2}}
    \put(42,24.4){-- bosonic root: $(\boldsymbol{\alpha}_{r}\cdot \boldsymbol{\alpha}_{r})=\text{unfixed}~.$}
  \end{picture}
\end{equation*}
On the other hand, when the two representations are different we have $(\boldsymbol{\alpha}_{r}\cdot\boldsymbol{\alpha}_{r})=-1$, which we refer to as a fermionic root
\begin{equation*}
\begin{picture}(300,30)(340,115)
    \Thicklines
    \unitlength 5pt 
    \put(80,25){\circle{2}}
    \put(79.4,24,4){\line(1,1){1.2}}
    \put(79.4,25,6){\line(1,-1){1.2}}
    \put(82,24.4){-- fermionic root: $(\boldsymbol{\alpha}_{r}\cdot\boldsymbol{\alpha}_{r})=-1~.$}
  \end{picture}
\end{equation*}
As one permutes the factors in \eqref{Yangian-generic-rep} some of the bosonic roots become fermionic and vice versa.
However, the information contained in the roots does not get lost: when going from one realization to the other some screening charges may become of Wakimoto type (or dressed) or vice versa.
We will not describe the full story here, but just list three basic properties that will be useful for us:
\paragraph{Bosonic root duality.} The bosonic roots always appear in pairs
\begin{equation}
    \boldsymbol{\alpha}\quad\text{and}\quad \boldsymbol{\alpha}^{\vee}=\frac{2\boldsymbol{\alpha}}{(\boldsymbol{\alpha}\cdot\boldsymbol{\alpha})}~.
\end{equation}
One can use either of these two roots to define the conformal algebra.
\paragraph{Dressed/Wakimoto bosonic screening.} Suppose that we have two fermionic roots $\boldsymbol{\alpha}_{1}$ and $\boldsymbol{\alpha}_{2}$, such that $(\boldsymbol{\alpha}_{1}\cdot\boldsymbol{\alpha}_{2})$ is arbitrary.
This corresponds to the case of three alternating representations, e.g. $\mathcal{F}_1\otimes\mathcal{F}_2\otimes\mathcal{F}_1$.
Then the same conformal algebra can be defined using the dressed bosonic screening 
\begin{equation}
   \mathcal{S}_{B}=\oint dz\,(\boldsymbol{\alpha}_{1}\cdot\partial\boldsymbol{\varphi})e^{(\boldsymbol{\beta}_{12}\cdot\boldsymbol{\varphi})}\quad\text{where}\quad
   \boldsymbol{\beta}_{12}=\frac{2(\boldsymbol{\alpha}_{1}+\boldsymbol{\alpha}_{2})}{(\boldsymbol{\alpha}_{1}+\boldsymbol{\alpha}_{2})^{2}}~.
\end{equation}
We will draw this situation as follows
\begin{equation}
\hspace*{-4cm}
\begin{picture}(200,50)(215,123)
    \Thicklines
    \unitlength 5pt
    \put(69,25){\circle{2}}
    \put(79,25){\circle{2}}
    \put(74,30){\circle{2}}
    \put(73.4,31.5){\line(1,0){1.2}}
    \put(68.4,24,4){\line(1,1){1.2}}
    \put(68.4,25,6){\line(1,-1){1.2}}
    \put(78.4,24,4){\line(1,1){1.2}}
    \put(78.4,25,6){\line(1,-1){1.2}}
    \put(70,25){\line(1,0){8}}
    \put(68.3,22){$\scriptstyle{\boldsymbol{\alpha}_{1}}$}
    \put(78.3,22){$\scriptstyle{\boldsymbol{\alpha}_{2}}$}
\end{picture}
\end{equation}
\paragraph{Dressed/Wakimoto fermionic screening.} Suppose that we have two fermionic roots $\boldsymbol{\alpha}_{1}$ and $\boldsymbol{\alpha}_{2}$, such that $(\boldsymbol{\alpha}_{1}\cdot\boldsymbol{\alpha}_{2})=-1$.
This corresponds to the case of three different representations following each other, e.g. $\mathcal{F}_1\otimes\mathcal{F}_2\otimes\mathcal{F}_3$.
Then the same conformal algebra can be defined using the dressed fermionic screening
\begin{equation}\label{dressed-fermionic}
   \mathcal{S}_{F}=\oint dz\,(\boldsymbol{\alpha}_{1}\cdot\partial\boldsymbol{\varphi})e^{(\boldsymbol{\beta}_{12}\cdot\boldsymbol{\varphi})}\quad\text{where}\quad
   \boldsymbol{\beta}_{12}=\nu\boldsymbol{\alpha}_{1}-(1+\nu)\boldsymbol{\alpha}_{2}~.
\end{equation}
We will draw this situation as follows
\begin{equation}
\hspace*{-4cm}
\begin{picture}(200,50)(215,113)
    \Thicklines
    \unitlength 5pt
    \put(69,25){\circle{2}}
    \put(79,25){\circle{2}}
    \put(74,30){\circle{2}}
    \put(73.4,31.5){\line(1,0){1.2}}
    \put(68.4,24,4){\line(1,1){1.2}}
    \put(68.4,25,6){\line(1,-1){1.2}}
    \put(78.4,24,4){\line(1,1){1.2}}
    \put(78.4,25,6){\line(1,-1){1.2}}
    \put(73.4,29,4){\line(1,1){1.2}}
    \put(73.4,30,6){\line(1,-1){1.2}}
    \put(70,25){\line(1,0){1}}
    \put(71.4,25){\line(1,0){1}}
    \put(72.8,25){\line(1,0){1}}
    \put(74.2,25){\line(1,0){1}}
    \put(75.6,25){\line(1,0){1}}
    \put(77,25){\line(1,0){1}}
    \put(68.3,22){$\scriptstyle{\boldsymbol{\alpha}_{1}}$}
    \put(78.3,22){$\scriptstyle{\boldsymbol{\alpha}_{2}}$}
\end{picture}
\end{equation}
The arbitrary parameter $\nu$ in \eqref{dressed-fermionic} reflects the fact that the Gram matrix of $\boldsymbol{\alpha}_{1}$ and $\boldsymbol{\alpha}_{2}$ is degenerate. It cannot be fixed if only these two roots are present; however, it is fixed if $\boldsymbol{\alpha}_1$ and $\boldsymbol{\alpha}_2$ are embedded in larger diagram.

In this paper we will consider $W$-algebras corresponding to the $OSP(N|2m)$ sigma model.
They do not belong to the class of algebras described above since, as follows from symmetry considerations, they have no spin $3$ current, but a non-vanishing spin $4$ current.
In \cite{Litvinov:2016mgi,Fateev:2018yos,Litvinov:2018bou} a class of such $W$-algebras corresponding to the deformed $O(N)$ sigma model was conjectured.
These are similar to the $W$-algebras considered above, depending on a continuous parameter $b$.
Since the cases of odd and even $N$ are slightly different, we focus on $N=2n+1$ for simplicity.
For the $O(N)$ sigma model we start from the CFT that corresponds to the following ``balalaika'' diagram
\begin{equation}\label{diagram-O(2n+1)}
\begin{picture}(300,100)(220,80)
    \Thicklines
    \unitlength 5pt
    \put(48,32){\circle{2}}
    \put(48,18){\circle{2}}
    \put(54.4,24,4){\line(-1,-1){7}}
    \put(54.4,25,6){\line(-1,1){7}}
    \put(47.4,31.4){\line(1,1){1.2}}
    \put(47.4,18.6){\line(1,-1){1.2}}
    \put(48,19){\line(0,1){12}}
    \put(55,25){\circle{2}}
    \put(54.4,24,4){\line(1,1){1.2}}
    \put(54.4,25,6){\line(1,-1){1.2}}
    \put(66,25){\line(1,0){8}}
    \put(56,25){\line(1,0){8}}
    \put(65,25){\circle{2}}
    \put(64.4,24,4){\line(1,1){1.2}}
    \put(64.4,25,6){\line(1,-1){1.2}}
    \put(75,25){\circle{2}}
    \put(74.4,24,4){\line(1,1){1.2}}
    \put(74.4,25,6){\line(1,-1){1.2}}
    \put(76,25){\line(1,0){2}}
    \put(80,25){\circle{.2}}
    \put(81,25){\circle{.2}}
    \put(82,25){\circle{.2}}
    \put(83,25){\circle{.2}}
    \put(84,25){\circle{.2}}
    \put(85,25){\circle{.2}}
    \put(87,25){\line(1,0){2}}
    \put(90,25){\circle{2}}
    \put(89.4,24,4){\line(1,1){1.2}}
    \put(89.4,25,6){\line(1,-1){1.2}}
    \put(91,25){\line(1,0){8}}
    \put(100,25){\circle{2}}
    \put(99.4,24,4){\line(1,1){1.2}}
    \put(99.4,25,6){\line(1,-1){1.2}}
    \put(51.5,20){$\scriptstyle{-b^{2}}$}
    \put(51.5,29){$\scriptstyle{-b^{2}}$}
    \put(43,24.7){$\scriptstyle{1+2b^{2}}$}
    \put(58.5,22.5){$\scriptstyle{1+b^{2}}$}
    \put(93.5,22.5){$\scriptstyle{1+b^{2}}$}
    \put(69,22.5){$\scriptstyle{-b^{2}}$}
    \put(45.5,16){$\scriptstyle{\boldsymbol{\alpha}_{1}}$}
    \put(45.5,33.5){$\scriptstyle{\boldsymbol{\alpha}_{2}}$}
     \put(54.3,27){$\scriptstyle{\boldsymbol{\alpha}_{3}}$}
     \put(64,27){$\scriptstyle{\boldsymbol{\alpha}_{4}}$}
     \put(74,27){$\scriptstyle{\boldsymbol{\alpha}_{5}}$}
      \put(88.6,27){$\scriptstyle{\boldsymbol{\alpha}_{2n-1}}$}
     \put(98.6,27){$\scriptstyle{\boldsymbol{\alpha}_{2n}}$}
  \end{picture}
\end{equation}
It is convenient to parameterize the vectors $\boldsymbol{\alpha}_r$ as
\begin{equation}\label{vectors-O(2n+1)-conformal}
\begin{aligned}
&  \boldsymbol{\alpha}_{1}=b\boldsymbol{E}_{1}+i\beta\boldsymbol{e}_{1} ~,\qquad && \boldsymbol{\alpha}_{2}=b\boldsymbol{E}_{1}-i\beta\boldsymbol{e}_{1}~,\\
&   \boldsymbol{\alpha}_{2k-1}=-b\boldsymbol{E}_{k-1}+i\beta\boldsymbol{e}_{k}~,\qquad  &&
\boldsymbol{\alpha}_{2k}=b\boldsymbol{E}_{k}-i\beta\boldsymbol{e}_{k}~,\qquad k=2,\dots,n~,
\end{aligned}
\end{equation}
where $(\boldsymbol{E}_{i},\boldsymbol{e}_{i})$, $i=1,\dots,n$, form an orthonormal basis of $\mathbb{R}^{2n}$.
According to \eqref{c-Virasoro} the CFT has central charge
\begin{equation}\label{c-O(2n+1)}
   c=2n+\frac{n(4n^{2}-1)}{b^{2}}-\frac{2n(n-1)(2n-1)}{1+b^{2}}~.
\end{equation}
In the limit $b\rightarrow\infty$, $c\rightarrow2n$, while the central charge diverges as $b\rightarrow0$.

The diagram \eqref{diagram-O(2n+1)} with the root $\boldsymbol{\alpha}_1$ removed corresponds to the alternating representation of the Yangian $Y\big(\widehat{\mathfrak{gl}}(1)\big)$
\begin{equation}\label{alternating-spin-chain}
 \underbrace{\mathcal{F}_2\otimes\mathcal{F}_1\otimes\dots\otimes\mathcal{F}_2\otimes\mathcal{F}_1}_{2n} ~.
\end{equation}
Therefore, the screening charge corresponding to $\boldsymbol{\alpha}_1$ can be understood as an integrable conformal perturbation.
One can show that the corresponding $W$-algebra will have a non-vanishing current of spin $4$.
The root $\boldsymbol{\alpha}_1$ plays the role of a ``boundary condition'' for the spin chain built on \eqref{alternating-spin-chain}.
All such boundary conditions for $Y\big(\widehat{\mathfrak{gl}}(1)\big)$ have been recently classified in \cite{feigin2020deformations}.

From symmetry considerations, it is clear that one can also perturb by the screening charge corresponding to the root
\begin{equation}\label{alpha-(2n+1)}
 \boldsymbol{\alpha}_{2n+1}=-b\boldsymbol{E}_{n}-i\beta\boldsymbol{e}_{n} ~,
\end{equation}
thus obtaining a similar $W$-algebra with $b\rightarrow i\beta$. Perturbing by both roots
\begin{equation}\label{diagram-O(2n+1)-affine}
\begin{picture}(300,90)(275,88)
    \Thicklines
    \unitlength 5pt
    \put(48,32){\circle{2}}
    \put(48,18){\circle{2}}
    \put(54.4,24,4){\line(-1,-1){7}}
    \put(54.4,25,6){\line(-1,1){7}}
    \put(47.4,31.4){\line(1,1){1.2}}
    \put(47.4,18.6){\line(1,-1){1.2}}
    \put(48,19){\line(0,1){12}}
    \put(55,25){\circle{2}}
    \put(54.4,24,4){\line(1,1){1.2}}
    \put(54.4,25,6){\line(1,-1){1.2}}
    \put(65.5,25){\line(1,0){7.5}}
    \put(56,25){\line(1,0){7.5}}
    \put(64.5,25){\circle{2}}
    \put(63.9,24,4){\line(1,1){1.2}}
    \put(63.9,25,6){\line(1,-1){1.2}}
    \put(74,25){\circle{2}}
    \put(73.4,24,4){\line(1,1){1.2}}
    \put(73.4,25,6){\line(1,-1){1.2}}
    \put(75,25){\line(1,0){2}}
    \put(79,25){\circle{.2}}
    \put(80,25){\circle{.2}}
    \put(81,25){\circle{.2}}
    \put(82,25){\circle{.2}}
    \put(83,25){\circle{.2}}
    \put(84,25){\circle{.2}}
    \put(86,25){\line(1,0){2}}
    \put(89,25){\circle{2}}
    \put(88.4,24,4){\line(1,1){1.2}}
    \put(88.4,25,6){\line(1,-1){1.2}}
    \put(90,25){\line(1,0){7.5}}
    \put(98.5,25){\circle{2}}
    \put(97.9,24,4){\line(1,1){1.2}}
    \put(97.9,25,6){\line(1,-1){1.2}}
    \put(99.5,25){\line(1,0){7.5}}
    \put(108,25){\circle{2}}
    \put(107.4,24,4){\line(1,1){1.2}}
    \put(107.4,25,6){\line(1,-1){1.2}}
    \put(107.4,24,4){\line(1,1){8.2}}
    \put(107.4,25,6){\line(1,-1){8.2}}
    \put(115,32){\circle{2}}
    \put(115,18){\circle{2}}
    \put(114.4,32.6){\line(1,-1){1.2}}
    \put(114.4,17.4){\line(1,1){1.2}}
    \put(115,19){\line(0,1){12}}
    \put(50.5,19){$\scriptstyle{-b^{2}}$}
    \put(50.5,30){$\scriptstyle{-b^{2}}$}
    \put(42,24.7){$\scriptstyle{1+2b^{2}}$}
    \put(58.25,22.5){$\scriptstyle{1+b^{2}}$}
    \put(67.75,22.5){$\scriptstyle{-b^{2}}$}
    \put(101.75,22.5){$\scriptstyle{-b^{2}}$}
    \put(92.25,22.5){$\scriptstyle{1+b^{2}}$}
    \put(108,19){$\scriptstyle{1+b^{2}}$}
    \put(108,30){$\scriptstyle{1+b^{2}}$}
    \put(115.5,24.7){$\scriptstyle{-1-2b^{2}}$}
    \put(45.5,16){$\scriptstyle{\boldsymbol{\alpha}_{1}}$}
    \put(45.5,33.5){$\scriptstyle{\boldsymbol{\alpha}_{2}}$}
     \put(54.4,27){$\scriptstyle{\boldsymbol{\alpha}_{3}}$}
     \put(63.6,27){$\scriptstyle{\boldsymbol{\alpha}_{4}}$}
     \put(73.1,27){$\scriptstyle{\boldsymbol{\alpha}_{5}}$}
     \put(95.7,27){$\scriptstyle{\boldsymbol{\alpha}_{2n-2}}$}
     \put(103.9,27){$\scriptstyle{\boldsymbol{\alpha}_{2n-1}}$}
    \put(116,16){$\scriptstyle{\boldsymbol{\alpha}_{2n+1}}$}
    \put(116,33.5){$\scriptstyle{\boldsymbol{\alpha}_{2n}}$}
  \end{picture}
  \vspace*{1cm}
\end{equation}
breaks the conformal symmetry, but still preserves an infinite tower of integrals of motion (see \cite{Litvinov:2018bou} for explicit expressions).

The prescription to go from $O(N)$ to $OSP(N|2m)$ at the level of screening charges in \eqref{diagram-O(2n+1)} is to drop the root $\boldsymbol{\alpha}_1$ and add $m$ Yangian representations $\mathcal{F}_3$ to \eqref{alternating-spin-chain}.
One way to do this is
\begin{equation}\label{alternating-spin-chain-F3}
\underbrace{\mathcal{F}_2\otimes\mathcal{F}_1\otimes\dots\otimes\mathcal{F}_2\otimes\mathcal{F}_1}_{2n}\,\,\longrightarrow\,\,
\underbrace{\mathcal{F}_2\otimes\mathcal{F}_1\otimes\dots\otimes\mathcal{F}_2\otimes\mathcal{F}_1}_{2n}\otimes\underbrace{\mathcal{F}_3\otimes\dots\otimes\mathcal{F}_3}_{m} ~.
\end{equation}
Then, in agreement with \cite{feigin2020deformations}, one can show that $\boldsymbol{\alpha}_1$ still defines a conformal integrable perturbation.
The central charge of the corresponding CFT is given by \eqref{c-O(2n+1)} with $n\rightarrow n-m$.

The ordering in \eqref{alternating-spin-chain-F3} does not lead to a QFT with a well-defined weak-coupling description.
By trial and error, we have found that ``good'' QFTs are obtained by the using the transformation $\mathfrak{J}$, which we call \emph{injection}, that acts as
\begin{equation}\label{Injection-action-on-F}
  \mathcal{F}_1\otimes\mathcal{F}_2\,\,\overset{\mathfrak{J}}{\longrightarrow}\,\,
  \mathcal{F}_1\otimes\mathcal{F}_3\otimes\mathcal{F}_2 ~.
\end{equation}
The injection transformation can be applied to both conformal diagram and its affine counterpart, meaning that the affine perturbation remains the same.
It acts on any root $\boldsymbol{\alpha}_{2k+1}$, $k=1,\dots,n-1$, in \eqref{diagram-O(2n+1)-affine} producing two fermionic roots from one.
On a general fermionic root $\boldsymbol{\alpha}$ it acts as
\begin{equation}\label{blow-up-transformation}
   \boldsymbol{\alpha}=-b\boldsymbol{E}+i\beta\boldsymbol{e}\overset{\mathfrak{J}_{\boldsymbol{\alpha}}}{\longrightarrow} \{\boldsymbol{\beta}_{1},\boldsymbol{\beta}_{2}\}=
   \Big\{-\frac{1}{b}\boldsymbol{E}+\frac{i\beta}{b}\boldsymbol{\epsilon},\frac{ib}{\beta}\boldsymbol{\epsilon}-\frac{i}{\beta}\boldsymbol{e}\Big\}~,
\end{equation}
where $\boldsymbol{\epsilon}$ is a new basis vector.
The two new roots $\boldsymbol{\beta}_{1}$ and $\boldsymbol{\beta}_{2}$ have scalar product $-1$ and hence there is an associated dressed fermionic screening with charge $\boldsymbol{\alpha}$.
If fermionic root $\boldsymbol{\alpha}$ is connected to the roots $\boldsymbol{\alpha}_{\pm}$, where
\begin{equation}
    \boldsymbol{\alpha}_{-}=b\boldsymbol{E}-i\beta\boldsymbol{e}_{-}~,\qquad
    \boldsymbol{\alpha}_{+}=b\boldsymbol{E}_{+}-i\beta\boldsymbol{e}~,
\end{equation}
then each pair $(\boldsymbol{\alpha}_{-},\boldsymbol{\beta}_{1})$ and $(\boldsymbol{\beta}_{1},\boldsymbol{\alpha}_{+})$ also has scalar product $-1$ and hence each has a corresponding dressed fermionic screening.
Diagrammatically, this can be shown as
\begin{equation}\label{Blow-up-picture}
\begin{picture}(300,40)(250,122)
    \Thicklines
    \unitlength 5pt 
    \put(45,25){\circle{2}}
    \put(55,25){\circle{2}}
    \put(46,25){\line(1,0){8}}
    \put(44.4,24,4){\line(1,1){1.2}}
    \put(44.4,25,6){\line(1,-1){1.2}}
    \put(54.4,24,4){\line(1,1){1.2}}
    \put(54.4,25,6){\line(1,-1){1.2}}
    \put(64.4,24,4){\line(1,1){1.2}}
    \put(64.4,25,6){\line(1,-1){1.2}}
    \put(65,25){\circle{2}}
    \put(56,25){\line(1,0){8}}
    \put(48.8,22){$\scriptstyle{-b^{2}}$}
    \put(58.5,22){$\scriptstyle{1+b^{2}}$}
    \put(44.3,22){$\scriptstyle{\boldsymbol{\alpha}_{-}}$}
    \put(54.5,22){$\scriptstyle{\boldsymbol{\alpha}}$}
    \put(64.3,22){$\scriptstyle{\boldsymbol{\alpha}_{+}}$}
     \put(71,24.5){$\overset{\mathfrak{J}_{\boldsymbol{\alpha}}}{\longrightarrow}$}
     \put(80,25){\circle{2}}
     \put(79.4,24,4){\line(1,1){1.2}}
     \put(79.4,25,6){\line(1,-1){1.2}}
     \put(81,25){\line(1,0){1}}
     \put(82.4,25){\line(1,0){1}}
     \put(83.8,25){\line(1,0){1}}
     \put(85.2,25){\line(1,0){1}}
     \put(86.6,25){\line(1,0){1}}
     \put(88,25){\line(1,0){1}}
     \put(90,25){\circle{2}}
     \put(89.4,24,4){\line(1,1){1.2}}
     \put(89.4,25,6){\line(1,-1){1.2}}
     \put(91,25){\line(1,0){1}}
     \put(92.4,25){\line(1,0){1}}
     \put(93.8,25){\line(1,0){1}}
     \put(95.2,25){\line(1,0){1}}
     \put(96.6,25){\line(1,0){1}}
     \put(98,25){\line(1,0){1}}
     \put(100,25){\circle{2}}
     \put(99.4,24,4){\line(1,1){1.2}}
     \put(99.4,25,6){\line(1,-1){1.2}}
     \put(101,25){\line(1,0){1}}
     \put(102.4,25){\line(1,0){1}}
     \put(103.8,25){\line(1,0){1}}
     \put(105.2,25){\line(1,0){1}}
     \put(106.6,25){\line(1,0){1}}
     \put(108,25){\line(1,0){1}}
     \put(110,25){\circle{2}}
     \put(109.4,24,4){\line(1,1){1.2}}
     \put(109.4,25,6){\line(1,-1){1.2}} 
     \put(79.3,22){$\scriptstyle{\boldsymbol{\alpha}_{-}}$}
    \put(89.4,22){$\scriptstyle{\boldsymbol{\beta}_{1}}$}
     \put(99.4,22){$\scriptstyle{\boldsymbol{\beta}_{2}}$}
    \put(109.3,22){$\scriptstyle{\boldsymbol{\alpha}_{+}}$}
    \put(60,30){\circle{2}}
    \put(59.4,31.5){\line(1,0){1.2}}
    \put(59.3,27){$\scriptstyle{\boldsymbol{\beta}}$}
    \put(85,30){\circle{2}}
    \put(84.4,31.5){\line(1,0){1.2}}
    \put(84.4,29,4){\line(1,1){1.2}}
    \put(84.4,30,6){\line(1,-1){1.2}}
    \put(84.3,27){$\scriptstyle{\boldsymbol{\beta}_{-}}$}
    \put(95,30){\circle{2}}
    \put(94.4,31.5){\line(1,0){1.2}}
    \put(94.4,29,4){\line(1,1){1.2}}
    \put(94.4,30,6){\line(1,-1){1.2}}
    \put(94.4,27){$\scriptstyle{\boldsymbol{\alpha}}$}
    \put(105,30){\circle{2}}
    \put(104.4,31.5){\line(1,0){1.2}}
    \put(104.4,29,4){\line(1,1){1.2}}
    \put(104.4,30,6){\line(1,-1){1.2}}
    \put(104.3,27){$\scriptstyle{\boldsymbol{\beta}_{+}}$}
  \end{picture}
\end{equation}
with
\begin{equation}\begin{split}
   \boldsymbol{\beta}_{-}& =-\frac{1}{1+b^{2}}\left(\boldsymbol{\alpha}_{-}+b^{2}\boldsymbol{\beta}_{1}\right)=\frac{i}{\beta}\boldsymbol{e}_{-}-\frac{ib}{\beta}\boldsymbol{\epsilon},\qquad
\\   \boldsymbol{\beta}_{+}&=\frac{1}{b^{2}}\left(\boldsymbol{\alpha}_{+}-(1+b^{2})\boldsymbol{\beta}_{2}\right)=\frac{1}{b}\boldsymbol{E}_{+}-\frac{i\beta}{b}\boldsymbol{\epsilon}.
\end{split}\end{equation}
In \eqref{Blow-up-picture} we have also shown the dressed root $\boldsymbol{\beta}=\frac{\boldsymbol{E}_{+}-\boldsymbol{E}}{b}$ which will play an important role in what follows.

It will also be important to understand how the injection $\mathfrak{J}$ acts on the weak-coupling ($b\rightarrow0$) and the strong-coupling ($b\rightarrow\infty$) screening charges
\begin{equation}
  \begin{aligned}
     &\text{weak coupling:}\qquad&& e^{-b\Phi+i\beta\varphi}\ \overset{\mathfrak{J}}{\longrightarrow} \ e^{-b\Phi+i\beta\varphi}\Big(\frac{1}{b}\partial\Phi-\frac{i\beta}{b}\partial\phi\Big)\\
     &\text{strong coupling:}\qquad&& e^{\frac{\Phi_{+}-\Phi}{b}}\left(b\partial\Phi_{+}-i\beta\partial\varphi\right) \ \overset{\mathfrak{J}}{\longrightarrow} \ \Big\{e^{-\frac{\Phi}{b}+\frac{i\beta}{b}\phi},e^{\frac{\Phi_{+}}{b}-\frac{i\beta}{b}\phi}\left(b\partial\Phi_{+}-i\beta\partial\varphi\right)\Big\}
  \end{aligned}
\end{equation}

To conclude this section, let us note that, as the action of the injection $\mathfrak{J}$ has been defined \eqref{Injection-action-on-F}, for $N=2n+1$ one can inject at most $n-1$ representations of the type $\mathcal{F}_3$, i.e. one is restricted to $OSP(N|2m)$ with $m<n$.
Thus the corresponding QFT always stays in asymptotically free regime.
\section{QFTs from screenings}\label{QFT}
In the previous section we constructed the system of screening charges for the $OSP(N|2m)$ model with $N=2n+1$ and $m<n$, i.e. in the asymptotically free region.
To do so we defined the injection transformation $\mathfrak{J}$, which when applied $m$ times to the $O(N)$ diagram \eqref{diagram-O(2n+1)-affine} gives a new diagram from which the screening charges can be read off.
In order to illustrate this procedure, let us consider various examples.
There is one diagram for the $OSP(5|2)$ theory, which is obtained by the injection of the root $\boldsymbol{\alpha}_3$ in the $O(5)$ diagram 
\begin{equation}\label{OSP(5|2)-diagram}
\begin{picture}(200,100)(125,80)
\Thicklines
\unitlength 5pt
\put(32,32){\circle{2}}
\put(32,18){\circle{2}}
\put(31.4,31.4){\line(1,1){1.2}}
\put(31.4,32.6){\line(1,-1){1.2}}
\put(31.4,17.4){\line(1,1){1.2}}
\put(31.4,18.6){\line(1,-1){1.2}}
\put(32,19){\line(0,1){12}}
\put(38.4,24,4){\line(-1,-1){0.6}}
\put(37.4,23,4){\line(-1,-1){0.6}}
\put(36.4,22,4){\line(-1,-1){0.6}}
\put(35.4,21,4){\line(-1,-1){0.6}}
\put(34.4,20,4){\line(-1,-1){0.6}}
\put(33.4,19,4){\line(-1,-1){0.6}}
\put(38.4,25,6){\line(-1,1){0.6}}
\put(37.4,26,6){\line(-1,1){0.6}}
\put(36.4,27,6){\line(-1,1){0.6}}
\put(35.4,28,6){\line(-1,1){0.6}}
\put(34.4,29,6){\line(-1,1){0.6}}
\put(33.4,30,6){\line(-1,1){0.6}}
\put(39,25){\circle{2}}
\put(38.4,24,4){\line(1,1){1.2}}
\put(38.4,25,6){\line(1,-1){1.2}}
\put(40,25){\line(1,0){1}}
\put(41.4,25){\line(1,0){1}}
\put(42.8,25){\line(1,0){1}}
\put(44.2,25){\line(1,0){1}}
\put(45.6,25){\line(1,0){1}}
\put(47,25){\line(1,0){1}}
\put(49,25){\circle{2}}
\put(48.4,24,4){\line(1,1){1.2}}
\put(48.4,25,6){\line(1,-1){1.2}}
\put(28,24.8){\circle{2}}
\put(27.4,26.3){\line(1,0){1.2}}
\put(27,22){$\scriptstyle{\boldsymbol{\beta}_{12}}$}
\put(60.1,24.8){\circle{2}}
\put(59.5,26.3){\line(1,0){1.2}}
\put(59.1,22){$\scriptstyle{\boldsymbol{\beta}_{45}}$}
\put(29.5,16){$\scriptstyle{\boldsymbol{\alpha}_{1}}$}
\put(29.5,33.5){$\scriptstyle{\boldsymbol{\alpha}_{2}}$}
\put(56.5,16){$\scriptstyle{\boldsymbol{\alpha}_{4}}$}
\put(56.5,33.5){$\scriptstyle{\boldsymbol{\alpha}_{5}}$}
\put(38.3,22){$\scriptstyle{\boldsymbol{\beta}_{1}}$}
\put(48.3,22){$\scriptstyle{\boldsymbol{\beta}_{2}}$}
\put(37.5,30.5){\circle{2}}
\put(36.9,32){\line(1,0){1.2}}
\put(36.9,29,9){\line(1,1){1.2}}
\put(36.9,31.1){\line(1,-1){1.2}}
\put(37.7,27.7){$\scriptstyle{\boldsymbol{\beta}_{-}^{+}}$}
\put(50.8,30.5){\circle{2}}
\put(50.2,32){\line(1,0){1.2}}
\put(50.2,29,9){\line(1,1){1.2}}
\put(50.2,31.1){\line(1,-1){1.2}}
\put(48.8,27.7){$\scriptstyle{\boldsymbol{\beta}_{+}^{+}}$}
\put(50.8,20.5){\circle{2}}
\put(50.2,22){\line(1,0){1.2}}
\put(50.2,19,9){\line(1,1){1.2}}
\put(50.2,21.1){\line(1,-1){1.2}}
\put(48.8,17.7){$\scriptstyle{\boldsymbol{\beta}_{+}^{-}}$}
\put(37.5,20.5){\circle{2}}
\put(36.9,22){\line(1,0){1.2}}
\put(36.9,19,9){\line(1,1){1.2}}
\put(36.9,21.1){\line(1,-1){1.2}}
\put(37.7,17.7){$\scriptstyle{\boldsymbol{\beta}_{-}^{-}}$}
\put(44,30){\circle{2}}
\put(43.4,31.5){\line(1,0){1.2}}
\put(43.4,29,4){\line(1,1){1.2}}
\put(43.4,30,6){\line(1,-1){1.2}}
\put(43.4,27){$\scriptstyle{\boldsymbol{\alpha}_{3}}$}
\put(56,32){\circle{2}}
\put(56,18){\circle{2}}
\put(55.4,31.4){\line(1,1){1.2}}
\put(55.4,32.6){\line(1,-1){1.2}}
\put(55.4,17.4){\line(1,1){1.2}}
\put(55.4,18.6){\line(1,-1){1.2}}
\put(56,19){\line(0,1){12}}
\put(49.6,24,4){\line(1,-1){0.6}}
\put(50.6,23,4){\line(1,-1){0.6}}
\put(51.6,22,4){\line(1,-1){0.6}}
\put(52.6,21,4){\line(1,-1){0.6}}
\put(53.6,20,4){\line(1,-1){0.6}}
\put(54.6,19,4){\line(1,-1){0.6}}
\put(49.6,25,6){\line(1,1){0.6}}
\put(50.6,26,6){\line(1,1){0.6}}
\put(51.6,27,6){\line(1,1){0.6}}
\put(52.6,28,6){\line(1,1){0.6}}
\put(53.6,29,6){\line(1,1){0.6}}
\put(54.6,30,6){\line(1,1){0.6}}
\end{picture}
\end{equation}
where $\boldsymbol{\alpha}_r$, $r=1,\dots,5$, are given by \eqref{vectors-O(2n+1)-conformal} and
\begin{equation}\label{alpha-for-OSP(5|2)}
\begin{gathered}
\boldsymbol{\beta}_{1}=-\frac{1}{b}\boldsymbol{E}_{1}+\frac{i\beta}{b}\boldsymbol{\epsilon}~,\qquad
\boldsymbol{\beta}_{2}=\frac{ib}{\beta}\boldsymbol{\epsilon}-\frac{i}{\beta}\boldsymbol{e}_{2}~,\qquad
\boldsymbol{\beta}_{12}=\frac{1}{b}\boldsymbol{E}_{1}~,\qquad
\boldsymbol{\beta}_{45}=\frac{i}{\beta}\boldsymbol{e}_{2}~,
\\
\boldsymbol{\beta}_{-}^{\pm}=\pm\frac{i}{\beta}\boldsymbol{e}_{1}-\frac{ib}{\beta}\boldsymbol{\epsilon}~,\qquad
\boldsymbol{\beta}_{+}^{\pm}=\pm\frac{1}{b}\boldsymbol{E}_{2}-\frac{i\beta}{b}\boldsymbol{\epsilon}~.
\end{gathered}
\end{equation}
For $OSP(7|2)$ there are two diagrams. The first corresponds to the injection of $\boldsymbol{\alpha}_3$
\begin{equation}\label{OSP(7|2)-diagram-1}
\begin{picture}(200,100)(165,80)
\Thicklines
\unitlength 5pt
\put(32,32){\circle{2}}
\put(32,18){\circle{2}}
\put(31.4,31.4){\line(1,1){1.2}}
\put(31.4,32.6){\line(1,-1){1.2}}
\put(31.4,17.4){\line(1,1){1.2}}
\put(31.4,18.6){\line(1,-1){1.2}}
\put(32,19){\line(0,1){12}}
\put(38.4,24,4){\line(-1,-1){0.6}}
\put(37.4,23,4){\line(-1,-1){0.6}}
\put(36.4,22,4){\line(-1,-1){0.6}}
\put(35.4,21,4){\line(-1,-1){0.6}}
\put(34.4,20,4){\line(-1,-1){0.6}}
\put(33.4,19,4){\line(-1,-1){0.6}}
\put(38.4,25,6){\line(-1,1){0.6}}
\put(37.4,26,6){\line(-1,1){0.6}}
\put(36.4,27,6){\line(-1,1){0.6}}
\put(35.4,28,6){\line(-1,1){0.6}}
\put(34.4,29,6){\line(-1,1){0.6}}
\put(33.4,30,6){\line(-1,1){0.6}}
\put(39,25){\circle{2}}
\put(38.4,24,4){\line(1,1){1.2}}
\put(38.4,25,6){\line(1,-1){1.2}}
\put(40,25){\line(1,0){1}}
\put(41.4,25){\line(1,0){1}}
\put(42.8,25){\line(1,0){1}}
\put(44.2,25){\line(1,0){1}}
\put(45.6,25){\line(1,0){1}}
\put(47,25){\line(1,0){1}}
\put(49,25){\circle{2}}
\put(48.4,24,4){\line(1,1){1.2}}
\put(48.4,25,6){\line(1,-1){1.2}}
\put(28,24.8){\circle{2}}
\put(27.4,26.3){\line(1,0){1.2}}
\put(27,22){$\scriptstyle{\boldsymbol{\beta}_{12}}$}
\put(29.5,16){$\scriptstyle{\boldsymbol{\alpha}_{1}}$}
\put(29.5,33.5){$\scriptstyle{\boldsymbol{\alpha}_{2}}$}
\put(38.3,22){$\scriptstyle{\boldsymbol{\beta}_{1}}$}
\put(48.3,22){$\scriptstyle{\boldsymbol{\beta}_{2}}$}
\put(37.5,30.5){\circle{2}}
\put(36.9,32){\line(1,0){1.2}}
\put(36.9,29,9){\line(1,1){1.2}}
\put(36.9,31.1){\line(1,-1){1.2}}
\put(37.7,27.7){$\scriptstyle{\boldsymbol{\beta}_{-}^{+}}$}
\put(37.5,20.5){\circle{2}}
\put(36.9,22){\line(1,0){1.2}}
\put(36.9,19,9){\line(1,1){1.2}}
\put(36.9,21.1){\line(1,-1){1.2}}
\put(37.7,17.7){$\scriptstyle{\boldsymbol{\beta}_{-}^{-}}$}
\put(44,30){\circle{2}}
\put(43.4,31.5){\line(1,0){1.2}}
\put(43.4,29,4){\line(1,1){1.2}}
\put(43.4,30,6){\line(1,-1){1.2}}
\put(43.4,27){$\scriptstyle{\boldsymbol{\alpha}_{3}}$}
\put(50,25){\line(1,0){1}}
\put(51.4,25){\line(1,0){1}}
\put(52.8,25){\line(1,0){1}}
\put(54.2,25){\line(1,0){1}}
\put(55.6,25){\line(1,0){1}}
\put(57,25){\line(1,0){1}}
\put(59,25){\circle{2}}
\put(58.4,24,4){\line(1,1){1.2}}
\put(58.4,25,6){\line(1,-1){1.2}}
\put(54,30){\circle{2}}
\put(53.4,31.5){\line(1,0){1.2}}
\put(53.4,29,4){\line(1,1){1.2}}
\put(53.4,30,6){\line(1,-1){1.2}}
\put(53.4,27){$\scriptstyle{\boldsymbol{\beta}_{+}}$}
\put(58.3,22){$\scriptstyle{\boldsymbol{\alpha}_{4}}$}
\put(60,25){\line(1,0){8}}
\put(69,25){\circle{2}}
\put(68.4,24,4){\line(1,1){1.2}}
\put(68.4,25,6){\line(1,-1){1.2}}
\put(64,30){\circle{2}}
\put(63.4,31.5){\line(1,0){1.2}}
\put(63.4,27){$\scriptstyle{\boldsymbol{\beta}_{45}}$}
\put(68.3,22){$\scriptstyle{\boldsymbol{\alpha}_{5}}$}
\put(68.4,24,4){\line(1,1){8.2}}
\put(68.4,25,6){\line(1,-1){8.2}}
\put(76,32){\circle{2}}
\put(76,18){\circle{2}}
\put(75.4,32.6){\line(1,-1){1.2}}
\put(75.4,17.4){\line(1,1){1.2}}
\put(76,19){\line(0,1){12}}
\put(76.5,16){$\scriptstyle{\boldsymbol{\alpha}_{7}}$}
\put(76.5,33.5){$\scriptstyle{\boldsymbol{\alpha}_{6}}$}
\put(70.5,30.5){\circle{2}}
\put(69.9,32){\line(1,0){1.2}}
\put(68.7,27.7){$\scriptstyle{\boldsymbol{\beta}_{56}}$}
\put(71,20){\circle{2}}
\put(70.4,21.5){\line(1,0){1.2}}
\put(68.7,17.7){$\scriptstyle{\boldsymbol{\beta}_{57}}$}
\put(80,24.8){\circle{2}}
\put(79.4,26.3){\line(1,0){1.2}}
\put(79,22){$\scriptstyle{\boldsymbol{\beta}_{67}}$}
\end{picture}
\end{equation}
with
\begin{equation}\label{alpha-for-OSP(7|2)-1}
\begin{gathered}
\boldsymbol{\beta}_{1}=-\frac{1}{b}\boldsymbol{E}_{1}+\frac{i\beta}{b}\boldsymbol{\epsilon}~,\qquad
\boldsymbol{\beta}_{2}=\frac{ib}{\beta}\boldsymbol{\epsilon}-\frac{i}{\beta}\boldsymbol{e}_{2}~,\qquad
\boldsymbol{\beta}_{12}=\frac{1}{b}\boldsymbol{E}_{1}~,\qquad
\boldsymbol{\beta}_{67}=\frac{i}{\beta}\boldsymbol{e}_{3}~,
\\
\boldsymbol{\beta}_{-}^{\pm}=\pm\frac{i}{\beta}\boldsymbol{e}_{1}-\frac{ib}{\beta}\boldsymbol{\epsilon}~,\qquad
\boldsymbol{\beta}_{+}=\frac{1}{b}\boldsymbol{E}_{2}-\frac{i\beta}{b}\boldsymbol{\epsilon}~,
\\
\boldsymbol{\beta}_{57}=-\frac{1}{b}(\boldsymbol{E}_{2}+\boldsymbol{E}_{3})~,\qquad
\boldsymbol{\beta}_{45}=\frac{i}{\beta}(\boldsymbol{e}_{2}-\boldsymbol{e}_{3})~,\qquad
\boldsymbol{\beta}_{56}=\frac{1}{b}(\boldsymbol{E}_{3}-\boldsymbol{E}_{2})~,
\end{gathered}
\end{equation}
and the second to the injection of $\boldsymbol{\alpha}_5$
\begin{equation}\label{OSP(7|2)-diagram-3}
\begin{picture}(200,100)(165,80)
\Thicklines
\unitlength 5pt
\put(32,32){\circle{2}}
\put(32,18){\circle{2}}
\put(31.4,31.4){\line(1,1){1.2}}
\put(31.4,32.6){\line(1,-1){1.2}}
\put(31.4,17.4){\line(1,1){1.2}}
\put(31.4,18.6){\line(1,-1){1.2}}
\put(32,19){\line(0,1){12}}
\put(38.4,24,4){\line(-1,-1){7}}
\put(38.4,25,6){\line(-1,1){7}}
\put(39,25){\circle{2}}
\put(38.4,24,4){\line(1,1){1.2}}
\put(38.4,25,6){\line(1,-1){1.2}}
\put(40,25){\line(1,0){8}}
\put(49,25){\circle{2}}
\put(48.4,24,4){\line(1,1){1.2}}
\put(48.4,25,6){\line(1,-1){1.2}}
\put(28,24.8){\circle{2}}
\put(27.4,26.3){\line(1,0){1.2}}
\put(27,22){$\scriptstyle{\boldsymbol{\beta}_{12}}$}
\put(29.5,16){$\scriptstyle{\boldsymbol{\alpha}_{1}}$}
\put(29.5,33.5){$\scriptstyle{\boldsymbol{\alpha}_{2}}$}
\put(38.3,22){$\scriptstyle{\boldsymbol{\alpha}_{3}}$}
\put(48.3,22){$\scriptstyle{\boldsymbol{\alpha}_{4}}$}
\put(37.5,30.5){\circle{2}}
\put(36.9,32){\line(1,0){1.2}}
\put(37.7,27.7){$\scriptstyle{\boldsymbol{\beta}_{23}}$}
\put(37.5,20.5){\circle{2}}
\put(36.9,22){\line(1,0){1.2}}
\put(37.7,17.7){$\scriptstyle{\boldsymbol{\beta}_{13}}$}
\put(44,30){\circle{2}}
\put(43.4,31.5){\line(1,0){1.2}}
\put(43.4,27){$\scriptstyle{\boldsymbol{\beta}_{34}}$}
\put(50,25){\line(1,0){1}}
\put(51.4,25){\line(1,0){1}}
\put(52.8,25){\line(1,0){1}}
\put(54.2,25){\line(1,0){1}}
\put(55.6,25){\line(1,0){1}}
\put(57,25){\line(1,0){1}}
\put(59,25){\circle{2}}
\put(58.4,24,4){\line(1,1){1.2}}
\put(58.4,25,6){\line(1,-1){1.2}}
\put(54,30){\circle{2}}
\put(53.4,31.5){\line(1,0){1.2}}
\put(53.4,29,4){\line(1,1){1.2}}
\put(53.4,30,6){\line(1,-1){1.2}}
\put(53.4,27){$\scriptstyle{\boldsymbol{\beta}_{-}}$}
\put(58.3,22){$\scriptstyle{\boldsymbol{\beta}_{1}}$}
\put(60,25){\line(1,0){1}}
\put(61.4,25){\line(1,0){1}}
\put(62.8,25){\line(1,0){1}}
\put(64.2,25){\line(1,0){1}}
\put(65.6,25){\line(1,0){1}}
\put(67,25){\line(1,0){1}}
\put(69,25){\circle{2}}
\put(68.4,24,4){\line(1,1){1.2}}
\put(68.4,25,6){\line(1,-1){1.2}}
\put(64,30){\circle{2}}
\put(63.4,31.5){\line(1,0){1.2}}
\put(63.4,29,4){\line(1,1){1.2}}
\put(63.4,30,6){\line(1,-1){1.2}}
\put(63.4,27){$\scriptstyle{\boldsymbol{\alpha}_{5}}$}
\put(68.3,22){$\scriptstyle{\boldsymbol{\beta}_{2}}$}
\put(68.4,24,4){\line(1,1){0.6}}
\put(69.4,25,4){\line(1,1){0.6}}
\put(70.4,26,4){\line(1,1){0.6}}
\put(71.4,27,4){\line(1,1){0.6}}
\put(72.4,28,4){\line(1,1){0.6}}
\put(73.4,29,4){\line(1,1){0.6}}
\put(74.4,30,4){\line(1,1){0.6}}
\put(75.4,31,4){\line(1,1){1.2}}
\put(68.4,25,6){\line(1,-1){0.6}}
\put(69.4,24,6){\line(1,-1){0.6}}
\put(70.4,23,6){\line(1,-1){0.6}}
\put(71.4,22,6){\line(1,-1){0.6}}
\put(72.4,21,6){\line(1,-1){0.6}}
\put(73.4,20,6){\line(1,-1){0.6}}
\put(74.4,19,6){\line(1,-1){0.6}}
\put(75.4,18,6){\line(1,-1){1.2}}
\put(76,32){\circle{2}}
\put(76,18){\circle{2}}
\put(75.4,32.6){\line(1,-1){1.2}}
\put(75.4,17.4){\line(1,1){1.2}}
\put(76,19){\line(0,1){12}}
\put(76.5,16){$\scriptstyle{\boldsymbol{\alpha}_{7}}$}
\put(76.5,33.5){$\scriptstyle{\boldsymbol{\alpha}_{6}}$}
\put(70.5,30.5){\circle{2}}
\put(69.9,32){\line(1,0){1.2}}
\put(69.9,29.9){\line(1,1){1.2}}
\put(69.9,31.1){\line(1,-1){1.2}}
\put(68.7,27.7){$\scriptstyle{\boldsymbol{\beta}_{+}^{+}}$}
\put(71,20){\circle{2}}
\put(70.4,21.5){\line(1,0){1.2}}
\put(70.4,19.4){\line(1,1){1.2}}
\put(70.4,20.6){\line(1,-1){1.2}}
\put(68.7,17.7){$\scriptstyle{\boldsymbol{\beta}_{+}^{-}}$}
\put(80,24.8){\circle{2}}
\put(79.4,26.3){\line(1,0){1.2}}
\put(79,22){$\scriptstyle{\boldsymbol{\beta}_{67}}$}
\end{picture}
\end{equation}
with
\begin{equation}\label{alpha-for-OSP(7|2)-3}
\begin{gathered}
\boldsymbol{\beta}_{1}=-\frac{1}{b}\boldsymbol{E}_{2}+\frac{i\beta}{b}\boldsymbol{\epsilon}~,\qquad
\boldsymbol{\beta}_{2}=\frac{ib}{\beta}\boldsymbol{\epsilon}-\frac{i}{\beta}\boldsymbol{e}_3~,\qquad
\boldsymbol{\beta}_{12}=\frac{1}{b}\boldsymbol{E}_{1}~,\qquad
\boldsymbol{\beta}_{67}=\frac{i}{\beta}\boldsymbol{e}_{3}~,
\\
\boldsymbol{\beta}_{-}=\frac{i}{\beta}\boldsymbol{e}_{2}-\frac{ib}{\beta}\boldsymbol{\epsilon}~,\qquad
\boldsymbol{\beta}_{+}^{\pm}=\pm\frac{1}{b}\boldsymbol{E}_{3}-\frac{i\beta}{b}\boldsymbol{\epsilon}~,
\\
\boldsymbol{\beta}_{23}=\frac{i}{\beta}(\boldsymbol{e}_{1}-\boldsymbol{e}_{2})~,\qquad
\boldsymbol{\beta}_{13}=-\frac{i}{\beta}(\boldsymbol{e}_{1}+\boldsymbol{e}_{2})~,\qquad
\boldsymbol{\beta}_{34}=\frac{1}{b}(\boldsymbol{E}_{2}-\boldsymbol{E}_{1})~.
\end{gathered}
\end{equation}

Thus far we have limited ourselves to considering chiral fields.
In order to build a QFT one needs to glue both chiralities in a consistent way.
Since the roots $\boldsymbol{\alpha}_r$ can be complex there are at least two options for the set of roots $\bar{\boldsymbol{\alpha}}_r$ defining the anti-holomorphic screenings
\begin{equation}\label{conjugation-two-options}
   \bar{\boldsymbol{\alpha}}_r=\boldsymbol{\alpha}_r\quad\text{or}\quad\bar{\boldsymbol{\alpha}}_r=\boldsymbol{\alpha}_r^* ~.
\end{equation}
In general, the resulting perturbed model does not define a self-consistent CFT and requires counterterms.
We will see that in the weak-coupling regime $b\rightarrow0$ only finitely many counterterms are needed, whereas in the strong-coupling regime $b\rightarrow\infty$, i.e. the sigma model regime, infinitely many are required.
\subsection{Weak coupling: Toda QFT}\label{weak-coupling-QFT}
In the weak-coupling regime we take $\bar{\boldsymbol{\alpha}}_r=\boldsymbol{\alpha}_r$, i.e. the first option in \eqref{conjugation-two-options}.
For $N>2m+2$ we choose an allowed diagram, such as \eqref{OSP(5|2)-diagram}, \eqref{OSP(7|2)-diagram-1} or  \eqref{OSP(7|2)-diagram-3}, and perturb the free theory by the fields corresponding to roots $\boldsymbol{\alpha}_r$.
These can be either exponentials or dressed exponentials, depending on the particular choice of diagram.
Then using the boson-fermion \cite{Coleman:1974bu,Mandelstam:1975hb} or boson-boson correspondence (see appendix \ref{boson-fermion-correspondence}) we rewrite all the bosonic fields with imaginary exponents or that come from dressed screenings as
\begin{equation}\label{boson-fermion-boson-correspondence}
\begin{aligned}
&\left\{\frac{1}{8\pi}(\partial_\mu\varphi)^2,e^{i\beta\varphi},e^{-i\beta\varphi}\right\}
\rightarrow
\left\{i\bar{\psi}\gamma^{\mu}\partial_\mu\psi+\frac{\pi b^2}{2(1+b^2)}(\bar{\psi}\gamma^{\mu}\psi)^2,\bar{\psi}\gamma_{+}\psi,\bar{\psi}\gamma_{-}\psi\right\}~,
\\
&\left\{\frac{1}{8\pi}(\partial_\mu\Phi)^2,e^{b\Phi},e^{-b\Phi}\big(\partial\Phi-i\beta\partial\phi\big)\big(\bar{\partial}\Phi-i\beta\bar{\partial}\phi\big)\right\}
\\ &\hspace{119pt}
\rightarrow\left\{i\bar{\uppsi}\gamma^{\mu}\partial_\mu\uppsi-\frac{\pi b^2}{2(1+b^2)}(\bar{\uppsi}\gamma^{\mu}\uppsi)^2,\bar{\uppsi}\gamma_{+}\uppsi,\bar{\uppsi}\gamma_{-}\uppsi\right\}~,
\end{aligned}
\end{equation}
where $\psi$ and $\uppsi$ are fermionic and bosonic Dirac spinors respectively and
\begin{equation}
  \gamma_{\pm}=\frac{1\pm\gamma_5}{2} ~.
\end{equation}

In addition to the interaction terms coming from the screening charges, counterterms regularizing the UV behaviour also need to be added.
This amounts to contracting fermionic or bosonic loops
\begin{equation}\begin{gathered}\label{contact-terms}
   \Lambda_1(\bar{\psi}\gamma_{+}\psi) A+\Lambda_2(\bar{\psi}\gamma_{-}\psi)B\longrightarrow \lambda_{\psi}\Lambda_1\Lambda_2 AB~,
\\
   \Lambda_1\bar{\uppsi}\gamma_{+}\uppsi A+\Lambda_2\bar{\uppsi}\gamma_{-}\uppsi B\longrightarrow\lambda_{\uppsi}\Lambda_1\Lambda_2 AB~,
\end{gathered}\end{equation} 
for any two local fields $A$ and $B$.
The precise form of the numerical factors $\lambda_{\psi}$ and $\lambda_{\uppsi}$ in \eqref{contact-terms} depends on the regularization scheme and it is hard to determine them from first principles.
We will leave these coefficients arbitrary and fix them using alternative arguments.
Note that the counterterm $AB$ could also lead to divergent integrals and hence a second generation of counterterms may be required.
Remarkably, one finds that for asymptotically free theories this process terminates and only finitely many counterterms are needed.
Let us illustrate this on the examples of $OSP(5|2)$ and $OSP(7|2)$.

For the $OSP(5|2)$ case the Lagrangian with all possible counterterms takes the form
\begin{equation}\begin{split}\label{OSP(5|2)-Lagrangian}
\mathcal{L}_{\scriptscriptstyle{OSP(5|2)}}
& =\frac{1}{8\pi}(\partial_{\mu}\Phi)^{2}+i\bar{\psi}_{1}\gamma^{\mu}\partial_{\mu}\psi_{1}+i\bar{\psi}_{2}\gamma^{\mu}\partial_{\mu}\psi_{2}+i\bar{\uppsi}\gamma^{\mu}\partial_{\mu}\uppsi
\\ & \ \
+ \frac{\pi b^{2}}{2(1+b^{2})}\Bigl((\bar{\psi}_{1}\gamma^{\mu}\psi_{1})^{2}+
(\bar{\psi}_{2}\gamma^{\mu}\psi_{2})^{2}-(\bar{\uppsi}\gamma^{\mu}\uppsi)^2\Bigr)
\\ & \ \
+ \Lambda_{1}(\bar{\psi}_{1}\psi_{1})(\bar{\uppsi}\gamma_{+}\uppsi)+\Lambda_{2}(\bar{\uppsi}\gamma_{-}\uppsi)(\bar{\psi}_{2}\gamma_{+}\psi_{2})+\Lambda_{3}(\bar{\psi}_{2}\gamma_{-}\psi_{2})\bigl(e^{b\Phi}+e^{-b\Phi}\bigr)
\\ & \ \
+ \lambda_1\Lambda_{1}^{2}(\bar{\uppsi}\gamma_{+}\uppsi)^{2}+\lambda_2\Lambda_{1}\Lambda_{2}(\bar{\psi}_{1}\psi_{1})(\bar{\psi}_{2}\gamma_{+}\psi_{2})
+\lambda_3\Lambda_{2}\Lambda_{3}(\bar{\uppsi}\gamma_{-}\uppsi)\bigl(e^{b\Phi}+e^{-b\Phi}\bigr)\quad \ \
\\ & \ \
+ \lambda_4\Lambda_{1}^{2}\Lambda_{2}(\bar{\uppsi}\gamma_{+}\uppsi)(\bar{\psi}_{2}\gamma_{+}\psi_{2})+
\lambda_5\Lambda_{1}\Lambda_{2}\Lambda_{3}(\bar{\psi}_{1}\psi_{1})\bigl(e^{b\Phi}+e^{-b\Phi}\bigr)
\\ & \ \
+ \lambda_6\Lambda_{1}^{2}\Lambda_{2}\Lambda_{3}(\bar{\uppsi}\gamma_{+}\uppsi)\bigl(e^{b\Phi}+e^{-b\Phi}\bigr)
+ \lambda_7\Lambda_{1}^{2}\Lambda_{2}^2\Lambda_{3}(\bar{\psi}_{2}\gamma_{+}\psi_{2})\bigl(e^{b\Phi}+e^{-b\Phi}\bigr)
\\ & \ \
+ \lambda_8\Lambda_{1}^{2}\Lambda_{2}^{2}\Lambda_{3}^{2}\bigl(e^{b\Phi}+e^{-b\Phi}\bigr)^{2} ~.
\end{split}\end{equation}
Using the freedom to rescale
\begin{equation}
(\bar{\uppsi}\gamma_{\pm}\uppsi)\rightarrow\rho_1^{\pm1}(\bar{\uppsi}\gamma_{\pm}\uppsi)~,\qquad
(\bar{\psi}_{2}\gamma_{\pm}\psi_{2})\rightarrow
\rho_2^{\pm1}(\bar{\psi}_{2}\gamma_{\pm}\psi_{2})~,
\end{equation}
and properly choosing the scheme-dependent parameters $\lambda_k$, we expect that the Lagrangian \eqref{OSP(5|2)-Lagrangian} should describe the weak-coupling expansion of a certain trigonometric $S$-matrix, which will be defined in section \ref{S-matrix}.
Our conjecture is
\vspace*{-0.3cm}
\begin{equation}\begin{split}\label{OSP(5|2)-Lagrangian-scaled}
\mathcal{L}_{\scriptscriptstyle{OSP(5|2)}} & = 
\frac{1}{8\pi}(\partial_{\mu}\Phi)^{2}+i\sum_{k=1}^2\bar{\psi}_{k}\gamma^{\mu}\partial_{\mu}\psi_{k}+i\bar{\uppsi}\gamma^{\mu}\partial_{\mu}\uppsi
\\ & \ \ +\frac{\pi b^{2}}{2(1+b^{2})}\Bigl(\sum_{k=1}^2(\bar{\psi}_{k}\gamma^{\mu}\psi_{k})^{2}-(\bar{\uppsi}\gamma^{\mu}\uppsi)^2\Bigr)
\\ & \ \ 
+
\big(4\pi b^{2}+\dots\big)\Big(\big(\bar{\psi}_{1}\psi_{1}\big)\big(\bar{\uppsi}\gamma_{+}\uppsi\big)+\frac{1}{2}\big(\bar{\uppsi}\gamma_{+}\uppsi\big)^2+
\big(\bar{\psi}_{1}\psi_{1}+\bar{\uppsi}\uppsi\big)\bar{\psi}_{2}\gamma_{+}\psi_{2}\Big)
\\ & \ \ - M\cosh b\Phi\left(\bar{\psi}_{1}\psi_{1}+\bar{\psi}_{2}\psi_{2}+\bar{\uppsi}\uppsi\right)+
\frac{M^{2}}{8\pi b^{2}}\sinh^{2}(b\Phi)~.
\end{split}\end{equation}
The logic leading to \eqref{OSP(5|2)-Lagrangian-scaled} is the following.
The structure of the ``four-spinor'' terms in the second line of \eqref{OSP(5|2)-Lagrangian-scaled} is fixed by the boson-fermion/boson-boson correspondence \eqref{boson-fermion-boson-correspondence}.
The Toda term and Yukawa-like terms in the last line of \eqref{OSP(5|2)-Lagrangian-scaled} are fixed such that the fields $(\psi_1,\uppsi,\psi_2,\Phi,\psi_2^*,\uppsi^*,\psi_1^*)$ have the same mass in the limit $b\rightarrow0$ and form an $OSP(5|2)$ multiplet.
The ``four-spinor'' terms in the third line of \eqref{OSP(5|2)-Lagrangian-scaled} are more subtle.
We have fixed their leading $b\rightarrow0$ behaviour by requiring that the tree-level $S$-matrix satisfies the classical YB equation.
However, we have not found a convincing argument to fix these coefficients from first principles and we cannot exclude the possibility that they can get modified by loop corrections.

Applying a similar logic to the $OSP(7|2)$ case gives two Lagrangians corresponding to the diagrams \eqref{OSP(7|2)-diagram-1}
\vspace*{-0.3cm}
\begin{align}
\mathcal{L}_{\scriptscriptstyle{OSP(7|2)}}^{(1)}
& =\frac{1}{8\pi}\sum_{k=1}^2(\partial_{\mu}\Phi_k)^{2}+i\sum_{k=1}^3\bar{\psi}_{k}\gamma^{\mu}\partial_{\mu}\psi_{k}+i\bar{\uppsi}\gamma^{\mu}\partial_{\mu}\uppsi\notag
\\ & \ \ 
+\frac{\pi b^{2}}{2(1+b^{2})}\Bigl(\sum_{k=1}^3(\bar{\psi}_{k}\gamma^{\mu}\psi_{k})^{2}-(\bar{\uppsi}\gamma^{\mu}\uppsi)^2\Bigr)\notag
\\ & \ \ 
+
\left(4\pi b^{2}+\dots\right)\Big(\big(\bar{\psi}_{1}\psi_{1}\big)\big(\bar{\uppsi}\gamma_{+}\uppsi\big)+\frac{1}{2}\big(\bar{\uppsi}\gamma_{+}\uppsi\big)^2+
\left(\bar{\psi}_{1}\psi_{1}+\bar{\uppsi}\uppsi\right)\bar{\psi}_{2}\gamma_{+}\psi_{2}\Big) \qquad\notag
\\ & \ \
- M\Big(e^{b\Phi_1}\left(\bar{\psi}_{1}\psi_{1}+\bar{\psi}_{2}\psi_{2}+\bar{\uppsi}\uppsi\right)+e^{-b\Phi_1}\bar{\psi}_3\gamma_{+}\psi_3+\cosh b\Phi_2\big(\bar{\psi}_3\gamma_{-}\psi_3\big)\Big)\notag
\\ & \ \ +\frac{M^2}{8\pi b^2}\left(e^{2b\Phi_1}+2e^{-b\Phi_1}\cosh b\Phi_2\right),\label{OSP(7|2)-Lagrangian-1}
\end{align}
\vspace*{-0.3cm}
and \eqref{OSP(7|2)-diagram-3}
\begin{align}
\mathcal{L}_{\scriptscriptstyle{OSP(7|2)}}^{(2)}
& =\frac{1}{8\pi}\sum_{k=1}^2(\partial_{\mu}\Phi_k)^{2}+i\sum_{k=1}^3\bar{\psi}_{k}\gamma^{\mu}\partial_{\mu}\psi_{k}+i\bar{\uppsi}\gamma^{\mu}\partial_{\mu}\uppsi\notag
\\ & \ \  +\frac{\pi b^{2}}{2(1+b^{2})}\Bigl(\sum_{k=1}^3(\bar{\psi}_{k}\gamma^{\mu}\psi_{k})^{2}-(\bar{\uppsi}\gamma^{\mu}\uppsi)^2\Bigr)\notag
\\ & \ \
+
\left(4\pi b^{2}+\dots\right)\Big(\big(\bar{\uppsi}\gamma_{+}\uppsi\big)\big(\bar{\psi}_2\gamma_{-}\psi_2\big)+\big(\bar{\uppsi}\gamma_{-}\uppsi\big)\big(\bar{\psi}_3\gamma_{+}\psi_3\big)+\big(\bar{\psi}_2\gamma_{-}\psi_2\big)\big(\bar{\psi}_3\gamma_{+}\psi_3\big)\Big)\notag
\\ & \ \ -
M\Big(e^{b\Phi_1}\bar{\psi}_{1}\psi_{1}+e^{-b\Phi_1}\big(\bar{\uppsi}\gamma_{+}\uppsi+\bar{\psi}_2\gamma_{+}\psi_2+\bar{\psi}_3\gamma_{+}\psi_3\big)\notag
\\ & \hspace{90pt} +\cosh b\Phi_2\big(\bar{\uppsi}\gamma_{-}\uppsi+\bar{\psi}_2\gamma_{-}\psi_2+\bar{\psi}_3\gamma_{-}\psi_3\big)\Big)\notag
\\ & \ \
+ \frac{M^2}{8\pi b^2}\left(e^{2b\Phi_1}+2e^{-b\Phi_1}\cosh b\Phi_2\right),\label{OSP(7|2)-Lagrangian-2}
\end{align}
where all the Dirac spinors and one of the scalars, namely $\Phi_2$, which is the lightest of $\Phi_1$ and $\Phi_2$, have the same mass.
Let us also note that setting the fermions to zero we find the affine $B^{\vee}(2)$ Toda QFT.

In order to write down the Lagrangians \eqref{OSP(5|2)-Lagrangian-scaled}, \eqref{OSP(7|2)-Lagrangian-1} and \eqref{OSP(7|2)-Lagrangian-2} we were guided by the following logic, which we describe for general $N=2n+1$ and $m$ with $N>m+2$.
The conjectured Lagrangian should take the following general form
\begin{equation}\begin{split}\label{OSP(N|2m)-Lagrangian}
\mathcal{L}_{\scriptscriptstyle{OSP(2n+1|2m)}}& =\frac{1}{8\pi}\sum_{k=1}^{n-m}(\partial_{\mu}\Phi_k)^{2}+i\sum_{k=1}^n\bar{\psi}_{k}\gamma^{\mu}\partial_{\mu}\psi_{k}+i\sum_{k=1}^{m}\bar{\uppsi}_k\gamma^{\mu}\partial_{\mu}\uppsi_k
\\ & \ \ + \frac{\pi b^{2}}{2(1+b^{2})}\Bigl(\sum_{k=1}^n(\bar{\psi}_{k}\gamma^{\mu}\psi_{k})^{2}-\sum_{k=1}^m(\bar{\uppsi}_k\gamma^{\mu}\uppsi_k)^2\Bigr)
+\text{``four-spinor'' terms }\qquad \,
\\ & \ \ + \text{``Yukawa-like'' massive terms}\, + \text{``Toda-like'' massive terms}~.
\end{split}\end{equation}
While the ``four-spinor''  and ``Yukawa-like'' terms depend on the choice of diagram, the ``Toda-like'' term is universal and corresponds to the affine $B^{\vee}(n-m)$ Toda QFT
\begin{equation}\begin{split}\label{Toda-coupling}
& \text{``Toda-like'' massive terms} \\ &\qquad =
\frac{M^2}{8\pi b^2}\Big(e^{2b\Phi_1}+2\sum_{k=2}^{n-m-1}e^{b(\Phi_k-\Phi_{k-1})}+2e^{-b\Phi_{n-m-1}}\cosh\Phi_{n-m}\Big) ~.
\end{split}\end{equation}
For the ``Yukawa-like'' terms we make the following conjecture based on the explicit examples that we have studied
\begin{equation}\begin{split}\label{Yukawa-coupling}
&    \text{``Yukawa-like'' massive terms} \\ & \qquad
=-M\Big(e^{b\Phi_1}\big(J_+^{(1)}+J_-^{(1)}\big)+\sum_{k=2}^{n-m-1}\big(e^{-b\Phi_{k-1}}J_+^{(k)}+e^{b\Phi_{k}}J_-^{(k)}\big)
\\ & \hspace{115pt}+e^{-b\Phi_{n-m-1}}J_+^{(n-m)}+\cosh b\Phi_{n-m}J_-^{(n-m)}\Big)~,
\end{split}\end{equation}
where
\begin{equation}
   J_{\pm}^{(k)}=\bar{\psi}\gamma_{\pm}\psi+\sum_{j=1}^k\big(\bar{\uppsi}_j\gamma_{\pm}\uppsi_j+\bar{\psi}_j\gamma_{\pm}\psi_j\big)~,
\end{equation}
The form of the terms \eqref{Toda-coupling} and \eqref{Yukawa-coupling} immediately implies that all the Dirac spinors and the scalar $\Phi_{n-m}$ have the same mass in the $b\rightarrow0$ limit.
Moreover, one can check that the sum of squared mass of the fermionic particles is equal to that of the bosonic particles (including the heavy particles)
\begin{equation}
    \sum_f m_f^2=\sum_b m_b^2 ~,
\end{equation}
which may signal for the UV finiteness of the theory \cite{Fateev:2019xuq}.

The structure of  the ``four-spinor'' terms in \eqref{OSP(N|2m)-Lagrangian} appears to be more complicated and, in general, is unknown.
We have found these terms explicitly in various cases, although we have only determined the corresponding coupling constants at leading order in $b$ by requiring that the tree-level $S$-matrix satisfies the classical YB equation.
This is in contrast to the $O(N)$ case \cite{Fateev:2018yos,Litvinov:2018bou} for which the ``four-spinor'' terms are absent and the dual Toda action is expected to be exact in $b$.
\subsection{Strong coupling: sigma model}\label{strong-coupling-QFT}
In the strong-coupling regime we perturb the theory by the screening charges that are light in the limit $b\rightarrow\infty$.
This perturbation is of sigma-model type and requires infinitely many counterterms.
Note that, compared to the $O(N)$ case, there are screening charges that contain imaginary exponents $e^{\pm\frac{i\beta}{b}\phi}$.
We fermionize these exponents according to the rule
\begin{equation}\label{sigma model-boson-fermion-correspondence}
\left\{\frac{1}{8\pi}(\partial_\mu\phi)^2,e^{\pm\frac{i\beta}{b}\phi}\right\}
\rightarrow
\left\{i\bar{\vartheta}\gamma^{\mu}\partial_{\mu}\vartheta+\frac{\pi}{2(1+b^2)}(\bar{\vartheta}\gamma^{\mu}\vartheta)^2,\bar{\vartheta}\gamma_{\pm}\vartheta\right\}~.
\end{equation}
The resulting theory takes the form of a bosonic sigma model coupled to first order fermions, i.e. their kinetic term is first order in derivatives.
However, the fermions in the sigma model on a supermanifold are second order.
To recover the latter from the former we integrate over half of the fermionic degrees of freedom in the first-order Lagrangian.
The starting point for this procedure always has the following form
\begin{equation}\label{L-SM-first-order-fermions}
    \mathcal{L}_{\scriptscriptstyle{\text{1st-order}}}=i\bar{\vartheta}\gamma^{\mu}\partial_{\mu}\vartheta+\frac{\pi}{2(1+b^2)}(\bar{\vartheta}\gamma^{\mu}\vartheta)^2+\bar{\vartheta}\gamma_{+}\vartheta A_{+}+\bar{\vartheta}\gamma_{-}\vartheta A_{-}+\frac{b^2}{4\pi}A_+A_- ~,
\end{equation}
where $A_{\pm}$ are local fields independent of $\vartheta$ and the final term corresponds to the usual contact term.
Now setting (we use the conventions $\gamma^1=\sigma_1$, $\gamma^2=\sigma_2$, $\partial = \partial_1 - i \partial_2$)
\begin{equation}
    \vartheta=\begin{pmatrix}
    i\chi\\
    \theta^*
    \end{pmatrix}~,\qquad
    \bar{\vartheta}=\begin{pmatrix}
    -i\chi^*& -\theta
    \end{pmatrix}~,
\end{equation}
the Lagrangian \eqref{L-SM-first-order-fermions} takes the form
\begin{equation}
    \mathcal{L}_{\scriptscriptstyle{\text{1st-order}}}=\chi^*\partial\theta^*+\chi\bar{\partial}\theta+\frac{2\pi}{1+b^2}\chi\chi^*\theta\theta^*+\chi\chi^*A_++\theta\theta^*A_-+\frac{b^2}{4\pi}A_+A_- ~,
\end{equation}
up to total derivatives.
At this point we can either integrate over $\chi$ and $\chi^*$ or $\theta$ and $\theta^*$.
Integrating out $\chi$ and $\chi^*$ and dropping the determinant contribution, we find
\begin{equation}\label{SM-fermion-terms-integrated}
    \mathcal{L}^{\textrm{eff}}=2\partial\theta^*\bar{\partial}\theta \, \Big(A_+^{-1}-\frac{2\pi}{1+b^2}A_+^{-2}\theta\theta^*\Big)+\theta\theta^{*}A_-+\frac{b^2}{4\pi}A_+A_- ~.
\end{equation}
Rescaling $\theta\rightarrow \frac{b}{4\sqrt{\pi}}\theta$, the leading term as $b\to\infty$ is
\begin{equation}
\mathcal{L}^{\textrm{eff}}=\frac{b^2}{8\pi}\Big(\partial\theta^*\bar{\partial}\theta \,\Big(A_+^{-1}-\frac{1}{8}A_+^{-2}\theta\theta^*\Big)+\frac{1}{2}\theta\theta^{*}A_-+2A_+A_-\Big)+O(1) ~,
\end{equation}
We disregard the $O(1)$ terms, which can be understood as quantum corrections, since they are of the same order as the dropped determinant and, in general, are beyond our control.

Let us now consider the $OSP(5|2)$ case \eqref{OSP(5|2)-diagram}, \eqref{alpha-for-OSP(5|2)} in more detail.
We take the screenings corresponding to the roots $\boldsymbol{\beta}_{12}$, $\boldsymbol{\beta}_1$ and $\boldsymbol{\beta}_{\pm}^{+}$ as our perturbing fields.
The resulting Lagrangian has the form
\begin{equation}\begin{split}\label{dual-Lagrangian-OSP(5|2)}
    \mathcal{L}_{\scriptscriptstyle{OSP(5|2)}}^{\text{dual}} & =
\frac{1}{8\pi}\sum_{k=1}^2
    \left(\big(\partial_{\mu}\Phi_k\big)^2+\big(\partial_{\mu}\varphi_k\big)^2\right)+i\bar{\vartheta}\gamma^{\mu}\partial_{\mu}\vartheta+\frac{\pi}{2(1+b^2)}(\bar{\vartheta}\gamma^{\mu}\vartheta)^2
\\ & \ \ +
 \Lambda e^{\frac{\Phi_1}{b}}\big|\partial\Phi_1+\frac{i\beta}{b}\partial\varphi_1\big|^2+e^{-\frac{\Phi_1}{b}}\big(\bar{\vartheta}\gamma_{+}\vartheta\big)-4\pi\Lambda\big(\bar{\vartheta}\gamma_{+}\vartheta\big)
\\ & \ \ +
\Lambda b^{-2}\Big(e^{\frac{\Phi_2}{b}}\big|\partial\Phi_2+\frac{i\beta}{b}\partial\varphi_2\big|^2+e^{-\frac{\Phi_2}{b}}\big|\partial\Phi_2-\frac{i\beta}{b}\partial\varphi_2\big|^2\Big)\big(\bar{\vartheta}\gamma_{-}\vartheta\big)
\\ & \ \ +\frac{\Lambda}{4\pi}\Big(e^{\frac{\Phi_2-\Phi_1}{b}}\big|\partial\Phi_2+\frac{i\beta}{b}\partial\varphi_2\big|^2+e^{-\frac{\Phi_2+\Phi_1}{b}}\big|\partial\Phi_2-\frac{i\beta}{b}\partial\varphi_2\big|^2\Big)+O(\Lambda^2) ~,
\end{split}\end{equation}
where the $O(\Lambda^2)$ terms are UV counterterms, of which infinitely many are typically required \cite{Litvinov:2018bou}.
Working to $O(\Lambda)$ we take
\begin{equation}
    A_+=e^{-\frac{\Phi_1}{b}}-4\pi\Lambda~,\qquad A_-=\Lambda\Big(e^{\frac{\Phi_2}{b}}\big|\partial\Phi_2+\frac{i\beta}{b}\partial\varphi_2\big|^2+e^{-\frac{\Phi_2}{b}}\big|\partial\Phi_2-\frac{i\beta}{b}\partial\varphi_2\big|^2\Big)~,
\end{equation}
in \eqref{L-SM-first-order-fermions} and integrate over half of the fermionic degrees of freedom following the general prescription outlined above.
Defining
\begin{equation}
    \Phi_1+i\varphi_1=bX_1=b(x_1+iy_1) ~, \Phi_2+i\varphi_2=bX_2=b(x_2+iy_2) ~,
\end{equation}
the leading asymptotic of the effective Lagrangian obtained from \eqref{dual-Lagrangian-OSP(5|2)}
\begin{equation}
    \mathcal{L}_{\scriptscriptstyle{OSP(5|2)}}=b^2\mathcal{L}^{\textrm{eff}}_{\scriptscriptstyle{OSP(5|2)}}+O(1) ~,
\end{equation}
finally takes the form
\begin{equation}\begin{split}\label{dual-Lagrangian-OSP(5|2)-eff}
    \mathcal{L}^{\textrm{eff}}_{\scriptscriptstyle{OSP(5|2)}} & =\frac{1}{8\pi}\Big(\partial X_1\bar{\partial}X_1^*+\partial X_2\bar{\partial}X_2^*+\big(e^{x_1}-\frac{1}{8}e^{2x_1}\theta\theta^*\big)\partial\theta\bar{\partial}\theta^*\Big)
\\& \ \ +2\Lambda\Big(e^{x_1}\partial X_1\bar{\partial}X_1^*+e^{x_2}\big(e^{-x_1}+\frac{1}{4}\theta\theta^*\big)\partial X_2\bar{\partial}X_2^*\\& \ \
+e^{-x_2}\big(e^{-x_1}+\frac{1}{4}\theta\theta^*\big)\partial X_2^*\bar{\partial}X_2
+e^{x_1}\big(e^{x_1}-\frac{1}{4}e^{2x_1}\theta\theta^*\big)\partial\theta\bar{\partial}\theta^*\Big)+O(\Lambda^2) ~. \quad
\end{split}\end{equation}
We will see in section \ref{SM} that \eqref{dual-Lagrangian-OSP(5|2)-eff} coincides with the leading UV asymptotic of the $\eta$-deformed $OSP(5|2)$ sigma model for a certain choice of $\mathcal{R}$ solving the classical YB equation.
We have also checked that similarly calculating the effective actions for the two possible $OSP(7|2)$ diagrams \eqref{OSP(7|2)-diagram-1} and \eqref{OSP(7|2)-diagram-3} we again find agreement with the UV expansion of the $\eta$-deformed $OSP(7|2)$ sigma model for different choices of $\mathcal{R}$. In section \ref{concl} we will give a general  conjecture. 
\section{Deformed \texorpdfstring{$OSP(N|2m)$}{OSP(N|2m)} sigma model}\label{SM}

In this section we review the $\eta$-deformed $G/H$ symmetric space sigma model and study the various deformations of the $OSP(N|2m)$ sigma model that follow from different Dynkin diagrams.
For various examples we investigate the one-loop renormalizability of the deformed model and show that the leading UV asymptotic of one of the three deformations of the $OSP(5|2)$ sigma model coincides with \eqref{dual-Lagrangian-OSP(5|2)-eff} as claimed in section \ref{QFT}.
\subsection{YB deformation of \texorpdfstring{$OSP(N|2m)$}{OSP(N|2m)} sigma model}
The $OSP(N|2m)$ sigma model can be written as a symmetric space sigma model on the supercoset
\begin{equation}\label{eq:symspace}
\frac{OSP(N|2m)}{OSP(N-1|2m)} ~.
\end{equation}
The Lorentzian action for the supergroup-valued field $\mathbf{g} \in OSP(N|2m)$ is
\begin{equation}\label{eq:sssm}
\mathcal{S}_0 = - \frac{R^2}{8\pi} \int d^2\xi \, \operatorname{STr}[\mathbf{J}_+ P \mathbf{J}_-] ~,
\end{equation}
where $\mathbf{J}_\pm = \mathbf{g}^{-1} \partial_\pm \mathbf{g}$ takes values in the Grassmann envelope of the Lie superalgebra $\mathfrak{osp}(N|2m;\mathds{R})$ and $\operatorname{STr}$ is the invariant bilinear form.
As we are considering the symmetric space \eqref{eq:symspace} we have the $\mathbb{Z}_2$ grading
\begin{equation}
\mathfrak{g} \equiv \mathfrak{osp}(N|2m;\mathds{R}) = \mathfrak{g}^{(0)}\oplus\mathfrak{g}^{(1)} ~, \qquad \mathfrak{g}^{(0)} = \mathfrak{osp}(N-1|2m;\mathds{R}) ~,
\end{equation}
with $P$ projecting onto the grade 1 subspace, referred to as the ``coset space'' in the Introduction.

In order to fix conventions we introduce the $(N+2m) \times (N+2m)$ supermatrix realization of the complexified superalgebra.
For this we define the $(N+2m) \times (N+2m)$ matrix
\begin{equation}
G = \begin{pmatrix} \operatorname{antidiag}(1,\ldots,1)_{N\times N} & 0_{N \times 2m} \\ 0_{2m \times N} & \operatorname{antidiag}(-1,1,\ldots,-1,1)_{2m\times 2m} \end{pmatrix}~,
\end{equation}
where $\operatorname{antidiag}$ represents a matrix whose non-zero entries lie on the antidiagonal
(the first argument denotes the entry in the top right corner of the matrix).
The Grassmann envelope of the Lie superalgebra $\mathfrak{osp}(N|2m;\mathds{C})$ is given by those supermatrices $M \in \operatorname{Mat}(N|2m;\Lambda)$ satisfying
\begin{equation}
M^{\operatorname{st}} G + G M = 0 ~.
\end{equation}
The explicit expressions for the supertranspose and supertrace are
\begin{equation}
\begin{pmatrix}a & b \\ c & d\end{pmatrix}^{\operatorname{st}}
= \begin{pmatrix}a^{\operatorname{t}} & c^{\operatorname{t}} \\ - b^{\operatorname{t}} & d^{\operatorname{t}} \end{pmatrix} ~, \qquad
\operatorname{STr}\begin{pmatrix}a & b \\ c & d\end{pmatrix} = \operatorname{Tr}a - \operatorname{Tr}d ~.
\end{equation}
Introducing the involutive antilinear antiautomorphism
\begin{equation}\begin{gathered}
M^\star = \Sigma M^\dagger \Sigma^{-1} ~, \qquad \begin{pmatrix} a & b \\ c & d \end{pmatrix}^\star = \begin{pmatrix}a^\dagger & - i c^\dagger \\ - i b^\dagger & d^\dagger\end{pmatrix} ~, \\ \Sigma = \operatorname{diag}(\underbrace{1,\dots,1}_{N},\underbrace{1,\dots,1}_{m},\underbrace{-1,\dots,-1}_{m}) ~,
\end{gathered}\end{equation}
we consider the real form $\mathfrak{osp}(N|2m;\mathds{R})$ whose Grassmann envelope is given by those supermatrices that additionally satisfy
\begin{equation}\label{eq:reality}
M^\star = - M ~.
\end{equation}
For this real form the bosonic subalgebra is $\mathfrak{so}(N)\oplus\mathfrak{sp}(2m,\mathds{R})$.

To define the projection operator $P$ we introduce the rotation matrix $\Lambda$ such that
\begin{equation}
\Lambda G \Lambda^{\operatorname{t}} = \tilde{G} = \begin{pmatrix} \operatorname{diag}(\underbrace{1,\ldots,1}_{\lfloor\frac{N+1}{2}\rfloor},\underbrace{-1,\ldots,-1}_{\lfloor\frac{N}{2}\rfloor}) & 0_{N \times 2m} \\ 0_{2m \times N} & \operatorname{antidiag}(-1,1,\ldots,-1,1) \end{pmatrix}~.
\end{equation}
To be explicit we take
\begin{equation}
\Lambda = \begin{pmatrix} \Lambda_a & 0_{N \times 2m} \\ 0_{2m \times N} & \operatorname{diag}(1,\ldots,1) \end{pmatrix}~,
\end{equation}
where
\begin{equation}\begin{split}
\Lambda_a & = \tfrac{1}{\sqrt{2}}\big(\operatorname{diag}(1,\ldots,1)
+ \operatorname{antidiag}(\underbrace{1,\ldots,1}_{\frac{N}{2}},\underbrace{-1,\ldots,-1}_{\frac{N}{2}})\big) ~, \qquad \text{$N$ even} ~,
\\
\Lambda_a & = \tfrac{1}{\sqrt{2}}\big(
\operatorname{diag}(\underbrace{1,\ldots,1}_{\frac{N-1}{2}},\sqrt{2},\underbrace{1,\ldots,1}_{\frac{N-1}{2}})\big)
+ \operatorname{antidiag}(\underbrace{1,\ldots,1}_{\frac{N-1}{2}},0,\underbrace{-1,\ldots,-1}_{\frac{N-1}{2}})\big) ~, \qquad \text{$N$ odd} ~.
\end{split}\end{equation}
In this basis the grade 0 subalgebra, $\mathfrak{osp}(N-1|2m;\mathds{R})$, is taken to be the bottom right $(N-1+2m)\times(N-1+2m)$ block, and hence the rotated projection operator $\tilde{P}$ simply sets entries in this block to zero.
Undoing the rotation we have
\begin{equation}
P = \operatorname{Ad}_\Lambda^{-1} \tilde{P} \operatorname{Ad}_\Lambda^{\vphantom{-1}} ~.
\end{equation}

The Lorentzian action of the YB deformed model is \cite{Klimcik:2002zj,Delduc:2013fga}
\begin{equation}\label{eq:deformedaction}
\mathcal{S}_\eta = - \frac{\eta}{8\pi\nu} \int d^2\xi \, \operatorname{STr}[\mathbf{J}_+ P \frac{1}{1-\eta \mathcal{R}_{\mathbf{g}} P} \mathbf{J}_-] ~,
\end{equation}
where $\eta$ is the deformation parameter and $\nu$ plays the role of the sigma model coupling.
The extra factor of $-\frac{1}{2}$ compared to \eqref{Coset-action-deformed} comes from the normalisation of the supertrace in the representation defined above.
Setting $\nu = \eta R^{-2}$ and taking $\eta \to 0$ we recover the undeformed model \eqref{eq:sssm}.
The operator $\mathcal{R}_\mathbf{g}$ is defined in terms of the linear operator $\mathcal{R}:\mathfrak{g}\to\mathfrak{g}$ through
\begin{equation}
\mathcal{R}_\mathbf{g} = \operatorname{Ad}_\mathbf{g}^{-1} \mathcal{R} \operatorname{Ad}_\mathbf{g}^{\vphantom{-1}} ~,
\end{equation}
where $\mathcal{R}$ is an antisymmetric solution of the (non-split) modified classical YB equation
\begin{equation}\begin{gathered}
[\mathcal{R}X,\mathcal{R}Y]-\mathcal{R}([X,\mathcal{R}Y]+[\mathcal{R}X,Y])=[X,Y] ~, \\ \operatorname{STr}[X(\mathcal{R}Y)] = -\operatorname{STr}[(\mathcal{R}X)Y] ~, 
\end{gathered} 
\qquad X,Y \in \mathfrak{g} ~.\end{equation}
Denoting the Cartan generators of $\mathfrak{g}$ as $h_i$ and the positive and negative roots as $e_m$ and $f_m$ respectively, the Drinfel'd-Jimbo r-matrix \cite{Drinfeld:1985rx,Jimbo:1985zk,Belavin-Drinfeld} in operator form can be written as
\begin{equation}\label{eq:rmatrix}
\mathcal{R}X = i \sum_m \big( \operatorname{STr}[X f_m] e_m - \operatorname{STr}[e_m X] f_m\big) ~, \qquad \operatorname{STr}[e_m f_n] = \delta_{mn} ~.
\end{equation}
We take the Cartan generators to span the diagonal matrices of the matrix realization introduced above, while the positive and negative roots span those matrices whose non-vanishing entries lie in the light and dark grey regions respectively of the following diagram
\begin{equation}\label{eq:posnegroots}
\begin{tikzpicture}[scale=0.4]
\node at (0,10) {$0$};
\node at (1,9) {$0$};
\node[rotate=-7] at (2,8.25) {$\ddots$};
\node at (3,7) {$0$};
\node at (4,6) {$0$};
\node at (5,5) {$0$};
\node[rotate=-7] at (6,4.25) {$\ddots$};
\node at (7,3) {$0$};
\node at (8,2) {$0$};
\node[rotate=-7] at (9,1.25) {$\ddots$};
\node at (10,0) {$0$};
\draw (-0.5,5.5)--(10.5,5.5);
\draw (4.5,-0.5)--(4.5,10.5);
\draw[dashed] (-0.5,2.5)--(10.5,2.5);
\draw[dashed] (7.5,-0.5)--(7.5,10.5);
\draw[gray,fill=gray,opacity=0.25] (1,10)--(4,10)--(4,7)--(1,10);
\draw[gray,fill=gray,opacity=0.25] (6,5)--(10,5)--(10,1)--(6,5);
\draw[gray,fill=gray,opacity=0.25] (8,10)--(10,10)--(10,6)--(8,6)--(8,10);
\draw[gray,fill=gray,opacity=0.25] (0,3)--(4,3)--(4,5)--(0,5)--(0,3);
\draw[gray,fill=gray,opacity=0.75] (0,9)--(0,6)--(3,6)--(0,9);
\draw[gray,fill=gray,opacity=0.75] (5,4)--(5,0)--(9,0)--(5,4);
\draw[gray,fill=gray,opacity=0.75] (5,10)--(7,10)--(7,6)--(5,6)--(5,10);
\draw[gray,fill=gray,opacity=0.75] (0,0)--(4,0)--(4,2)--(0,2)--(0,0);
\draw[decorate,decoration={brace,amplitude=4pt},yshift=-24pt,thick] (4.4,-0.5) -- (-0.6,-0.5) node [black,midway,yshift=-12pt] {$N$};
\draw[decorate,decoration={brace,amplitude=4pt},yshift=-24pt,thick] (7.4,-0.5) -- (4.6,-0.5) node [black,midway,yshift=-12pt] {$m$};
\draw[decorate,decoration={brace,amplitude=4pt},yshift=-24pt,thick] (10.4,-0.5) -- (7.6,-0.5) node [black,midway,yshift=-12pt] {$m$};
\draw[decorate,decoration={brace,amplitude=4pt},xshift=48pt,thick] (10.5,10.4) -- (10.5,5.6) node [black,midway,xshift=12pt] {$N$};
\draw[decorate,decoration={brace,amplitude=4pt},xshift=48pt,thick] (10.5,5.4) -- (10.5,2.6) node [black,midway,xshift=12pt] {$m$};
\draw[decorate,decoration={brace,amplitude=4pt},xshift=48pt,thick] (10.5,2.4) -- (10.5,-0.6) node [black,midway,xshift=12pt] {$m$};
\draw[thick] (-0.5,10.5) .. controls (-1.5,7) and (-1.5,3) .. (-0.5,-0.6);
\draw[thick] (10.5,10.5) .. controls (11.5,7) and (11.5,3) .. (10.5,-0.6);
\end{tikzpicture}
\end{equation}
As usual, the positive and negative roots are not themselves elements of the real form $\mathfrak{g} = \mathfrak{osp}(N|2m,\mathds{R})$, but appropriate linear combinations can be constructed that do satisfy the reality condition \eqref{eq:reality}.
Furthermore, it is straightforward to check that the operator $\mathcal{R}$ defined in equation \eqref{eq:rmatrix} preserves this real form as required.
The choice of root system in equation \eqref{eq:posnegroots} corresponds to the following Dynkin diagrams
\begin{equation}\begin{aligned}
& N=1 &\quad&
\begin{tikzpicture}[baseline=-0.1cm]
\node (n1) at (0,0) {};
\node (n2) at (1,0) {};
\node (n3) at (2,0) {};
\node (n4) at (3,0) {};
\draw (n1) -- (n2);
\draw[dotted] (n2) -- (n3);
\draw[->-,double distance=2.2pt] (n3) -- (n4);
\draw[fill=white] (n1) circle (0.15cm);
\draw[fill=white] (n2) circle (0.15cm);
\draw[fill=white] (n3) circle (0.15cm);
\draw[fill=black] (n4) circle (0.15cm);
\draw[decorate,decoration={brace,amplitude=4pt},thick] (2,-0.5) -- (0,-0.5) node [black,midway,yshift=-12pt] {$m-1$};
\end{tikzpicture} ~,
\\
& N=2 &\quad&
\begin{tikzpicture}[baseline=-0.1cm]
\node (n1) at (0,0) {};
\node (n2) at (1,0) {};
\node (n3) at (2,0) {};
\node (n4) at (3,-0.5) {};
\node (n5) at (3,0.5) {};
\draw (n1) -- (n2);
\draw[dotted] (n2) -- (n3);
\draw (n3) -- (n4);
\draw (n3) -- (n5);
\draw[double distance=2.2pt] (n4) -- (n5);
\draw[fill=white] (n1) circle (0.15cm);
\draw[fill=white] (n2) circle (0.15cm);
\draw[fill=white] (n3) circle (0.15cm);
\draw[fill=white,cross] (n4) circle (0.15cm);
\draw[fill=white,cross] (n5) circle (0.15cm);
\draw[decorate,decoration={brace,amplitude=4pt},thick] (2,-0.5) -- (0,-0.5) node [black,midway,yshift=-12pt] {$m-1$};
\end{tikzpicture} ~,
\\
& N=2n+1>1 &\quad&
\begin{tikzpicture}[baseline=-0.1cm]
\node (n1) at (0,0) {};
\node (n2) at (1,0) {};
\node (n3) at (2,0) {};
\node (n4) at (3,0) {};
\node (n5) at (4,0) {};
\node (n6) at (5,0) {};
\draw (n1) -- (n2);
\draw[dotted] (n2) -- (n3);
\draw[dotted] (n3) -- (n4);
\draw (n4) -- (n5);
\draw[->-,double distance=2.2pt] (n5) -- (n6);
\draw[fill=white] (n1) circle (0.15cm);
\draw[fill=white] (n2) circle (0.15cm);
\draw[fill=white,cross] (n3) circle (0.15cm);
\draw[fill=white] (n4) circle (0.15cm);
\draw[fill=white] (n5) circle (0.15cm);
\draw[fill=white] (n6) circle (0.15cm);
\draw[decorate,decoration={brace,amplitude=4pt},thick] (1.5,-0.5) -- (0,-0.5) node [black,midway,yshift=-12pt] {$m-1$};
\draw[decorate,decoration={brace,amplitude=4pt},thick] (4,-0.5) -- (2.5,-0.5) node [black,midway,yshift=-12pt] {$n-1$};
\end{tikzpicture} ~,
\\
& N=2n+2>2 &\quad&
\begin{tikzpicture}[baseline=-0.1cm]
\node (n1) at (0,0) {};
\node (n2) at (1,0) {};
\node (n3) at (2,0) {};
\node (n4) at (3,0) {};
\node (n5) at (4,0) {};
\node (n6) at (5,-0.5) {};
\node (n7) at (5,0.5) {};
\draw (n1) -- (n2);
\draw[dotted] (n2) -- (n3);
\draw[dotted] (n3) -- (n4);
\draw (n4) -- (n5);
\draw (n5) -- (n6);
\draw (n5) -- (n7);
\draw[fill=white] (n1) circle (0.15cm);
\draw[fill=white] (n2) circle (0.15cm);
\draw[fill=white,cross] (n3) circle (0.15cm);
\draw[fill=white] (n4) circle (0.15cm);
\draw[fill=white] (n5) circle (0.15cm);
\draw[fill=white] (n6) circle (0.15cm);
\draw[fill=white] (n7) circle (0.15cm);
\draw[decorate,decoration={brace,amplitude=4pt},thick] (1.5,-0.5) -- (0,-0.5) node [black,midway,yshift=-12pt] {$m-1$};
\draw[decorate,decoration={brace,amplitude=4pt},thick] (4,-0.5) -- (2.5,-0.5) node [black,midway,yshift=-12pt] {$n-1$};
\end{tikzpicture} ~.
\end{aligned}
\end{equation}
Inequivalent choices of roots, e.g. corresponding to different Dynkin diagrams, can lead to different operators $\mathcal{R}$ and hence different deformations \cite{Delduc:2014kha,Hoare:2016ibq,Hoare:2018ngg}.

In terms of coordinates on the target superspace, the action \eqref{eq:deformedaction} can be written in the following form
\begin{equation}
\mathcal{S}_\eta = \frac{1}{4\pi} \int d^2\xi \, \big(G_\ind{MN}(z) + B_\ind{MN}(z)\big) 
\partial_+ z^\ind{N} \partial_- z^\ind{M} ~,
\qquad z^\ind{M} = (x^\mu, \psi^\alpha) ~,
\end{equation}
where $x^\mu$ and $\psi^{\alpha}$ are the bosonic and fermionic coordinates respectively.
The metric $G_\ind{MN}$ and Kalb-Ramond field $B_\ind{MN}$ parametrize the parity-even and parity-odd part of the Lagrangian and hence have the following symmetry properties
\begin{equation}
G_\ind{MN} = (-1)^{\ind{M}\ind{N}} G_\ind{NM} ~,
\qquad
B_\ind{MN} = - (-1)^{\ind{M}\ind{N}} B_\ind{NM} ~,
\end{equation}
where in the sign factors $M,N,\ldots=0$ for bosonic coordinates and $M,N,\ldots=1$ for fermionic.

For $m=1$ the deformed sigma model is parametrized by $N-1$ bosons and a symplectic fermion, $\psi^a$, $a = 1,2$.
When $N$ is even we denote the bosons as $r_i$, $\phi_i$, $\phi_{\frac{N}{2}}$, $i=1,\dots,\frac{N}{2}-1$, while when $N$ is odd we have $r_i$, $\phi_i$, $i = 1,\dots\frac{N}{2}-\frac{1}{2}$.

For even $N$ we conjecture that the non-vanishing components of the metric are given by
\begin{equation}\label{eq:defmet62}
\begin{aligned}
G_{r_ir_i} & = \frac{\eta  \big(\prod_{j=1}^{i-1} r_j^2 \big)(1+\eta^2 \big(\prod_{j=1}^{i-1} r_j^4 \big) r_i^2 + (1-\eta^2  \big(\prod_{j=1}^{i-1} r_j^4 \big) r_i^2)\psi\cdot\psi)}{\nu(1+\eta^2  \big(\prod_{j=1}^{i-1} r_j^4 \big) r_i^2)^2(1-r_i^2)} ~,
\\
G_{\phi_i\phi_i} & = \frac{\eta \big(\prod_{j=1}^{i-1} r_j^2 \big) (1+\eta^2  \big(\prod_{j=1}^{i-1} r_j^4 \big) r_i^2 + (1-\eta^2  \big(\prod_{j=1}^{i-1} r_j^4 \big) r_i^2)\psi\cdot\psi)(1-r_i^2)}{\nu(1+\eta^2  \big(\prod_{j=1}^{i-1} r_j^4 \big) r_i^2)^2} ~,
\\
G_{\phi_{\frac{N}{2}}\phi_{\frac{N}{2}}} & = \frac{\eta \big(\prod_{j=1}^{\frac{N}{2}-1} r_j^2 \big)(1+\psi\cdot\psi)}{\nu} ~,
\quad \ \
G_{\psi^1\psi^2} =
- G_{\psi^2\psi^1} = \frac{\eta(1+\eta^2+\tfrac12(1-\eta^2)\psi\cdot\psi)}{\nu(1+\eta^2)^2} ~,
\end{aligned}\end{equation}
while for the Kalb-Ramond field we have
\begin{equation}\begin{aligned}\label{eq:defbfi62}
B_{r_i\phi_i} & =
- B_{\phi_ir_i} =
- \frac{\eta^2  \big(\prod_{j=1}^{i-1} r_j^4 \big)r_i(1+\eta^2 \big(\prod_{j=1}^{i-1} r_j^4 \big)r_i^2+2\psi\cdot\psi)}{\nu(1+\eta^2  \big(\prod_{j=1}^{i-1} r_j^4 \big) r_i^2)^2} ~,
\\
B_{\psi^1\psi^1} & =
B_{\psi^2\psi^2} = \frac{\eta^2(1+\eta^2+\psi\cdot\psi)}{\nu(1+\eta^2)^2} ~,
\end{aligned}\end{equation}
where $\psi\cdot\psi = 2\psi^1\psi^2$.
The isometries of this background are $\frac{N}{2}$ $U(1)$ shift symmetries and $SO(2)$ rotations of the symplectic fermion.
\begin{equation}
\phi_i \to \phi_i + c_i ~, \qquad \psi^a \to (\delta^a{}_b \cos \tilde c + \epsilon^a{}_b \sin \tilde c) \psi^b ~.
\end{equation}
These symmetries correspond to the Cartan subgroup of $OSP(N|2)$ associated with the root system introduced above.
Indeed, the rank of $OSP(N|2)$ for even $N$ is $\frac{N}{2}+1$.

For odd $N$ our conjecture for the metric and Kalb-Ramond field is simply given by that for even $N$ in one dimension higher with $\phi_{\frac{N}{2}}$ set to zero.
In this case the isometries of the background are $\frac{N}{2}-\frac{1}{2}$ shift symmetries of $\phi_i$ and $SO(2)$ rotations of the symplectic fermion.
With the rank of $OSP(N|2)$ equalling $\frac{N}{2}+\frac{1}{2}$ for odd $N$ these symmetries again correspond to the Cartan subgroup associated with the root system introduced above.

Setting $\psi^a$ to zero in \eqref{eq:defmet62} and \eqref{eq:defbfi62} provides us also with a conjecture for the metric and Kalb-Ramond field of the deformed $O(N)$ sigma model, i.e. $m=0$.
For $m=1$ we have verified that the metrics and Kalb-Ramond fields explicitly match \eqref{eq:defmet62} and \eqref{eq:defbfi62} up to $N=8$, while for $m=0$ we have checked up to $N=32$.
Let us also observe that setting $\nu = \eta R^{-2}$ and taking $\eta \to 0$ the Kalb-Ramond field vanishes, while the metric \eqref{eq:defmet62} becomes
\begin{equation}\begin{aligned}\label{eq:undefmet62}
G_{\phi_i\phi_i} & = R^2 \big(\textstyle\prod_{j=1}^{i-1} r_j^2 \big) (1+\psi\cdot\psi)(1-r_i^2) ~,
 \ \  &
G_{r_ir_i} & = \frac{R^2 \big(\prod_{j=1}^{i-1} r_j^2 \big)(1+\psi\cdot\psi)}{1-r_i^2} ~,
\\
G_{\phi_{\frac{N}{2}}\phi_{\frac{N}{2}}} & = R^2 \big(\textstyle\prod_{j=1}^{\frac{N}{2}-1} r_j^2 \big) (1+\psi\cdot\psi) ~,
 \ \  &
G_{\psi^1\psi^2}& =
- G_{\psi^2\psi^1} = R^2(1+\tfrac12\psi\cdot\psi) ~,
\end{aligned}\end{equation}
which, as expected, is the metric of the $OSP(N|2m)$ sigma model
\begin{equation}
ds^2 = G_{\ind{MN}}dz^\ind{N}dz^\ind{M} = R^2\big(\sum_{I=1}^N dy_\ind{I} dy_\ind{I} - \sum_{I=1}^m d\psi_\ind{I} \cdot d \psi_\ind{I} \big) ~,
\qquad
\sum_{I=1}^N y_\ind{I}y_\ind{I} - \sum_{I=1}^m \psi_\ind{I} \cdot \psi_\ind{I} = 1 ~,
\end{equation}
for $m=1$. This can be seen explicitly for even $N$ by setting
\begin{equation}\begin{split}
y_{2i-1} + i y_{2i} & = (1+\tfrac12\psi\cdot\psi) \big(\textstyle\prod_{j=1}^{i-1} r_j \big)\sqrt{1-r_i^2}e^{i\phi_i} ~,
\\
y_{2N-1} + i y_{2N} & = (1+\tfrac12\psi\cdot\psi) \big(\textstyle\prod_{j=1}^{\frac{N}{2}-1} r_j \big) e^{i\phi_{\frac{N}{2}}} ~,
\end{split}\end{equation}
to solve the constraint.
As before, for odd $N$ a similar analysis holds and is simply given by the analysis for even $N$ in one dimension higher with $\phi_{\frac{N}{2}}$ set to zero.
Furthermore, the $m=0$ case is again recovered by setting $\psi^a$ to zero.

\subsection{Ricci flow}
On general grounds \cite{Valent:2009nv,Sfetsos:2009dj,Squellari:2014jfa,Sfetsos:2015nya} the YB deformation of the $OSP(N|2m)$ sigma model is expected to be renormalizable at one loop.
To confirm this, we need to know the Ricci tensor for supermanifolds with torsion.
We follow the conventions of \cite{DeWitt:1992cy}, in particular a comma denotes differentiation from the right.
The inverse metric is defined through
\begin{equation}
(-1)^\ind{M}G_\ind{MN} G^\ind{NP} = \delta^\ind{P}{}_\ind{M} ~,
\end{equation}
where we recall that $M,N,\ldots$ run over all coordinates, bosonic and fermionic.
The Christoffel symbols and the torsion are given by
\begin{equation}\begin{split}
\bar\Gamma^\ind{M}{}_\ind{NP} & = \frac12 (-1)^\ind{Q} G^\ind{MQ}\big(G_{\ind{QN},\ind{P}} + (-1)^\ind{NP} G_{\ind{QP},\ind{N}} + (-1)^{\ind{Q}(\ind{N}+\ind{P})} G_{\ind{NP},\ind{Q}}\big) ~,
\\
H^\ind{M}{}_\ind{NP} & = (-1)^\ind{Q} G^\ind{MQ}\big(B_{\ind{QN},\ind{P}} + (-1)^{\ind{Q}(\ind{N}+\ind{P})} B_{\ind{NP},\ind{Q}} + (-1)^{\ind{P}(\ind{Q}+\ind{N})} B_{\ind{PQ},\ind{N}}\big) ~,
\end{split}\end{equation}
such that the torsionful connection and its Riemann curvature are
\begin{equation}\begin{split}
\Gamma^\ind{M}{}_\ind{NP} & = \bar\Gamma^\ind{M}{}_\ind{NP} - \frac12 H^\ind{M}{}_\ind{NP} ~,
\\
R^\ind{M}{}_\ind{NPQ} & = - \Gamma^\ind{M}{}_{\ind{NP},\ind{Q}} + (-1)^{\ind{PQ}} \Gamma^\ind{M}{}_{\ind{NQ},\ind{P}}
\\ & \qquad
+(-1)^{\ind{P}(\ind{N}+\ind{R})}\Gamma^\ind{M}{}_\ind{RP}\Gamma^\ind{R}{}_\ind{NQ}
-(-1)^{\ind{Q}(\ind{N}+\ind{P}+\ind{R})}\Gamma^\ind{M}{}_\ind{RQ}\Gamma^\ind{R}{}_\ind{NP} ~. 
\end{split}\end{equation}
The Ricci tensor is then given by
\begin{equation}
R_\ind{MN} = (-1)^{\ind{P}(\ind{M}+1)}R^\ind{P}{}_\ind{MPN} ~.
\end{equation}

The condition for one-loop renormalizability is
\begin{equation}\label{eq:ricci}
R_\ind{MN} + \frac{d}{dt} E_\ind{MN} + (\mathcal{L}_Z E)_\ind{MN} + (dY)_\ind{MN} = 0 ~, \qquad E_\ind{MN} = G_\ind{MN} + B_\ind{MN} ~.
\end{equation}
Here $t=\log\frac{\Lambda^{*}}{\Lambda}$ is the RG flow time with $\Lambda$ the running scale.
The second term therefore accounts for the renormalization of the parameters of the model.
The third term is the Lie derivative of the tensor $E_\ind{MN}$ with respect to an arbitrary vector $Z$
\begin{equation}
(\mathcal{L}_ZE)_\ind{MN} = (-1)^{(\ind{M}+\ind{N}+1)\ind{P}} Z^\ind{P} E_{\ind{MN},\ind{P}} + (-1)^{(\ind{M}+1)\ind{P}} Z^\ind{P}{}_{,\ind{M}} E_\ind{PN}
+ E_\ind{MP} Z^\ind{P}{}_{,\ind{N}} ~.
\end{equation}
Pulling back to the worldsheet, divergences of this type can be removed via wavefunction renormalization.
The final term is the exterior derivative of an arbitrary one-form $Y$
\begin{equation}
(dY)_\ind{MN} = - Y_{\ind{M},\ind{N}} + (-1)^{\ind{MN}} Y_{\ind{N},\ind{M}} ~,
\end{equation}
and becomes a total derivative upon pulling back to the worldsheet.

Substituting the metric and Kalb-Ramond field of the deformed $OSP(N|2m)$ sigma model for $m=1$ with $N=1,\ldots,6$ into the Ricci flow equation \eqref{eq:ricci} we indeed find that it is one-loop renormalizable, with the parameters $\nu$ and $\eta$ satisfying
\begin{equation}\label{eq:rgflow}
\frac{d\nu}{dt} = 0 ~, \qquad \frac{d\eta}{dt} = - \nu (N-2m-2) (1+\eta^2) ~.
\end{equation}
which is also the expected result for general $N$ and $m$.
In particular, it agrees with the known result for $m=0$ \cite{Squellari:2014jfa,Litvinov:2018bou}, i.e. the deformed $O(N)$ sigma model, and furthermore, taking the undeformed limit, $\nu = \eta R^{-2}$ with $\eta \to 0$, we find the familiar RG flow equation for the radius of the $OSP(N|2m)$ sigma model
\begin{equation}
\frac{dR^2}{dt} = - (N-2m-2) ~.
\end{equation}
The expressions for the non-vanishing components of the vector $Z$ and one-form $Y$ are given in appendix \ref{app:vecofocomp}.

Solving the RG flow equations \eqref{eq:rgflow} for real $\eta$ we find cyclic solutions.
This motivates us to consider the analytically-continued regime
\begin{equation}\label{eq:etakapac}
\nu \to i \nu ~, \qquad \eta \to i \kappa ~,
\end{equation}
in which we have ancient solutions and a UV fixed point, i.e. $\kappa = 1$.
In this regime the solution to \eqref{eq:rgflow} is
\begin{equation}\label{eq:flownukap}
\nu = \mathrm{constant} ~, \qquad \kappa = - \tanh\big(\nu(N-2m-2) t\big) ~.
\end{equation}
In the following subsection we will consider the expansion around this UV fixed point.
\subsection{UV limit of deformed sigma models}
Let us now turn to the example of $OSP(5|2)$.
The deformed sigma model is parametrized by four bosons, $\phi_1$, $\phi_2$, $r_1$ and $r_2$, and a symplectic fermion, $\psi^a$, where $a=1,2$.
The Lagrangian is given by
\unskip\footnote{
Here $\tilde{\mathcal{L}}$ is normalised such that $\mathcal{S} = \frac{1}{4\pi}\int d^2\xi \, \tilde{\mathcal{L}}$, while for $\mathcal{L}$ in section \ref{QFT} we have $\mathcal{S} = \int d^2\xi \, \mathcal{L}$.
}
\par\vspace{-10pt}
\begingroup\small
\begin{equation}\begin{split}\label{eq:deflag52}
\tilde{\mathcal{L}}^{(0)}_{\scriptscriptstyle{OSP(5|2)}} & =
\frac{\kappa(1-\kappa^2r_1^2+(1+\kappa^2r_1^2)\psi \cdot \psi)}{\nu(1-\kappa^2 r_1^2)^2}\big[
\frac{\partial_+ r_1\partial_- r_1}{1-r_1^2}
+ (1-r_1^2)\partial_+\phi_1 \partial_-\phi_1
\\ & \hspace{175pt}
+ i\kappa r_1(1+\psi\cdot\psi)( \partial_+ r_1 \partial_-\phi_1 - \partial_+ \phi_1 \partial_- r_1)
\big]
\\
& \ \  + \frac{\kappa r_1^2(1-\kappa^2 r_1^4 r_2^2 +(1+\kappa^2r_1^4r_2^2)\psi\cdot\psi)}{\nu(1-\kappa^2r_1^4 r_2^2)^2}\big[
\frac{\partial_+ r_2 \partial_- r_2}{1-r_2^2}
+ (1-r_2^2)\partial_+ \phi_2 \partial_- \phi_2
\\ & \hspace{175pt}
+ i\kappa r_1^2 r_2(1+\psi\cdot\psi) ( \partial_+ r_2 \partial_-\phi_2 - \partial_+ \phi_2\partial_- r_2 )
\big] 
\\
& \ \  - \frac{\kappa(1-\kappa^2+\tfrac12(1+\kappa^2)\psi\cdot\psi )}{\nu(1-\kappa^2)^2}\big[
\partial_+\psi\cdot\partial_-\psi
- i\kappa (1+\tfrac12\psi\cdot\psi)\partial_+ \psi \wedge \partial_- \psi\big] ~,
\end{split}\end{equation}
\endgroup
where we have implemented the analytic continuation \eqref{eq:etakapac} and introduced the following contractions of the symplectic fermion
\begin{equation}
\chi\cdot\chi' = \epsilon_{ab} \chi^a \chi'{}^b ~, \qquad
\chi \wedge \chi' = \delta_{ab} \chi^a \chi'{}^b ~.
\end{equation}
We are interested in the expansion around the UV fixed point, i.e. $\kappa = 1$.
The specific limit we consider \cite{Litvinov:2018bou} is given by first setting
\begin{equation}
r_1 = \exp(-\epsilon e^{-2 x_1}) ~, \qquad r_2 = \tanh x_2 ~, \qquad \psi^a = \epsilon \theta^a ~, \qquad \kappa = 1-\frac{\epsilon^2}{2} ~,
\end{equation}
and subsequently expanding around $\epsilon = 0$.
Introducing the complex fields
\begin{equation}
X_1 = x_1-i\phi_1 ~, \qquad X_2 = x_2 - i\phi_2 ~, \qquad \Theta = \theta^1 - i \theta^2 ~,
\end{equation}
we find the following expansion
\begin{equation}\begin{split}
\tilde{\mathcal{L}}^{(0)}_{\scriptscriptstyle{OSP(5|2)}} & = \frac{1}{\nu}\Big(\partial_+X_1\partial_- X_1^* + \partial_+ X_2 \partial_- X_2^*  +i(1-i\Theta\Theta^*)\partial_+\Theta \partial_- \Theta^*\Big)
\\ & \ \ - \frac{\epsilon}{\nu} \Big(\frac12e^{2x_1}(1 + 2 i \Theta\Theta^*)\partial_+ X_1 \partial_- X_1^*
\\ & \qquad\qquad\qquad\ \ + e^{-2x_1+2x_2}\partial_+ X_2\partial_-X_2^*
+ e^{-2x_1-2x_2}\partial_+X_2^*\partial_- X_2 \Big) + O(\epsilon^2) ~,
\end{split}\end{equation}
up to total derivatives.
This does not match the effective Lagrangian \eqref{dual-Lagrangian-OSP(5|2)-eff} found from the screening charge construction in section \ref{QFT}, and it appears that it is not possible to recover \eqref{dual-Lagrangian-OSP(5|2)-eff} starting from \eqref{eq:deflag52}.

To resolve this mismatch we recall that in the case of supergroups the $\eta$-deformation is not unique.
So far we have been working with the operator $\mathcal{R}$ associated to the root system of the distinguished Dynkin diagram
\begin{equation}\label{eq:alternativedynkin}
\begin{tikzpicture}[baseline=-0.1cm]
\node (n1) at (0,0) {};
\node (n2) at (1,0) {};
\node (n3) at (2,0) {};
\draw (n1) -- (n2);
\draw[->-,double distance=2.2pt] (n2) -- (n3);
\draw[fill=white,cross] (n1) circle (0.15cm);
\draw[fill=white] (n2) circle (0.15cm);
\draw[fill=white] (n3) circle (0.15cm);
\end{tikzpicture} ~.
\end{equation}
However, $OSP(5|2)$ has two other Dynkin diagrams:
\begin{equation}\begin{split}\label{eq:alternativedynkins}
\begin{tikzpicture}[baseline=-0.1cm]
\node (n1) at (0,0) {};
\node (n2) at (1,0) {};
\node (n3) at (2,0) {};
\draw (n1) -- (n2);
\draw[->-,double distance=2.2pt] (n2) -- (n3);
\draw[fill=white,cross] (n1) circle (0.15cm);
\draw[fill=white,cross] (n2) circle (0.15cm);
\draw[fill=white] (n3) circle (0.15cm);
\end{tikzpicture} ~,
\qquad \qquad \qquad
\begin{tikzpicture}[baseline=-0.1cm]
\node (n1) at (0,0) {};
\node (n2) at (1,0) {};
\node (n3) at (2,0) {};
\draw (n1) -- (n2);
\draw[->-,double distance=2.2pt] (n2) -- (n3);
\draw[fill=white] (n1) circle (0.15cm);
\draw[fill=white,cross] (n2) circle (0.15cm);
\draw[fill=black] (n3) circle (0.15cm);
\end{tikzpicture} ~,
\end{split}\end{equation}
and each of these has a corresponding solution $\mathcal{R}$ of the modified classical YB equation.
For these operators $\mathcal{R}$, after implementing the analytic continuation \eqref{eq:etakapac}, the deformed Lagrangians are given by
\unskip\footnote{The explicit forms of the operators $\mathcal{R}$ can be found by interchanging a subset of the positive and negative roots in \eqref{eq:rmatrix} according to
\begin{equation*}
	e_\mu \to f_\mu ~, \qquad f_\mu \to (-1)^{[\mu]} e_\mu ~, \qquad \{\mu\} \subset \{m\} ~.
\end{equation*}
where $[\mu] = 0$ for a bosonic root and $1$ for a fermionic root.
For the first Dynkin diagram in \eqref{eq:alternativedynkins} we take $\{\mu\} = \{10\}$ while for the second we take $\{\mu\} = \{9,10\}$ where
\begin{equation*}\begin{aligned}
	e_9 & = E_{4,7} + E_{6,2} ~, & \qquad f_9 & = - E_{2,6} + E_{7,4} ~,
	\\
	e_{10} & = E_{5,7} + E_{6,1} ~, & \qquad f_{10} & = - E_{1,6} + E_{7,5} ~,
\end{aligned}\end{equation*}
with $(E_{i,j})_{kl} = \delta_{ik}\delta_{jl}$ denoting the unit matrix.}\par\vspace{-10pt}
\begingroup\small
\begin{align}
&& \begin{split}\label{eq:deflag52alt1}
\tilde{\mathcal{L}}^{(1)}_{\scriptscriptstyle{OSP(5|2)}} & =
\frac{\kappa}{\nu(1-\kappa^2 r_1^2)}\big[
\frac{\partial_+ r_1\partial_- r_1}{1-r_1^2}
+ (1-r_1^2)\partial_+\phi_1 \partial_-\phi_1
+ i\kappa r_1( \partial_+ r_1 \partial_-\phi_1 - \partial_+ \phi_1 \partial_- r_1)
\big]
\\
& \ \  + \frac{\kappa r_1^2(1-\kappa^2 r_1^4 r_2^2 +(1+\kappa^2r_1^4r_2^2)\psi\cdot\psi)}{\nu(1-\kappa^2r_1^4 r_2^2)^2}\big[
\frac{\partial_+ r_2 \partial_- r_2}{1-r_2^2}
+ (1-r_2^2)\partial_+ \phi_2 \partial_- \phi_2
\\ & \hspace{175pt}
+ i\kappa r_1^2 r_2(1+\psi\cdot\psi) ( \partial_+ r_2 \partial_-\phi_2 - \partial_+ \phi_2 \partial_- r_2)
\big]
\\
& \ \  - \frac{\kappa r_1^2(1-\kappa^2 r_1^4+\tfrac12(1+\kappa^2r_1^4)\psi\cdot\psi )}{\nu(1-\kappa^2r_1^4)^2}\big[
\partial_+\psi\cdot\partial_-\psi
- i\kappa r_1^2 (1+\tfrac12\psi\cdot\psi)\partial_+ \psi \wedge \partial_- \psi\big] ~,
\end{split}
\end{align}
\endgroup
and\par\vspace{-10pt}
\begingroup\small
\begin{align} 
&& \begin{split}\label{eq:deflag52alt2}
\tilde{\mathcal{L}}^{(2)}_{\scriptscriptstyle{OSP(5|2)}} & =
\frac{\kappa}{\nu(1-\kappa^2 r_1^2)}\big[
\frac{\partial_+ r_1\partial_- r_1}{1-r_1^2}
+ (1-r_1^2)\partial_+\phi_1 \partial_-\phi_1
+ i\kappa r_1( \partial_+ r_1 \partial_-\phi_1 - \partial_+ \phi_1 \partial_- r_1)
\big]
\\
& \ \  + \frac{\kappa r_1^2}{\nu(1-\kappa^2r_1^4 r_2^2)}\big[
\frac{\partial_+ r_2 \partial_- r_2}{1-r_2^2}
+ (1-r_2^2)\partial_+ \phi_2 \partial_- \phi_2
+ i\kappa r_1^2 r_2 ( \partial_+ r_2 \partial_-\phi_2- \partial_+ \phi_2 \partial_- r_2)
\big]
\\
& \ \  - \frac{\kappa r_1^2 r_2^2(1-\kappa^2r_1^4 r_2^4+\tfrac12(1+\kappa^2r_1^4 r_2^4)\psi\cdot\psi )}{\nu(1-\kappa^2r_1^4 r_2^4)^2}\big[
\partial_+\psi\cdot\partial_-\psi
\\ & \hspace{200pt}
- i\kappa r_1^2 r_2^2 (1+\tfrac12\psi\cdot\psi)\partial_+ \psi \wedge \partial_- \psi\big] ~.
\end{split}\end{align}
\endgroup
where the labels $(1)$ and $(2)$ refer to the first and second diagrams in \eqref{eq:alternativedynkins} respectively.
These sigma models are again renormalizable at one-loop with the parameters $\nu$ and $\kappa$ running in the same way as before \eqref{eq:flownukap}.
The expansions around the UV fixed point of \eqref{eq:deflag52alt1} and \eqref{eq:deflag52alt2} are
\begin{align}
\begin{split}
r_1 & = \exp(-\epsilon e^{-2x_1}) ~, \qquad r_2 = \tanh x_2 ~, \qquad  \psi^a = 2\epsilon^{\frac12} \theta ~, \qquad \kappa = 1-\frac{\epsilon^2}{2} ~,
\\
\tilde{\mathcal{L}}^{(1)}_{\scriptscriptstyle{OSP(5|2)}} & = \frac{1}{\nu}\big(\partial_+X_1\partial_-X_1^* + \partial_+ X_2 \partial_- X_2^*
+ie^{2x_1}(1-ie^{2x_1}\Theta\Theta^*)\partial_+\Theta\partial_-\Theta^*\big)
\\ & \ \  - \frac{\epsilon}{\nu} \big(
e^{2x_1}\partial_+X_1\partial_-X_1^*
+ e^{-2x_1+2x_2}(1+2ie^{2x_1}\Theta\Theta^*)\partial_+X_2\partial_-X_2^*
\\ & \qquad \qquad \qquad \qquad \quad \ \
+ e^{-2x_1-2x_2}(1+2ie^{2x_1}\Theta\Theta^*)\partial_+X_2^*\partial_-X_2
\\ & \qquad \qquad \qquad\qquad \quad \ \
+\frac{i}{4}e^{4x_1}(1-2ie^{2x_1}\Theta\Theta^*)\partial_+\Theta\partial_-\Theta^*
\big) + O(\epsilon^2) ~,
\end{split}
\end{align}
and
\begin{align}
\begin{split}
r_1 & = \exp(-\epsilon e^{-2x_2}) ~, \qquad r_2 = \exp(-\epsilon^{\frac12} e^{-2x_2}) ~, \qquad  \psi^a = 2\epsilon^{\frac14} \theta ~, \qquad \kappa = 1-\frac{\epsilon^2}{2} ~,
\\
\tilde{\mathcal{L}}^{(2)}_{\scriptscriptstyle{OSP(5|2)}} & = \frac{1}{\nu}\big(\partial_+X_1\partial_-X_1^* + \partial_+ X_2 \partial_- X_2^*\big)
+ie^{2x_2}(1-ie^{2x_2}\Theta\Theta^*)\partial_+\Theta\partial_-\Theta^*\big)
\\ &  \ \  - \frac{\epsilon^{\frac12}}{\nu} \big(
2e^{-2x_1+2x_2}\partial_+ X_2\partial_- X_2^*
+i e^{-2 x_1 + 4 x_2}(1-2ie^{2x_2}\Theta\Theta^*)\partial_+\Theta\partial_-\Theta^*
\big) + O(\epsilon) ~,
\end{split}
\end{align}
again up to total derivatives and with
\begin{equation}
X_1 = x_1-i\phi_1 ~, \qquad X_2 = x_2 - i\phi_2 ~, \qquad \Theta = \theta^1 - i \theta^2 ~.
\end{equation}

Up to the normalizations of the fields, $\tilde{\mathcal{L}}^{(1)}_{\scriptscriptstyle{OSP(5|2)}}$ indeed matches the effective Lagrangian \eqref{dual-Lagrangian-OSP(5|2)-eff} as claimed.
Therefore, we conjecture that it is the deformed sigma model \eqref{eq:deflag52alt1} that is dual to the Toda QFT \eqref{OSP(5|2)-Lagrangian-scaled}, which is constructed from the screening charges associated to the diagram \eqref{OSP(5|2)-diagram}.
It is not clear if it is possible to construct similar duals for the other two deformations \eqref{eq:deflag52} and \eqref{eq:deflag52alt2}.
\subsection{\texorpdfstring{$OSP(N|2m)$}{OSP(N|2m)} from \texorpdfstring{$O(N+2m)$}{O(N+2m)}}\label{OSP-fermionization-trick}
To conclude this section, let us comment on an interesting observation.
The three Lagrangians \eqref{eq:deflag52}, \eqref{eq:deflag52alt1} and \eqref{eq:deflag52alt2} can all be found from the deformed $O(7)$ sigma model using a trick that is reminiscent of the analytic continuations relating the three inequivalent deformations of the $O(2,4)$ sigma model to the $O(6)$ sigma model \cite{Hoare:2016ibq}.
Starting from the Lagrangian of the deformed $O(7)$ sigma model written in the form\par\vspace{-10pt}
\begingroup\small
\begin{equation}\begin{split}
\tilde{\mathcal{L}}_{\scriptscriptstyle{O(7)}} & =
\frac{\kappa}{\nu(1-\kappa^2 r_1^2)}\big[
\frac{\partial_+ r_1\partial_- r_1}{1-r_1^2}
+ (1-r_1^2)\partial_+\phi_1 \partial_-\phi_1
+ i\kappa r_1( \partial_+ r_1 \partial_-\phi_1 -\partial_+ \phi_1 \partial_- r_1)
\big]
\\
& \ \  + \frac{\kappa r_1^2}{\nu(1-\kappa^2r_1^4 r_2^2)}\big[
\frac{\partial_+ r_2 \partial_- r_2}{1-r_2^2}
+ (1-r_2^2)\partial_+ \phi_2 \partial_- \phi_2
+ i\kappa r_1^2 r_2 ( \partial_+ r_2 \partial_-\phi_2- \partial_+ \phi_2 \partial_- r_2)
\big]
\\
& \ \  + \frac{\kappa r_1^2 r_2^2}{\nu(1-\kappa^2 r_1^4 r_2^4 r_3^2)}\big[
\frac{\partial_+ r_3 \partial_- r_3}{1-r_3^2}
+ (1-r_3^2)\partial_+ \phi_3 \partial_- \phi_3
+ i\kappa r_1^2 r_2^2 r_3 ( \partial_+ r_3 \partial_-\phi_3- \partial_+ \phi_3 \partial_- r_3)\big] ~,
\end{split}\end{equation}
\endgroup
we pick one of the three two-spheres parametrized by $(r_i,\phi_i)$ and change to stereographic coordinates
\begin{equation}
\frac{z}{\sqrt{2}} = \sqrt{\frac{1-r_i}{1+r_i}} e^{i\phi_i} ~.
\end{equation}
We then formally replace
\begin{multline}\label{eq:replacement}
z \bar z \to - \tfrac12 \psi \cdot \psi ~, \\
\partial_+ z \partial_- \bar z + \partial_+ \bar z \partial_- z \to - \partial_+\psi \cdot \partial_- \psi ~,
\quad \partial_+ z \partial_- \bar z - \partial_+ \bar z \partial_- z \to i \partial_+\psi \wedge \partial_- \psi ~.
\end{multline}
Implementing this trick with each of the three two-spheres in $\tilde{\mathcal{L}}_{\scriptscriptstyle{O(7)}}$ gives the three Lagrangians \eqref{eq:deflag52}, \eqref{eq:deflag52alt1} and \eqref{eq:deflag52alt2} that follow from the three inequivalent Dynkin diagrams of $OSP(5|2)$.

It is therefore natural to conjecture that starting from the metric and Kalb-Ramond field of the deformed $O(N+2m)$ sigma model ($i=1,\dots,\frac{N}{2}+m-1$ for even $N$ and $i=1,\dots,\frac{N}{2}+m-\frac{1}{2}$ for odd $N$)
\begin{equation}\begin{aligned}\label{eq:defmeton}
G_{r_ir_i} & = \frac{\eta \big(\prod_{j=1}^{i-1} r_j^2\big)}{\nu(1+\eta^2  \big(\prod_{j=1}^{i-1} r_j^4 \big) r_i^2)(1-r_i^2)} ~,
\qquad
G_{\phi_i\phi_i} = \frac{\eta \big(\prod_{j=1}^{i-1} r_j^2 \big) (1-r_i^2)}{\nu(1+\eta^2  \big(\prod_{j=1}^{i-1} r_j^4 \big) r_i^2)} ~,
\\
B_{r_i\phi_i} & =
- B_{\phi_ir_i} =
- \frac{\eta^2  \big(\prod_{j=1}^{i-1} r_j^4 \big)r_i}{\nu(1+\eta^2  \big(\prod_{j=1}^{i-1} r_j^4 \big) r_i^2)} ~,
\end{aligned}\end{equation}
and, for even $N$,
\begin{equation}
G_{\phi_{\frac{N}{2}+m}\phi_{\frac{N}{2}+m}} = \frac{\eta \big(\prod_{j=1}^{\frac{N}{2}+m-1} r_j^2 \big)}{\nu} ~,
\end{equation}
we can find a Lagrangian for the deformed $O(N|2m)$ model by choosing $m$ of the two-spheres $(r_i,\phi_i)$ and implementing the trick outlined above, i.e. for the chosen two-spheres we change to stereographic coordinates and formally make the replacement \eqref{eq:replacement}.
We then expect that different choices of two-spheres will give deformations of $OSP(N|2m)$ based on operators $\mathcal{R}$ built from root systems associated to different Dynkin diagrams.
For odd $N$ there are $\frac{N}{2}+m-\frac{1}{2}$ two-spheres, and hence $\binom{\frac{N}{2}+m-\frac{1}{2}}{m}$ ways to choose $m$ two-spheres.
This matches the number of Dynkin diagrams for $OSP(N|2m)$ with $N$ odd.
For even $N$ there are $\frac{N}{2}+m-1$ two-spheres, and hence $\binom{\frac{N}{2}+m-1}{m}$ ways to choose of $m$ two-spheres.
However, the number of Dynkin diagrams for $OSP(N|2m)$ with $N$ even is $\binom{\frac{N}{2}+m}{m}$.
Therefore, either some deformations cannot be found via this trick, or they are equivalent, e.g. related by field redefinitions, to those that can be found.
The simplest case with $N$ even is $OSP(2|2)$, which has two Dynkin diagrams,
\begin{equation}
\begin{tikzpicture}[baseline=-0.1cm]
\node (n1) at (0,0) {};
\node (n2) at (1,0) {};
\draw[double distance=2.2pt] (n1) -- (n2);
\draw[fill=white,cross] (n1) circle (0.15cm);
\draw[fill=white,cross] (n2) circle (0.15cm);
\end{tikzpicture} ~,
\qquad
\begin{tikzpicture}[baseline=-0.1cm]
\node (n1) at (0,0) {};
\node (n2) at (1,0) {};
\draw[->-,double distance=2.2pt] (n2) -- (n1);
\draw[fill=white,cross] (n1) circle (0.15cm);
\draw[fill=white] (n2) circle (0.15cm);
\end{tikzpicture} ~.
\end{equation}
In this case the two corresponding Lagrangians indeed turn out to be related by a field redefinition up to a total derivative.
\section{Orthosymplectic trigonometric \texorpdfstring{$S$}{S}-matrix}\label{S-matrix}

In the Introduction we outlined the three steps for checking the duality between the deformed sigma model and the Toda QFT.
The third of these is the comparison between the perturbative $S$-matrix of the Toda QFT and the trigonometric deformation of the rational $S$-matrix corresponding to the undeformed sigma model.
This requires knowledge of solutions to the YB equation with $U_q(\widehat{\mathfrak{osp}}(N|2m))$ symmetry.
In this section we will consider such a solution based on the $U_q(\widehat{\mathfrak{osp}}(N|2m))$ $R$-matrix first found in \cite{Bazhanov:1986av}.
Subsequently, such quantum $R$-matrices were investigated in a variety of works \cite{Deguchi:1989gq,Maassarani:1994ac,Martins:1994vh,Gould_1997,Bassi:1999ua,Arnaudon:2003zw,Galleas:2004zz,dancer2005solutions,Galleas:2006kd,Karakhanyan:2009zz}.
Having fixed the overall scalar factor using braiding unitarity and crossing symmetry, we compare the expansion of the exact $S$-matrix with the perturbative $S$-matrix of the Toda QFT \eqref{OSP(5|2)-Lagrangian-scaled} for the $OSP(5|2)$ case and explore the rational limit \cite{Saleur:2001cw}.
\subsection{\texorpdfstring{$R$}{R}-matrix with \texorpdfstring{$U_q(\widehat{\mathfrak{osp}}(N=2n+1|2m))$}{Uq(osp(N|2m))} symmetry}
In this subsection we recast the non-graded version of the $U_q(\widehat{\mathfrak{osp}}(N|2m))$ $R$-matrix in a form that will be used later for the comparison that follows. $R$-matrices based on superalgebras are known to satisfy the graded YB equation
\begin{equation}\begin{split}\label{YBE_R}
& R_{i_1 i_2}^{k_1 k_2}(\theta_1)R_{k_1 i_3}^{j_1 k_3}(\theta_1+\theta_2)R_{k_2 k_3}^{j_2 j_3}(\theta_2)(-1)^{p_{i_1}p_{i_2}+p_{k_1}p_{i_3}+p_{k_2}p_{k_3}}
\\ & \qquad \qquad \qquad
= R_{i_2 i_3}^{k_2 k_3}(\theta_2)R_{i_1 k_3}^{k_1 j_3}(\theta_1+\theta_2)R_{k_1 k_2}^{j_1 j_2}(\theta_1)(-1)^{p_{i_2}p_{i_3}+p_{i_1}p_{k_3}+p_{k_1}p_{k_2}}\;,
\end{split}\end{equation}
where the indices $i_l$, $j_l$ and $k_l$ run from $1$ to $N+2m$ and $p_i$ is the grading of the component labelled by index $i$, i.e. it is $0$ for even components and $1$ for odd components.
In our case we have $N$ even and $2m$ odd components.

Using the graded permutation operator
\begin{equation}
P_{i_1 i_2}^{j_1 j_2}=(-1)^{p_{i_1}p_{i_2}}\delta_{i_1}^{j_2}\delta_{i_2}^{j_1}\;,
\end{equation}
we can introduce the non-graded $\check{R}$-matrix
\begin{equation}
\check{R}_{i_1 i_2}^{k_1 k_2}(\theta)=P_{i_1 i_2}^{j_1 j_2} R_{j_1 j_2}^{k_1 k_2}(\theta)\;,
\end{equation}
which satisfies the standard YB equation
\begin{equation}\label{YBE_R_check}
\check{R}_{i_1 i_2}^{k_2 k_1}(\theta_1)\check{R}_{k_1 i_3}^{k_3 j_1}(\theta_1+\theta_2)\check{R}_{k_2 k_3}^{j_3 j_2}(\theta_2)=\check{R}_{i_2 i_3}^{k_3 k_2}(\theta_2)\check{R}_{i_1 k_3}^{j_3 k_1}(\theta_1+\theta_2)\check{R}_{k_1 k_2}^{j_2 j_1}(\theta_1)\;.
\end{equation}

A solution of \eqref{YBE_R_check} with $U_q(\widehat{\mathfrak{osp}}(N|2m))$ symmetry in the fundamental representation was found for general $N$ and $m$ in \cite{Bazhanov:1986av}.
Here we will use the explicit expressions for this solution given in \cite{Galleas:2004zz,Galleas:2006kd}.
It will also be convenient to use a parametrization similar to that used in \cite{Fateev:2018yos} for the $O(N)$ case.
The details of this reparametrization are given in appendix \ref{app_R_matrix_parametrization}.
Restricting to the case of odd $N=2n+1$, the spectrum consists of $n$ even and $m$ odd charged particles, and one neutral particle, which is even.
We enumerate the particles as 
\begin{align}
& (A_1,\ldots,A_{n+m},A_{n+m+1},A_{n+m+2},\ldots,A_{N+2m})\;, \\ & \bar{A}_i=A_{\bar{\imath}}=A_{N+2m+1-i}\;, \quad  A_{n+m+1}=\bar{A}_{n+m+1}\;.     
\end{align}
where we have introduced $\bar{\imath}=N+2m+1-i$ such that $\bar{\imath}$ labels the conjugate of the particle labelled by $i$.
The $\check{R}$-matrix for this multiplet of particles then has the form ($k=N-2m-2=2n-2m-1$)\par\vspace{-10pt}
\begingroup\footnotesize
\begin{align}
& \check{R}_{ii}^{ii}(\theta)=\sinh k\lambda(\theta-i\pi) \sinh k\lambda\Big((-1)^{p_{i}}\theta-\frac{2i\pi}{k}\Big)\;, \quad && i \neq \bar{\imath}\;, \notag \\
& \check{R}_{\bar{\imath}i}^{i\bar{\imath}}(\theta)=\sinh k\lambda\theta \sinh k\lambda\Big((-1)^{p_{i}}(\theta-i\pi)+\frac{2i\pi}{k}\Big) \;, \quad && i \neq \bar{\imath}\;, \notag \\
& \check{R}_{ii}^{ii}(\theta)=\sinh k\lambda(\theta-i\pi) \sinh k\lambda \theta-\sin 2\pi\lambda \sin k\pi\lambda\;, \quad && i=\bar{\imath}\;, \notag \\
& \check{R}_{ji}^{ij}(\theta)=\sinh k\lambda(\theta-i\pi) \sinh k\lambda\theta\;, \quad && i \neq j\;, \quad i \neq \bar{\jmath}\;, \notag \\
& \check{R}_{ij}^{ij}(\theta)=-ie^{-(s_i-s_j+k)\lambda\theta}\sinh k\lambda(\theta-i\pi) \sin 2\pi\lambda\;, \quad && i>j\;, \quad i \neq \bar{\jmath}\;, \notag \\
& \check{R}_{ij}^{ij}(\theta)=-ie^{-(s_i-s_j-k)\lambda\theta}\sinh k\lambda(\theta-i\pi) \sin 2\pi\lambda\;, \quad && i<j\;, \quad i \neq \bar{\jmath}\;, \notag \\
& \check{R}_{\bar{\imath}i}^{j\bar{\jmath}}(\theta)=ie^{-(s_i-s_j+k)\lambda(i\pi-\theta)}\frac{\varepsilon_i}{\varepsilon_j}\sinh k\lambda\theta \sin 2\pi\lambda\;, \quad && i>j\;, \notag \\
& \check{R}_{\bar{\imath}i}^{j\bar{\jmath}}(\theta)=ie^{-(s_i-s_j-k)\lambda(i\pi-\theta)}\frac{\varepsilon_i}{\varepsilon_j}\sinh k\lambda\theta \sin 2\pi\lambda\;, \quad && i<j\;, \notag \\
& \check{R}_{i\bar{\imath}}^{i\bar{\imath}}(\theta)=i\big(e^{-(s_{\bar{\imath}}-s_{i}+k)\lambda(i\pi-\theta)}(-1)^{p_i}\sinh k\lambda\theta-e^{(s_{\bar{\imath}}-s_{i}+k)\lambda\theta}\sinh k\lambda(\theta-i\pi)\big) \sin 2\pi\lambda\;, \ && i<\bar{\imath}\;, \notag \\
& \check{R}_{i\bar{\imath}}^{i\bar{\imath}}(\theta)=i\big(e^{-(s_{\bar{\imath}}-s_{i}-k)\lambda(i\pi-\theta)}(-1)^{p_i}\sinh k\lambda\theta-e^{(s_{\bar{\imath}}-s_{i}-k)\lambda\theta}\sinh k\lambda(\theta-i\pi)\big) \sin 2\pi\lambda\;, \ && i>\bar{\imath}\;,
\label{R_check_matrix_GM}
\end{align}
\endgroup
where $\theta = \vartheta_1 - \vartheta_2$ is the difference of rapidities and
\begin{equation}\label{epsilon_parameter}
    \varepsilon_i=\left\{\begin{array}{ll}
    (-1)^{-\frac{p_i}{2}} & i<\bar{\imath}\;, \\
    1 & i=\bar{\imath}\;, \\
    (-1)^{\frac{p_i}{2}} \qquad & i>\bar{\imath}\;.
    \end{array}\right.
\end{equation}
If we have
\begin{equation}\label{s_parameter_conjugation}
    s_{\bar{\imath}}=-s_i \;,
\end{equation}
then the $\check{R}$-matrix \eqref{R_check_matrix_GM} is $\boldsymbol{P}\boldsymbol{T}$-invariant
\begin{equation}\label{PT_symmetry_R_check_GM}
\check{R}_{i_1 i_2}^{j_1 j_2}(\theta)=\check{R}_{j_1 j_2}^{i_1 i_2}(\theta)\;.
\end{equation}
Furthermore, if
\begin{align}\label{s_parameter_shift}
& s_1=k+2p_1\;, \notag \\
& s_{i+1}-s_{i}=-2+2(p_{i+1}+p_{i})\;, \quad i<\bar{\imath}\;,
\end{align}
then the $\check{R}$-matrix is also invariant under crossing symmetry\footnote{For the non-zero $\check{R}$-matrix elements the usual sign factors in the crossing symmetry relation turn out to be trivial.}
\begin{equation}\label{crossing_symmetry_R_check_GM}
\check{R}_{i_1 i_2}^{j_1 j_2}(\theta)=(C^{-1})_{i_1}^{k_1}\check{R}_{k_1 j_1}^{i_2 k_2}(i\pi-\theta)(C)_{k_2}^{j_2}
\end{equation}
with the non-vanishing elements of the charge conjugation matrix elements given by 
\begin{equation}
C_{i}^{\bar{\imath}}=\varepsilon_i\;,
\end{equation}
where $\varepsilon_i$ was defined in \eqref{epsilon_parameter}.

\subsection{\texorpdfstring{$OSP(N=2n+1|2m)$}{OSP(N=2m+1|2m)} \texorpdfstring{$q$}{q}-deformed \texorpdfstring{$S$}{S}-matrix: Toda QFT}

In the previous subsection we presented the $\boldsymbol{P}\boldsymbol{T}$-invariant and crossing symmetric solution \eqref{R_check_matrix_GM} of the YB equation \eqref{YBE_R_check}.
In order to compare with the perturbative $S$-matrix of the Toda QFT \eqref{OSP(5|2)-Lagrangian-scaled}, we take the scalar factor to be same as in the $O(N)$ case \cite{Fateev:2018yos} with the familiar replacement $N \to N-2m=k+2$.

We consider the following $S$-matrix
\begin{equation}\label{S_check_matrix_GM}
    S_{i_1 i_2}^{j_1 j_2}(\theta)=F(\theta)\frac{\check{R}_{i_1 i_2}^{j_1 j_2}(\theta)}{\sinh k\lambda(\theta-i\pi) \sinh k\lambda\left(\theta-\frac{2i\pi}{k}\right)}\;,
\end{equation}
where $\check{R}_{i_1 i_2}^{j_1 j_2}(\theta)$ is given by \eqref{R_check_matrix_GM} and the scalar factor $F(\theta)$ is
\begin{equation}\label{F_scalar}
F(\theta)=-\exp\left(i\int\limits_{-\infty}^{+\infty}d\omega\frac{\cosh\frac{\pi\omega(k-2)}{2k}\sinh\frac{\pi\omega}{k}\left(\frac{1}{2\lambda}-1\right)\sin\omega\theta}{\omega\cosh\frac{\pi\omega}{2}\sinh\frac{\pi\omega}{2k\lambda}}\right)\;.
\end{equation}
The scalar factor \eqref{F_scalar} satisfies
\begin{align}
    & F(\theta)F(-\theta)=1\;, \label{F_br_unit} \\
    & F(i\pi-\theta)=F(\theta)\frac{\sinh k\lambda\theta \sinh k\lambda\left(\theta-\frac{i\pi(k-2)}{k}\right)}{\sinh k\lambda(\theta-i\pi) \sinh k\lambda\left(\theta-\frac{2i\pi}{k}\right)}\;, \label{F_cr_sym}
\end{align}
ensuring that the $S$-matrix satisfies braiding unitarity
\begin{equation}\label{braiding_unitarity_S}
S_{i_1 i_2}^{j_1 j_2}(\theta)S_{j_1 j_2}^{k_1 k_2}(-\theta)=\delta_{i_1}^{k_1}\delta_{k_1}^{k_2}
\end{equation}
and is crossing symmetric \eqref{S_check_matrix_GM} assuming that \eqref{s_parameter_shift} holds.
If \eqref{s_parameter_conjugation} also holds then the $S$-matrix is also $\boldsymbol{P}\boldsymbol{T}$-invariant \eqref{PT_symmetry_R_check_GM}.
As a consistency check, if we take $m=0$ then the $S$-matrix \eqref{S_check_matrix_GM} coincides with the $O(N)$ $q$-deformed $S$-matrix as presented in \cite{Litvinov:2018bou} for odd $N=2n+1$ as expected.

Let us now consider the limit $\lambda \to \frac12$.
By the usual relation between $\lambda$ and $b$ we have $\lambda=\frac{1}{2}+O(b^2)$.
Therefore, this limit corresponds to $b\to 0$, i.e. the perturbative limit of the Toda QFT.
It turns out that for $OSP(5|2)$ the Lagrangian description of this model \eqref{OSP(5|2)-Lagrangian-scaled} corresponds to the grading choice $p_i = (0,1,0,0,0,1,0)_i$, while for $OSP(7|2)$, \eqref{OSP(7|2)-Lagrangian-1} corresponds to $p_i = (0,1,0,0,0,0,0,1,0)_i$ and \eqref{OSP(7|2)-Lagrangian-2} to $p_i = (0,0,1,0,0,0,1,0,0)_i$.
The computation of the perturbative $S$-matrix of these models is potentially complicated by the bosonic spinor $\uppsi$ and understanding how to properly continue to Lorentzian signature.
The approach we use is described in appendix \ref{app_OSP(5|2)_S_matrix} and involves first integrating over half the degrees of freedom in $\uppsi$.
While this allows us to compute the tree-level $S$-matrix, we drop a determinant contribution that will be relevant at higher orders.
We have calculated the full tree-level $S$-matrix for the $OSP(5|2)$ case as well as the functional dependence of the matrix elements not involving $\uppsi$ on the rapidity difference $\theta$ for both $OSP(7|2)$ Lagrangians.
In all cases we find agreement with the corresponding matrix elements of \eqref{S_check_matrix_GM}.

The precise statement for the $OSP(5|2)$ case with grading $p_i = (0,1,0,0,0,1,0)_i$ is as follows.
We first apply the gauge transformation
\unskip\footnote{See Lemma 10 in subsection 12.2.5 of \cite{essler_frahm_goehmann_kluemper_korepin_2005} for the definition of the gauge transformation.}
determined by the matrix
\begin{equation}
    M(\vartheta)=\begin{pmatrix}
    1 & 0 & 0 & 0 & 0 & 0 & 0 \\
    0 & e^{-\frac{\vartheta}{2}} & 0 & 0 & 0 & 0 & 0 \\
    0 & 0 & 1 & 0 & 0 & 0 & 0 \\
    0 & 0 & 0 & 1 & 0 & 0 & 0 \\
    0 & 0 & 0 & 0 & 1 & 0 & 0 \\
    0 & 0 & 0 & 0 & 0 & -e^{\frac{\vartheta}{2}} & 0 \\
    0 & 0 & 0 & 0 & 0 & 0 & 1
    \end{pmatrix}~,
\end{equation}
and the twist transformation
\unskip\footnote{See Lemma 11 and Lemma 12 in subsection 12.2.5 of \cite{essler_frahm_goehmann_kluemper_korepin_2005} for the definition of the twist transformation.}
that changes the sign of the mutual braiding of $\psi_1$ and $\psi_2$, i.e. the sign of the components $S_{13}^{31}$, $S_{31}^{13}$, $S_{15}^{51}$, $S_{51}^{15}$, $S_{37}^{73}$, $S_{73}^{37}$, $S_{57}^{75}$ and $S_{75}^{57}$.
The resulting matrix
\begin{equation}
    S_{\mathrm{tr}}{}_{i_1 i_2}^{j_1 j_2}(\vartheta_1-\vartheta_2)=(M(\vartheta_2)^{-1})_{i_1}^{k_1}(M(\vartheta_1)^{-1})_{i_2}^{k_2}S_{\mathrm{tw}_1}{}_{k_1 k_2}^{l_1 l_2}(\vartheta_1-\vartheta_2)M_{l_1}^{j_1}(\vartheta_1)M_{l_2}^{j_2}(\vartheta_2) ~,
\end{equation}
where $\textrm{tw}_1$ denotes the twist transformation,
reproduces the tree level $S$-matrix of \eqref{OSP(5|2)-Lagrangian-scaled} when expanded to $O(\lambda-1/2) \sim O(b^2)$.

\subsection{\texorpdfstring{$OSP(N=2n+1|2m)$}{OSP(N=2n+1|2m)} \texorpdfstring{$q$}{q}-deformed \texorpdfstring{$S$}{S}-matrix: rational limit}

In the limit $\lambda\to0$, or $b\to\infty$, we expect to recover the undeformed $OSP(N|2m)$ sigma model.
The rational $S$-matrices of these sigma models were first calculated in \cite{Saleur:2001cw} and subsequently studied in \cite{Saleur:2003zm,Saleur:2009bf}.
Therefore, we now check whether the $S$-matrix \eqref{S_check_matrix_GM}, which indeed becomes rational as $\lambda\to0$, matches these results in this limit.

To write the $S$-matrix of \cite{Saleur:2001cw} in a convenient form we recall the definition of the orthosymplectic metric tensor
\begin{equation}
    J_{ij}=(1-p_i)\delta_{ij}+p_i\delta_{i\bar{\jmath}}\,\textrm{sign}(i-\bar{\imath})\;.
\end{equation}
Explicitly for $OSP(5|2)$ and the grading choice $p_i = (0,1,0,0,0,1,0)_i$ we have
\begin{equation}
    J=\begin{pmatrix}
    1 & 0 & 0 & 0 & 0 & 0 & 0 \\
    0 & 0 & 0 & 0 & 0 & -1 & 0 \\
    0 & 0 & 1 & 0 & 0 & 0 & 0 \\
    0 & 0 & 0 & 1 & 0 & 0 & 0 \\
    0 & 0 & 0 & 0 & 1 & 0 & 0 \\
    0 & 1 & 0 & 0 & 0 & 0 & 0 \\
    0 & 0 & 0 & 0 & 0 & 0 & 1 \\
    \end{pmatrix}\;.
\end{equation}
Defining
\begin{equation}
    E_{i_1 i_2}^{j_1 j_2}=J_{i_1 i_2}J^{j_1 j_2}\;,
\end{equation}
where $J^{j_1 j_2}=(J^{-1})_{j_1 j_2}$ is the matrix inverse of the orthosymplectic metric tensor, the $S$-matrix \cite{Saleur:2001cw} of the $OSP(N|2m)$ sigma model is
\begin{equation}\label{S_matrix_rational}
S_{\mathrm{rat}}{}_{i_1 i_2}^{j_2 j_1}(\theta)=\sigma_1(\theta)E_{i_1 i_2}^{j_2 j_1}+\sigma_2(\theta)P_{i_1 i_2}^{j_2 j_1}+\sigma_3(\theta)I_{i_1 i_2}^{j_2 j_1}\;.
\end{equation}
The coefficients of the tensor structures satisfy
\begin{equation}\label{sigmas}
\sigma_1(\theta)=-\frac{2i\pi}{k(i\pi-\theta)}\sigma_2(\theta)\;, \quad \sigma_3(\theta)=-\frac{2i\pi}{k\theta}\sigma_2(\theta)\;,
\end{equation}
with $\sigma_2(\theta)$ the same as in the $O(N)$ case, again with the replacement $N \to N-2m=k+2$
\begin{equation}\label{sigma2}
\sigma_2(\theta)=\frac{\Gamma\left(1-\frac{\theta}{2i\pi}\right)}{\Gamma\left(\frac{\theta}{2i\pi}\right)}\frac{\Gamma\left(\frac{1}{2}+\frac{\theta}{2i\pi}\right)}{\Gamma\left(\frac{1}{2}-\frac{\theta}{2i\pi}\right)}\frac{\Gamma\left(\frac{1}{k}+\frac{\theta}{2i\pi}\right)}{\Gamma\left(1+\frac{1}{k}-\frac{\theta}{2i\pi}\right)}\frac{\Gamma\left(\frac{1}{2}+\frac{1}{k}-\frac{\theta}{2i\pi}\right)}{\Gamma\left(\frac{1}{2}+\frac{1}{k}+\frac{\theta}{2i\pi}\right)}\;.
\end{equation}

To establish the relation with the $S$-matrix \eqref{S_check_matrix_GM} we start with the scalar factor, noting that
\begin{equation}
    \left. F(\theta) \right|_{\lambda=0}=\frac{\theta-\frac{2i\pi}{k}}{\theta}\sigma_2(\theta)\;.
\end{equation}
Now if we apply the gauge transformation determined by the matrix
\begin{equation}\label{V_rotation}
    V_j^l=\frac{1}{\sqrt{2}}\Big(\delta_{jl}(-i)^{1-\Theta(\bar{\jmath}-j)}+\delta_{j\bar{l}}(1-p_j)i^{1-\Theta(\bar{l}-l)}\Big)\;,
\end{equation}
where $\Theta(x)$ is the Heaviside step function, and apply the twist transformation that changes the sign of the mutual braiding of the odd particles, denoted $\textrm{tw}_2$, to the $S$-matrix \eqref{S_check_matrix_GM}, then at leading order in the limit $\lambda\to0$, we indeed find the rational $S$-matrix \eqref{S_matrix_rational}
\begin{equation}
    S_{\mathrm{rat}}{}_{i_1 i_2}^{j_1 j_2}(\theta)=\left.\left((V^{-1})_{i_1}^{k_1}(V^{-1})_{i_2}^{k_2}S_{\mathrm{tw}_2}{}_{k_1 k_2}^{l_1 l_2}(\theta)V_{l_1}^{j_1}V_{l_2}^{j_2}\right)\right|_{\lambda=0}\;.
\end{equation}
For the $OSp(5|2)$ case the gauge transformation matrix takes the form
\begin{equation}
    V=\frac{1}{\sqrt{2}}\begin{pmatrix}
    1 & 0 & 0 & 0 & 0 & 0 & i \\
    0 & 1 & 0 & 0 & 0 & 0 & 0 \\
    0 & 0 & 1 & 0 & i & 0 & 0 \\
    0 & 0 & 0 & 1 & 0 & 0 & 0 \\
    0 & 0 & 1 & 0 & -i & 0 & 0 \\
    0 & 0 & 0 & 0 & 0 & -i & 0 \\
    1 & 0 & 0 & 0 & 0 & 0 & -i \\
    \end{pmatrix}\;,
\end{equation}
which illustrates that each pair of even and odd particles is rotated by the matrices
\begin{equation}
    \begin{pmatrix}
    1 & i \\
    1 & -i
    \end{pmatrix}\;, \quad
    \begin{pmatrix}
    1 & 0 \\
    0 & -i
    \end{pmatrix}\;,
\end{equation}
respectively.

Therefore, the proposed $S$-matrix \eqref{S_check_matrix_GM} agrees with the known rational $S$-matrices corresponding to the asymptotically free ($k>0$) undeformed $OSP(N|2m)$ sigma model considered in \cite{Saleur:2001cw}.
\section{Concluding remarks}\label{concl}

In this paper we investigated $\eta$-deformations of integrable sigma models and their dual Toda-like description.
In particular, we formulated and checked the conjecture of duality for sigma models on supermanifolds.
One of the key new features in the case of supermanifolds is the existence of inequivalent $\eta$-deformations corresponding to different choices of simple roots.
This freedom manifests itself in all aspects of the duality and is the source of a number of complications.
This meant that we were restricted to formulating the duality for only a certain class of $\eta$-deformations.
Let us now summarise this construction.
 
In section \ref{Screenings} we found the system of screening charges corresponding to the asymptotically free $OSP(2n+1|2m)$ sigma model ($n-m>0$).
To do this we introduced the injection transformation $\mathfrak{J}$, which was applied to any $m$ odd roots $\boldsymbol{\alpha}_{2k-1}$ with $k=2,\dots,n$ in the affine $O(2n+1)$ diagram \eqref{diagram-O(2n+1)-affine}.
Accordingly, we have $\binom{n-1}{m}$ different ways to choose the roots $\boldsymbol{\alpha}_{2k-1}$ and we claim that different choices lead to inequivalent field theories.
Any such choice corresponds to a sequence of the form
\begin{equation}\label{choice-of-order-of-the-roots}
        1<k_1<k_2<\dots<k_m\leq n ~.
\end{equation}
The number of choices $\binom{n-1}{m}$ is less than the number of Dynkin diagrams for $OSP(2n+1|2m)$, which is $\binom{n+m}{m}$, and thus many choices remain unidentified.
It remains an open problem to determine if there exists a dual Toda QFT for these unidentified cases and if so, what form it takes.

In section \ref{QFT} we described how to construct the weak-coupling Toda QFT ($b\rightarrow0$) and the strong-coupling sigma model ($b\rightarrow\infty$) from a given system of screening charges.
In the weak-coupling theory we used boson-fermion/boson-boson correspondence to rewrite the action in terms of suitable microscopic degrees of freedom and determined the counterterms required to improve the UV behaviour.
With these terms taken into account we can reproduce the tree-level $S$-matrix that agrees with the expansion of the exact $S$-matrix constructed in section \ref{S-matrix}.
Furthermore, in the strong-coupling sigma model we demonstrated that, after integrating over half of the fermionic degrees of freedom, the Lagrangian matches the UV expansion of a certain $\eta$-deformed sigma model.

In section \ref{SM} we studied the $\eta$-deformed $OSP(N|2m)$ sigma model.
In particular, we found that for inequivalent choices of Dynkin diagram we obtain different field theories.
The different deformed backgrounds can be found using a ``fermionization'' trick, described in detail in subsection \ref{OSP-fermionization-trick}.
In essence, for odd $N = 2n+1$, we parametrize the $O(2n+2m+1)$ sigma model in terms of $n+m$ nested two-spheres.
We then choose $m$ of these two-spheres and replace the bosonic coordinates by fermionic ones according to a given set of rules.
There are $\binom{n+m}{m}$ ways of doing this, which matches the number of Dynkin diagrams.
However, we found dual descriptions only for a subset of these.
The nesting of two-spheres in the $O(2n+2m+1)$ model gives a natural ordering.
In order to find a deformed sigma model whose dual description we know, we do not fermionize the first and last two-spheres or any two neighbouring two-spheres.
This setup can be represented by a sequence of $n+m$ circles, which we fill in if the corresponding two-sphere has been ``fermionized.''
A typical configuration will bes
    \begin{equation}\label{SM-fermionization-rule}
	    \psfrag{1}{$z_1$}
	    \psfrag{2}{$\psi_1$}
	    \psfrag{3}{$z_{k_1}$}
	    \psfrag{4}{$\psi_2$}
	    \psfrag{5}{$z_{k_2}$}
	    \psfrag{6}{$z_n$}
		\includegraphics[width=.7\textwidth]{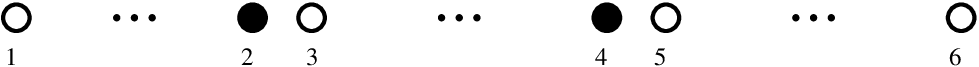} \qquad \ .
     \end{equation}
Note this sequence is of exactly the same form as \eqref{choice-of-order-of-the-roots}.

In section \ref{S-matrix} we investigated the trigonometric $OSP(N|2m)$ $S$-matrix ($N=2n+1$), obtained from Bazhanov and Shadrikov's solution to the YB equation \cite{Bazhanov:1986av} by a certain basis transformation and multiplication by an overall factor.
Again, the key difference with the case of classical Lie groups is that there are different $S$-matrices corresponding to inequivalent Dynkin diagrams.
The particle content consists of one neutral bosonic particle $\Phi$ and $n+m$ charged particles $(p_k^{\vphantom{*}},p_k^*)$, where $n$ are fermionic and $m$ are bosonic.
Together they form an $OSP(2n+1|2m)$ multiplet
\begin{equation}
    (p_1^{\vphantom{*}},\dots,p_{n+m}^{\vphantom{*}},\Phi,p_{n+m}^*,\dots,p_1^*)~,
\end{equation}
and for different choices of gradings, i.e. different choices of statistics for the particles $p_k$, we have different $S$-matrices.
However, we have only been able to find a perturbative interpretation of the $S$-matrix for special choices of gradings, which is in a sense the opposite of \eqref{SM-fermionization-rule}.
Namely, $p_1$  and $p_{n+m}$ should be fermionic (denoted by $\psi$) and any bosonic particle $p_k$ (denoted by $\uppsi$)  should be surrounded by two fermionic particles
\begin{equation}\label{grading-rule}
    (\psi_1,\dots,\uppsi_1,\psi_{k_1},\dots,\uppsi_2,\psi_{k_2},\dots,\psi_n,\Phi,\psi_n^*,\dots,\psi_{k_2}^*,\uppsi_2^*,\dots,\psi_{k_1}^*,\uppsi_1^*,\dots,\psi_1^*) ~,
\end{equation}
This sequence is again exactly the same as in \eqref{choice-of-order-of-the-roots} and \eqref{SM-fermionization-rule}.

We have focused on the case of odd $N$ in the asymptotically free regime; however, the case of even $N$ is similar.
Starting from the system of screening charges for the $O(2n)$ theory \cite{Fateev:2018yos,Litvinov:2018bou} we can obtain an $OSP(2n|2m)$ system by applying the injection transformation $\mathfrak{J}$ to any root $\boldsymbol{\alpha}_{2k-1}$ with $k=2,\dots,n-1$.
Using this procedure we cannot cross the line of asymptotically free theories since at most $n-2$ roots $\boldsymbol{\alpha}_{2k-1}$ can be injected, yielding the $OSP(2n|2n-4)$ theory, which is asymptotically free.
By counting the number of degrees of freedom, we cannot exclude that a dual description of the conformal $OSP(2n|2n-2)$ sigma models exists.
We leave this question for future investigation.
\section*{Acknowledgments}

We are grateful to Dmitri Bykov, Vladimir Fateev, Frank G\"ohmann, Nat Levine, Ruslan Metsaev
, Pavel Pyatov
, Fiona Seibold, Arkady Tseytlin, Stijn van Tongeren and Konstantin Zarembo, for fruitful and stimulating discussions.
M.A. would like to thank HU Berlin, University of Montpellier, ENS Paris and NORDITA, where part of this research was done, for their kind hospitality.
B.H. would like to thank HSE University and the Skolkovo Institute of Science and Technology, where part of this research was done, for their kind hospitality.
The research of B.F. was carried out within the HSE University Basic Research Program and funded by Russian Academic Excellence Project “5-100”.
The work of B.H. was supported by the Swiss National Science Foundation through the NCCR SwissMAP and a UKRI Future Leaders Fellowship (grant number MR/T018909/1).
A.L. has been supported by the Russian Science Foundation under the grant 18-12-00439. 
\Appendix
\section{Bosonization of the Thirring model}\label{boson-fermion-correspondence}
In this appendix, mostly following \cite{Naon:1984zp}, we derive the boson-fermion/boson-boson correspondence using the path integral approach.
The boson-fermion correspondence, i.e. the equivalence between the massive Thirring and sine-Gordon models, is well known since Coleman and Mandelstam \cite{Coleman:1974bu,Mandelstam:1975hb}; however, the boson-boson correspondence is less well known. 

Consider the Euclidean massive Thirring model given by the Lagrangian ($\gamma^{1}=\sigma_{1}$, $\gamma^{2}=\sigma_{2}$, $\partial = \partial_1 - i \partial_2$)
\begin{equation}\label{Thirring-Lagrangian-spinor}
\mathcal{L}_{\textrm{T}}=i\bar{\mathcal{B}}\gamma^{\mu}\partial_{\mu}\mathcal{B}-m\bar{\mathcal{B}}\mathcal{B}-\frac{\pi\lambda^{2}}{2}
\bigl(\bar{\mathcal{B}}\gamma^{\mu}\mathcal{B}\bigr)^{2}~,\qquad
\mathcal{B}=
\begin{pmatrix}
i\chi\\
\bar{\eta}
\end{pmatrix}~,\quad
\bar{\mathcal{B}}=
\begin{pmatrix}
-i\bar{\chi}&\mp \eta
\end{pmatrix}~,
\end{equation}
where we take $-$ for Fermi and $+$ for Bose statistics.
In components this Lagrangian (after integrating by parts) has the form
\begin{equation}\label{Thirring-Lagrangian}
\mathcal{L}_{\textrm{T}}=\chi\bar{\partial}\eta+\bar{\chi}\partial\bar{\eta}-m\left(\bar{\chi}\chi+\bar{\eta}\eta\right)-2\pi\lambda^{2}\chi\eta\bar{\chi}\bar{\eta}~.
\end{equation}
We apply the Hubbard-Stratonovich transformation 
\begin{equation}
\exp\left(2\pi\lambda^{2}\int\chi\eta\bar{\chi}\bar{\eta}\, d^{2}x\right)=
\int\mathcal{D}A\mathcal{D}\bar{A}\exp\left[-\int\Bigl(\frac{A\bar{A}}{2\pi}-\lambda(\bar{A}\chi\eta+A\bar{\chi}\bar{\eta})\Bigr)d^{2}x\right],
\end{equation}
to decouple the quartic interaction. It is convenient to represent
\begin{equation}
A=\partial\bar{X} ~,\qquad \bar{A}=\bar{\partial}X ~,
\end{equation}
such that, up to a constant factor, we have
\begin{multline}
\int\mathcal{D}A\mathcal{D}\bar{A}\,\exp\left[-\int\Bigl(\frac{A\bar{A}}{2\pi}-\lambda(\bar{A}\chi\eta+A\bar{\chi}\bar{\eta})\Bigr)d^{2}x\right]
\\ \sim
\int\mathcal{D}X\mathcal{D}\bar{X}\,\exp\left[-\int\Bigl(\frac{1}{8\pi}\partial_{a}X\partial_{a}\bar{X}-\lambda(\bar{\partial}X\chi\eta+\partial\bar{X}\bar{\chi}\bar{\eta})\Bigr)d^{2}x\right].
\end{multline}
Thus the effective Lagrangian has the form
\begin{equation}\label{L-eff-Thirring}
\mathcal{L}_{\textrm{eff}}=\frac{1}{8\pi}\partial_{a}X\partial_{a}\bar{X}+\chi\bigl(\bar{\partial}-\lambda\bar{\partial}X\bigr)\eta
+\bar{\chi}\bigl(\partial-\lambda\partial\bar{X}\bigr)\bar{\eta}-m\left(\bar{\chi}\chi+\bar{\eta}\eta\right)~.
\end{equation}
We now perform the gauge transformation
\begin{equation}
\chi=e^{-\lambda X}\chi_{0}~,\quad \bar{\chi}=e^{-\lambda \bar{X}}\bar{\chi_{0}}~,\quad
\eta=e^{\lambda X}\eta_{0}~,\quad \bar{\eta}=e^{\lambda \bar{X}}\bar{\eta}_{0}~,
\end{equation}
which brings the action to the form
\begin{equation}
\mathcal{L}_{\textrm{eff}}=\frac{1}{8\pi}\partial_{a}X\partial_{a}\bar{X}+\chi_{0}\bar{\partial}\eta_{0}
+\bar{\chi}_{0}\partial\bar{\eta}_{0}-m\left(e^{-\lambda(X+\bar{X})}\bar{\chi}_{0}\chi_{0}+e^{\lambda(X+\bar{X})}\bar{\eta}_{0}\eta_{0}\right)+\mathcal{L}_{\textrm{axial}}~,
\end{equation}
where $\mathcal{L}_{\textrm{axial}}$ is a contribution coming from the axial anomaly
\begin{equation}
\mathcal{L}_{\textrm{axial}}=\pm\frac{\lambda^{2}}{2\pi}\,|\partial_{a}X|^{2}~,
\end{equation}
with $+$ corresponding to Fermi and $-$ to Bose statistics. Setting $X=x+iy$ we see that $y$ decouples and effectively we have
\begin{equation}\label{L-eff-Thirring-2}
\mathcal{L}_{\textrm{eff}}=\frac{1}{8\pi}(\partial_{a}y)^{2}+\frac{1}{8\pi}(\partial_{a}x)^{2}+\chi_{0}\bar{\partial}\eta_{0}
+\bar{\chi}_{0}\partial\bar{\eta}_{0}-m\left(e^{-2\lambda x}\bar{\chi}_{0}\chi_{0}+e^{2\lambda x}\bar{\eta}_{0}\eta_{0}\right)+\mathcal{L}_{\textrm{axial}}~.
\end{equation}
Finally substituting
\begin{equation}
x=\frac{\omega}{\sqrt{1\pm4\lambda^{2}}}~,
\end{equation}
we obtain
\begin{equation}\label{L-eff-Thirring-3}
\mathcal{L}_{\textrm{eff}}=\frac{1\pm4\lambda^2}{8\pi}(\partial_{a}y)^{2}+
\frac{1}{8\pi}(\partial_{a}\omega)^{2}+\chi_{0}\bar{\partial}\eta_{0}
+\bar{\chi}_{0}\partial\bar{\eta}_{0}-m\left(e^{-\gamma\omega}\bar{\chi}_{0}\chi_{0}+e^{\gamma\omega}\bar{\eta}_{0}\eta_{0}\right)~,
\end{equation}
where
\begin{equation}
    \gamma=\frac{2\lambda}{\sqrt{1\pm4\lambda^{2}}}~.
\end{equation}
The theory with the Lagrangian \eqref{Thirring-Lagrangian} is equivalent to the theory with the effective Lagrangian \eqref{L-eff-Thirring-3}. In particular, the perturbative expansion in $m$ should be the same. We note that in  \eqref{L-eff-Thirring-3} $\chi_0$ and $\eta_0$ are not self-interacting  and hence in perturbation theory can replace \cite{Friedan:1985ge}
\begin{equation}\label{bosonization-rules}
\begin{aligned}
&\bar{\chi}_{0}\chi_{0}=e^{-iv}\quad &&\bar{\eta}_{0}\eta_{0}=e^{iv}&&\quad\text{for fermions}~,\\
&\bar{\chi}_{0}\chi_{0}=e^{-u-iv}\quad &&\bar{\eta}_{0}\eta_{0}=e^{u+iv}(i\partial v)\,(i\bar{\partial}v) &&\quad\text{for bosons}~,
\end{aligned}
\end{equation}
where $u$ and $v$ are bosonic fields with canonical normalization
\begin{equation}
\mathcal{L}_{u}=\frac{1}{8\pi}(\partial_{a}u)^{2}~,\qquad
\mathcal{L}_{u}=\frac{1}{8\pi}(\partial_{a}v)^{2}~.
\end{equation}
This implies that the bosonization of the fields $\chi\bar{\chi}$ and $\eta\bar{\eta}$ has the following rules
\begin{equation}\label{bosonization-rules-original}
\begin{aligned}
&\bar{\chi}\chi=e^{-\gamma\omega-iv}\quad &&\bar{\eta}\eta=e^{\gamma\omega+iv} && \quad\text{for fermions}~,\\
&\bar{\chi}\chi=e^{-\gamma\omega-u-iv}\quad &&\bar{\eta}\eta=e^{\gamma\omega+u+iv}(i\partial v)\,(i\bar{\partial}v) && \quad\text{for bosons}~.\\
\end{aligned}
\end{equation}

In our parametrization we have
\begin{equation}
\lambda^{2}=\mp\frac{b^{2}}{4(1+b^{2})}~,
\end{equation}
where $-$ corresponds to Fermi and $+$ to Bose statistics. We note that the bosonization formula \eqref{bosonization-rules-original} depends only on one linear combination of the bosonic fields $\gamma\omega+iv$ in the fermionic case and on two linear combinations $iv$ and $\gamma\omega+u+iv$ in the bosonic case. Using simple redefinition of the basis one can rewrite \eqref{bosonization-rules-original} as
\begin{equation}\label{bosonization-rules-original-2}
\begin{aligned}
&\bar{\chi}\chi=e^{i\beta\varphi}\quad &&\bar{\eta}\eta=e^{-i\beta\varphi}&&\quad\text{for fermions}~,\\
&\bar{\chi}\chi=e^{b\varphi}\quad &&\bar{\eta}\eta=e^{-b\varphi}\left(\frac{1}{b}\partial\varphi-\frac{i\beta}{b}\partial\phi\right)
\left(\frac{1}{b}\bar{\partial}\varphi-\frac{i\beta}{b}\bar{\partial}\phi\right)&&\quad\text{for bosons}~,\\
\end{aligned}
\end{equation}
where $\beta=\sqrt{1+b^2}$ and
\begin{equation}
\begin{aligned}
    &\varphi=-\frac{\gamma\omega+iv}{i\beta}\quad&&\text{for fermions}~,\\
    &\varphi=-\frac{\gamma\omega+u+iv}{b},\qquad \frac{1}{b}\varphi-\frac{i\beta}{b}\phi=iv\quad&&\text{for bosons}~.
\end{aligned}
\end{equation}
\section{The \texorpdfstring{$U_q(\widehat{\mathfrak{osp}}(N=2n+1|2m))$}{Uq(osp(N=2n+1|2m))} \texorpdfstring{$\check{R}$}{R}-matrix}\label{app_R_matrix_parametrization}

The solution to the YB equation \eqref{YBE_R_check} is given in \cite{Galleas:2004zz} where it is written as an $\check{R}$-matrix acting in the $(N|2m)$ graded spaces $a$ and $b$
\begin{multline}\label{R_check_quantum_OSP_group}
\check{R}_{ab}(\mu)=\sum\limits_{\atopfrac{\alpha=1}{\alpha=\alpha'}}^{N+2m}a_{\alpha}(\mu)\hat{e}^{(a)}_{\alpha\alpha}\otimes\hat{e}^{(b)}_{\alpha\alpha}+b(\mu)\sum\limits_{\atopfrac{\alpha,\beta=1}{\alpha\neq\beta, \alpha\neq\beta'}}^{N+2m}\hat{e}^{(a)}_{\alpha\beta}\otimes\hat{e}^{(b)}_{\beta\alpha}+\bar{c}(\mu)\sum\limits_{\atopfrac{\alpha,\beta=1}{\alpha<\beta, \alpha\neq\beta'}}^{N+2m}\hat{e}^{(a)}_{\alpha\alpha}\otimes\hat{e}^{(b)}_{\beta\beta}+ \\
+c(\mu)\sum\limits_{\atopfrac{\alpha,\beta=1}{\alpha>\beta, \alpha\neq\beta'}}^{N+2m}\hat{e}^{(a)}_{\alpha\alpha}\otimes\hat{e}^{(b)}_{\beta\beta}+\sum\limits_{\alpha,\beta=1}^{N+2m}d_{\alpha\beta}(\mu)\hat{e}^{(a)}_{\alpha'\beta}\otimes\hat{e}^{(b)}_{\alpha\beta'}\;,
\end{multline}
where $\alpha'=N+2m+1-\alpha$.
The matrix $\hat{e}^{(a)}_{\alpha\beta}$ acts in the space $a$ and has $1$ at the position $(\alpha,\beta)$ and $0$ otherwise, and the coefficients $a_{\alpha}(\mu)$, $b(\mu)$, $\bar{c}(\mu)$, $c(\mu)$ and $d_{\alpha\beta}(\mu)$ (Boltzmann weights) depend on $N$, $m$ and the deformation parameter $q$.

In this appendix we focus on odd $N=2n+1$, where $n$ is a non-negative integer.
In terms of $k = N-2m-2$, the expressions for the Boltzmann weights from \cite{Galleas:2004zz} are given by
\begin{align}
& a_{\alpha}(\theta)=4e^{2k\lambda\left(\theta+\frac{(k+2)i\pi}{2k}\right)}\sinh k\lambda(\theta-i\pi) \sinh k\lambda\left((-1)^{p_{\alpha}}\theta-\frac{2i\pi}{k}\right)\;, \\
& b(\theta)=4e^{2k\lambda\left(\theta+\frac{(k+2)i\pi}{2k}\right)}\sinh k\lambda(\theta-i\pi) \sinh k\lambda \theta\;, \\
& c(\theta)=-4ie^{k\lambda\left(\theta+\frac{(k+2)i\pi}{k}\right)}\sinh k\lambda(\theta-i\pi) \sin 2\pi\lambda\;, \\
& \bar{c}(\theta)=-4ie^{k\lambda\left(3\theta+\frac{(k+2)i\pi}{k}\right)}\sinh k\lambda(\theta-i\pi) \sin 2\pi\lambda\;,
\end{align}
where $p_{\alpha}$ is the parity of the particle labelled by $\alpha$ ($0$ for even and $1$ for odd).
The remaining Boltzmann weights can be reparametrized as\par\vspace{-10pt}
\begingroup\footnotesize
\begin{equation}\begin{split}
& d_{\alpha\beta}(\theta)
\\ & =
\left\{\begin{array}{ll}
4e^{2k\lambda\left(\theta+\frac{(k+2)i\pi}{2k}\right)}\left(\sinh k\lambda(\theta-i\pi) \sinh k\lambda \theta-\sin 2\pi\lambda \sin k\pi\lambda\right)\; & \alpha=\beta=\beta' \;,\\
4e^{2k\lambda\left(\theta+\frac{(k+2)i\pi}{2k}\right)}\sinh k\lambda\theta \sinh k\lambda\left((-1)^{p_{\alpha}}(\theta-i\pi)+\frac{2i\pi}{k}\right)\; & \alpha=\beta \neq \beta'\;, \\
4ie^{k\lambda\left(\theta+\frac{(k+2)i\pi}{k}\right)} e^{i\pi\lambda(k+2(t_{\alpha}-t_{\beta}))}\frac{\varepsilon_{\alpha}}{\varepsilon_{\beta}}\sinh k\lambda\theta \sin 2\pi\lambda\; & \alpha<\beta\;, \; \alpha \neq \beta'\;, \\
4ie^{k\lambda\left(\theta+\frac{(k+2)i\pi}{k}\right)}\sin 2\pi\lambda \left((-1)^{p_{\alpha}}e^{i\pi\lambda(k+2(t_{\alpha}-t_{\alpha'}))}\sinh k\lambda\theta-\sinh k\lambda(\theta-i\pi)\right)\; & \alpha<\beta=\alpha'\;, \\
4ie^{k\lambda\left(3\theta+\frac{(k+2)i\pi}{k}\right)} e^{i\pi\lambda(-k+2(t_{\alpha}-t_{\beta}))}\frac{\varepsilon_{\alpha}}{\varepsilon_{\beta}} \sinh k\lambda\theta \sin 2\pi\lambda\; & \alpha>\beta\;, \; \alpha \neq \beta'\;, \\
4ie^{k\lambda\left(3\theta+\frac{(k+2)i\pi}{k}\right)}\sin 2\pi\lambda \left((-1)^{p_{\alpha}}e^{i\pi\lambda(-k+2(t_{\alpha}-t_{\alpha'}))}\sinh k\lambda\theta-\sinh k\lambda(\theta-i\pi)\right)\; \quad & \alpha>\beta=\alpha'\;.
\end{array}\right.
\end{split}\end{equation}
\endgroup
where, as argued in \cite{Galleas:2006kd}, \eqref{R_check_quantum_OSP_group} solves \eqref{YBE_R_check} provided the parameters $\varepsilon_\alpha$ and $t_\alpha$ satisfy
\begin{align}
    & \varepsilon_{\alpha}=(-1)^q{p_{\alpha}}\varepsilon_{\alpha'} && \alpha < \alpha' \;, \\
    & t_{\alpha}=t_{\alpha'}-2\Big(p_{\alpha}+\frac{N}{2}+m-\alpha-2\sum\limits_{\beta=\alpha}^{n+m+1}p_{\beta}\Big) && \alpha < \alpha' \;.
\label{epsilon_t_conditions}
\end{align}
A possible choice of parameters $\varepsilon_{\alpha}$ and $t_{\alpha}$ solving \eqref{epsilon_t_conditions} is
\begin{align}
& \varepsilon_{\alpha}=e^{-i\pi p_{\alpha}} && \alpha<\alpha'\;, \\
& t_{\alpha}=\alpha+\frac{1}{2}-p_{\alpha}+2\sum\limits_{\beta=\alpha}^{n+m+1}p_{\beta} \; && \alpha<\alpha'~,
\label{epsilon_t_solution}
\end{align}
with the remaining parameters determined by \eqref{epsilon_t_conditions}.

Dividing \eqref{R_check_quantum_OSP_group} by $e^{2k\lambda(\theta+\frac{(k+2)i\pi}{2k})}$ and applying the $\theta$-dependent gauge transformation defined by the diagonal matrix (we now use $i$ and $\bar{\imath}=N+2m+1-i$ instead of $\alpha$ and $\alpha'$)
\begin{equation}
    K_i^j(\vartheta)=e^{s_i k\lambda\vartheta}\delta_i^j\;, \quad 1 \leq i,j \leq N+2m\;,
\end{equation}
subject to $s_{\bar{\imath}}=-s_i$ and $s_i=0$ for $i=\bar{\imath}$, which ensures that the result only depends on the difference of rapidities $\theta = \vartheta_1 - \vartheta_2$, we find the following non-zero matrix elements\par\vspace{-10pt}
\begingroup\footnotesize
\begin{align}
& \check{R}_{ii}^{ii}(\theta)=\sinh k\lambda(\theta-i\pi) \sinh k\lambda\big((-1)^{p_{i}}\theta-\frac{2i\pi}{k}\big) \quad && i \neq \bar{\imath}\;, \notag \\
& \check{R}_{\bar{\imath}i}^{i\bar{\imath}}(\theta)=\sinh k\lambda\theta \sinh k\lambda\big((-1)^{p_{i}}(\theta-i\pi)+\frac{2i\pi}{k}\big) \quad && i \neq \bar{\imath}\;, \notag \\
& \check{R}_{ii}^{ii}(\theta)=\sinh k\lambda(\theta-i\pi) \sinh k\lambda \theta-\sin 2\pi\lambda \sin k\pi\lambda \quad && i=\bar{\imath}\;, \notag \\
& \check{R}_{ji}^{ij}(\theta)=\sinh k\lambda(\theta-i\pi) \sinh k\lambda\theta \quad && i \neq j\;, \quad i \neq \bar{\jmath}\;, \notag \\
& \check{R}_{ij}^{ij}(\theta)=-ie^{-(s_i-s_j+k)\lambda\theta}\sinh k\lambda(\theta-i\pi) \sin 2\pi\lambda \quad && i>j\;, \quad i \neq \bar{\jmath}\;, \notag \\
& \check{R}_{ij}^{ij}(\theta)=-ie^{-(s_i-s_j-k)\lambda\theta}\sinh k\lambda(\theta-i\pi) \sin 2\pi\lambda \quad && i<j\;, \quad i \neq \bar{\jmath}\;, \notag \\
& \check{R}_{\bar{\imath}i}^{j\bar{\jmath}}(\theta)=ie^{(s_i-s_j+k)\lambda\theta}e^{i\pi\lambda(-k+2(t_i-t_j))}\frac{\varepsilon_i}{\varepsilon_j} \sinh k\lambda\theta \sin 2\pi\lambda \quad && i>j\;, \quad i \neq \bar{\jmath}\;, \notag \\
& \check{R}_{\bar{\imath}i}^{j\bar{\jmath}}(\theta)=ie^{(s_i-s_j-k)\lambda\theta}e^{i\pi\lambda(k+2(t_i-t_j))}\frac{\varepsilon_i}{\varepsilon_j} \sinh k\lambda\theta \sin 2\pi\lambda \quad && i<j\;,  \quad i \neq \bar{\jmath}\;, \notag \\
& \check{R}_{i\bar{\imath}}^{i\bar{\imath}}(\theta)=ie^{(s_{\bar{\imath}}-s_{i}+k)\lambda\theta} \big((-1)^{p_i}e^{i\pi\lambda(-k+2(t_{\bar{\imath}}-t_{i}))}\sinh k\lambda\theta-\sinh k\lambda(\theta-i\pi)\big)\sin 2\pi\lambda \quad && i<\bar{\imath}\;, \notag \\
& \check{R}_{i\bar{\imath}}^{i\bar{\imath}}(\theta)=ie^{(s_{\bar{\imath}}-s_{i}-k)\lambda\theta}\big((-1)^{p_i}e^{i\pi\lambda(k+2(t_{\bar{\imath}}-t_{i}))}\sinh k\lambda\theta-\sinh k\lambda(\theta-i\pi)\big)\sin 2\pi\lambda \quad && i>\bar{\imath}\;.
\label{R_check_matrix_transformed}
\end{align}
\endgroup

Demanding $\boldsymbol{P}\boldsymbol{T}$ invariance
\begin{equation}\label{PT_symmetry_app}
\check{R}_{i_1 i_2}^{j_1 j_2}(\theta)=\check{R}_{j_1 j_2}^{i_1 i_2}(\theta) \;,
\end{equation}
implies that
\begin{equation}\label{PT_epsilon_t_constraint}
e^{(s_i+s_{\bar{\imath}}-s_j-s_{\bar{\jmath}})\lambda\theta}e^{2i\pi\lambda(t_i-t_j+t_{\bar{\imath}}-t_{\bar{\jmath}})}\frac{\varepsilon_{i} \varepsilon_{\bar{\imath}}}{\varepsilon_{j} \varepsilon_{\bar{\jmath}}}=1\;.
\end{equation}
If we assume that $\varepsilon_i$ and $t_i$, $i=1,\ldots,2n+2m+1$ are determined by \eqref{epsilon_t_conditions} and \eqref{epsilon_t_solution} and that $s_{\bar{\imath}}=-s_i$, then $\boldsymbol{P}\boldsymbol{T}$-symmetry does not impose any additional constraints.

For crossing symmetry we require that
\begin{equation}\label{crossing_symmetry_app}
\check{R}_{i_1 i_2}^{j_1 j_2}(\theta)=(C^{-1})_{i_1}^{k_1}\check{R}_{k_1 j_1}^{i_2 k_2}(i\pi-\theta)(C)_{k_2}^{j_2}\;,
\end{equation}
where the non-zero elements of the charge conjugation matrix $C$ are given by
\begin{equation}\label{C_matrix_app}
    C_i^{\bar{\imath}}=\left\{\begin{array}{ll}
    e^{-i\pi p_i} \; & i<\bar{\imath} \\
    1 \; & i=\bar{\imath} \\
    e^{i\pi p_i} \; & i>\bar{\imath}
    \end{array}\right.\;,
\end{equation}
which leads to the following condition
\begin{equation}
e^{-i\pi\lambda(s_i-s_j)}=e^{2i\pi\lambda(t_i-t_j)}\frac{\varepsilon_i}{\varepsilon_j}(C^{-1})_{i}^{\bar{\imath}}C_{j}^{\bar{\jmath}}\;.
\end{equation}
Therefore, using that $s_{n+m+1}=0$, $(C^{-1})_{i}^{\bar{\imath}}=(C_{i}^{\bar{\imath}})^{-1}$ together with \eqref{C_matrix_app}, we find that crossing symmetry implies
\begin{equation}
e^{-i\pi\lambda s_i}=e^{2i\pi\lambda(t_{i}-n-m-1)}\frac{\varepsilon_{i}}{C_{i}^{\bar{\imath}}}=e^{2i\pi\lambda(t_{i}-n-m-1)}\;,
\end{equation}
which has the solution
\begin{equation}\label{s_t_relation}
s_i=2n+2m+2-2t_{i}+\frac{2l_i}{\lambda}\;, \qquad l_i \in \mathbb{Z}\;, \quad i<\bar{\imath}\;.
\end{equation}
Substituting the $t_i$ from \eqref{epsilon_t_solution}, we obtain
\begin{equation}
s_i=2n+2m+1-2i+2p_i-4\sum\limits_{j=i}^{\frac{r-1}{2}+m}p_j+\frac{4l_i}{\lambda}\;, \quad l_i \in \mathbb{Z}\;, \quad i<\bar{\imath}\;.
\end{equation}
Since we are interested in the $\check{R}$-matrix that becomes rational in the $\lambda\to 0$ limit, we can fix $l_i = 0$ leading to
\begin{equation}
s_i=2n+2m+1-2i+2p_i-4\sum\limits_{j=i}^{\frac{N-1}{2}+m}p_j\;, \qquad i<\bar{\imath}\;.
\end{equation}
Equivalently, we may reformulate this as
\begin{align}\label{s_coefficients_solution}
& s_1=N-2m-2+2p_1\;, \\
& s_{i+1}-s_{i}=-2+2(p_{i+1}+p_{i}) && i<\bar{\imath}\;, \\
& s_{\bar{\imath}}=-s_{i} && i<\bar{\imath}\;, \\
& s_i=0 &&  i=\bar{\imath}\;.
\end{align}

Therefore, the $\check{R}$-matrix \eqref{R_check_matrix_transformed} with $s_i$, $\epsilon_i$ and $t_i$ given by \eqref{s_coefficients_solution}, \eqref{epsilon_t_solution} and \eqref{epsilon_t_conditions} satisfies the YB equation \eqref{YBE_R_check} and is $\boldsymbol{P}\boldsymbol{T}$-invariant \eqref{PT_symmetry_app} and crossing-symmetric \eqref{crossing_symmetry_app} with the charge conjugation matrix \eqref{C_matrix_app}.
Furthermore, as a consequence of setting $l_i=0$, the $\lambda\to 0$ limit yields a rational $\check{R}$-matrix.
In section \ref{S-matrix} we use the $\check{R}$-matrix \eqref{R_check_matrix_transformed} in this form with $t_i$ in \eqref{s_t_relation} given by
\begin{equation}
t_i=n+m+1-\frac{s_i}{2}\;.
\end{equation}
\section{Tree-level \texorpdfstring{$OSP(5|2)$}{OSP(5|2)} \texorpdfstring{$S$}{S}-matrix for \texorpdfstring{$2\to 2$}{2->2} scattering}\label{app_OSP(5|2)_S_matrix}
In this appendix we outline the computation of the tree-level S-matrix for $2 \to 2$ scattering in the $OSP(5|2)$ Toda QFT.
Our starting point is the Euclidean Lagrangian \eqref{OSP(5|2)-Lagrangian-scaled}.
It is not immediately clear how we should treat the bosonic spinor $\uppsi$ in the perturbative scattering theory.
Here we present a prescription for this, in part guided by physical considerations and the requirement of integrability; however, it remains to understand this from first principles.
We first set
\begin{equation}
\uppsi = \begin{pmatrix}-i\gamma \\ \beta^*\end{pmatrix}~,\qquad
\bar\uppsi = \begin{pmatrix}i\gamma^* & \beta\end{pmatrix}~.
\end{equation}
Recalling our conventions, $\gamma^1 = \sigma_1$, $\gamma^2 = \sigma_2$, $\partial = \partial_1 - i \partial_2$, \eqref{OSP(5|2)-Lagrangian-scaled} can then be written as
\begin{equation}\begin{aligned}\label{eq:c2}
\mathcal{L} & =
\frac12 \partial \Phi \bar\partial \Phi
+ i \bar \psi_1 \gamma^\mu \partial_\mu \psi_1
+ i \bar \psi_2 \gamma^\mu \partial_\mu \psi_2
+ \beta \bar\partial \gamma + \beta^* \partial \gamma^*
\\ & \quad
+
\big(\hat{b}^2+\dots\big) \big( \frac{1}{8}(\bar\psi_1 \gamma^\mu \psi_1)^2
+\frac{1}{8} (\bar\psi_2 \gamma^\mu \psi_2)^2
-\frac{1}{2} \beta \beta^* \gamma \gamma^* \big)
\\ & \quad
+ \big(\hat{b}^2 +\dots\big) \big(\bar\psi_1 \psi_1 \gamma \gamma^*
+ \frac12 (\gamma \gamma^*)^2
+ \bar\psi_1 \psi_1 \bar\psi_2 \gamma_+ \psi_2
+ (\beta \beta^* +\gamma \gamma^*) \bar\psi_2 \gamma_+ \psi_2 \big)
\\ & \quad
- M \bar\psi_1 \psi_1 \cosh \hat{b} \Phi
- M \bar\psi_2 \psi_2 \cosh \hat{b} \Phi
- M (\beta \beta^* + \gamma \gamma^*) \cosh \hat{b} \Phi
+ \frac{M^2}{2\hat{b}^2} \sinh^2 \hat{b} \Phi ~,
\end{aligned}\end{equation}
where we have rescaled $\Phi \to 2\sqrt{\pi}\Phi$ and set $\hat{b} = 2\sqrt{\pi}{b}$.
Note that in the limit $\hat b \to 0$, the Lagrangian for the bosonic spinor is the familiar (massive) $\beta\gamma$-system.

We now perform a four steps: first we replace $\gamma\gamma^* \to - \gamma\gamma^*$ 
in the final two lines of \eqref{eq:c2}
;
\unskip\footnote{This can be understood as replacing the scalar $\bar\uppsi\uppsi$ 
by the pseudoscalar $
-
\bar\uppsi\gamma_5\uppsi$ 
($\bar\uppsi\gamma_\pm\uppsi \to \mp\bar\uppsi\gamma_\pm\uppsi$)
in 
the final two lines of
\eqref{OSP(5|2)-Lagrangian-scaled}, 
and
leads to the complex scalar $\Upsilon$ having a physical mass $M$ 
in agreement with the spin-statistics theorem
.}
second, we integrate over the field $\beta$ and set $\gamma = \sqrt{M}\Upsilon$; third,
we continue to Lorentzian signature;
\unskip\footnote{
Our continuation to Lorentzian signature is given by $\xi^2 = i \xi^0$, $\partial_2 = - i \partial_0$, $\mathcal{L} \to - \mathcal{L}$, $M \to - M$ 
and $\gamma^\mu \to \gamma^\mu_\ind{M}$
.
Therefore, $\partial = \partial_1 - i \partial_2 = \partial_1 - \partial_0 \equiv - \partial_-$ and
$\bar\partial = \partial_1 + i \partial_2 = \partial_1 + \partial_0 \equiv \partial_+$.
In addition, we use $(+,-)$ signature in Lorentzian space and hence $A^\mu B_\mu \to - A^\mu B_\mu$.}
and finally we expand to $O(\hat{b}^2)$ to give
\begin{equation}\begin{aligned}\label{lagscatter}
\mathcal{L} & =
\frac12 \partial_- \Phi \partial_+ \Phi - \frac{M^2}{2} \Phi^2
+ \bar \psi_1 (i\gamma_\ind{M}^\mu \partial_\mu - M) \psi_1
+ \bar \psi_2 (i\gamma_\ind{M}^\mu \partial_\mu - M) \psi_2
+ \partial_- \Upsilon^* \partial_+ \Upsilon - M^2 \Upsilon^*\Upsilon
\\ & \quad
- \frac{\hat{b}^2}{6} M^2 \Phi^4
- \frac{\hat{b}^2}{2} M \bar\psi_1 \psi_1 \Phi^2
- \frac{\hat{b}^2}{2} M \bar\psi_2 \psi_2 \Phi^2
- \frac{\hat b^2}{2} (\partial_- \Upsilon^* \partial_+ \Upsilon + M^2 \Upsilon^* \Upsilon) \Phi^2
\\ & \quad
+
\frac{\hat{b}^2}{8} (\bar\psi_1 \gamma_\ind{
M
}^\mu\psi_1)^2
+
\frac{\hat{b}^2}{8} (\bar\psi_2 \gamma_\ind{
M
}^\mu\psi_2)^2
- \hat{b}^2 \bar\psi_1 \psi_1 \bar \psi_2 \gamma_{\ind{
M
}+} \psi_2
\\ & \quad\vphantom{\frac{1}{2}}
- \hat{b}^2 M \bar\psi_1 \psi_1 \Upsilon^* \Upsilon
- \hat{b}^2 M^{-1} \bar\psi_2 \gamma_{\ind{
M
}+} \psi_2 (\partial_- \Upsilon^* \partial_+ \Upsilon + M^2 \Upsilon^* \Upsilon)
\\ & \quad
- \frac{\hat{b}^2}{2} \Upsilon^* \Upsilon (\partial_- \Upsilon^* \partial_+ \Upsilon + M^2 \Upsilon^* \Upsilon)
+ O(\hat b^4)\ .
\end{aligned}\end{equation}
From this Lagrangian it is straightforward to compute the tree-level $S$-matrix for $2\to2$ scattering.
Our conventions in Lorentzian signature are
\begin{equation}\begin{gathered}
\partial_\pm = \partial_0 \pm \partial_1 \ , \quad
\{\gamma_\ind{
M
}^\mu , \gamma_\ind{
M
}^\nu \} = 2 \eta^{\mu\nu}I \ ,
\quad \gamma_\ind{
M
}^0 = \begin{pmatrix} 0 & -i \\ i & 0 \end{pmatrix} \ ,
\ \ \gamma_\ind{
M
}^1 = \begin{pmatrix} 0 & i \\ i & 0 \end{pmatrix} \ ,
\\
\gamma_{\ind{
M
}5} = \gamma_\ind{
M
}^0 \gamma_\ind{
M
}^1 = \begin{pmatrix} 1 & 0 \\ 0 & -1 \end{pmatrix}
\ , \ \
\gamma_{\ind{
M
}+} = \tfrac12(1+\gamma_{\ind{
M
}5}) = \begin{pmatrix} 1 & 0 \\ 0 & 0 \end{pmatrix}
\ , \ \
\gamma_{\ind{
M
}-} = \tfrac12(1-\gamma_{\ind{
M
}5}) = \begin{pmatrix} 0 & 0 \\ 0 & 1 \end{pmatrix} \ .
\end{gathered}\end{equation}
The mass-shell condition for all the fields in the free Lagrangian is
\begin{equation}
(\partial_+ \partial_- + M^2) F = 0 \qquad \Rightarrow \qquad p_+ p_- = M^2 \ ,
\end{equation}
and we introduce the usual relativistic rapidity
\begin{equation}
p_\pm = M e^{\pm \vartheta} \ ,
\end{equation}
such that the on-shell fermion wave-functions are given by
\begin{equation}\begin{aligned}
u(p) & = \sqrt{M} \begin{pmatrix} e^{-\frac{\vartheta}{2}} \\ i e^{\frac{\vartheta}{2}} \end{pmatrix} \ , \qquad
& \bar u(p) & = \sqrt{M} \begin{pmatrix} e^{\frac{\vartheta}{2}} & -i e^{-\frac{\vartheta}{2}} \end{pmatrix} \ , \\
v(p) & = \sqrt{M} \begin{pmatrix} i e^{-\frac{\vartheta}{2}} \\ e^{\frac{\vartheta}{2}} \end{pmatrix} \ , \qquad
& \bar v(p) & = \sqrt{M} \begin{pmatrix} i e^{\frac{\vartheta}{2}} & - e^{-\frac{\vartheta}{2}} \end{pmatrix} \ ,
\end{aligned}\end{equation}
for an incoming particle, outgoing particle, outgoing antiparticle and incoming antiparticle respectively.
The resulting tree-level S-matrix precisely matches (up to a rapidity-dependent gauge transformation) the expansion of the exact S-matrix constructed in section \ref{S-matrix} and hence we do not give the explicit expression here.
Furthermore, it follows that the tree-level S-matrix is $\boldsymbol{P}\boldsymbol{T}$-invariant, crossing symmetric and satisfies braiding unitarity.

We conclude this appendix by noting that when we integrated over $\beta$ we dropped a determinant contribution that will play a role at higher orders.
In order to avoid this issue one could attempt to compute the perturbative S-matrix starting directly from the Lagrangian \eqref{OSP(5|2)-Lagrangian-scaled} for the bosonic spinor $\uppsi$.
This should be possible although it remains to be understood how to properly continue this action to Lorentzian signature.

\section{The diffeomorphism vector and one-form}\label{app:vecofocomp}
For the deformed $OSP(N|2)$ sigma model, $N=1,\dots,6$ (the metric and Kalb-Ramond field of which are given in equations \eqref{eq:defmet62} and \eqref{eq:defbfi62} respectively), the non-vanishing components of the vector $Z$ and one-form $Y$ that solve the Ricci flow equation \eqref{eq:ricci} with the parameters $\nu$ and $\eta$ flowing as in \eqref{eq:rgflow} are\par\vspace{-10pt}
\begingroup\footnotesize
\allowdisplaybreaks
\begin{align}
&
\begin{aligned}
N=6: \qquad & Z^{r_1} = \nu\eta r_1(1-r_1^2) \Big[1+\frac{2(1-r_1^2r_2^2)}{1+\eta^2r_1^4 r_2^2}
+\Big(1+\frac{2(1-r_1^2r_2^2)(1-\eta^2 r_1^4 r_2^2)}{(1+\eta^2 r_1^4 r_2^2)^2}\Big) \psi\cdot\psi \Big] ~,
\\ & Z^{r_2} = \nu\eta r_1^2 r_2(1-r_2^2)(1+\psi\cdot\psi) ~,
\qquad Z^{\psi^a} = -\nu\eta\Big(1+\frac{2(1-r_1^2)}{1+\eta^2r_1^2} + \frac{2(1-r_1^4 r_2^2)}{1+\eta^2r_1^4r_2^2}\Big)\psi^a ~,
\\ & Y^{\phi_1} = \eta(1-r_1^2)\Big[1+\frac{1}{1+\eta^2r_1^2}+\frac{2}{1+\eta^2r_1^4r_2^2}
+\Big(1+\frac{1-\eta^2r_1^2}{(1+\eta^2r_1^2)^2} + \frac{2(1-\eta^2r_1^4r_2^2)}{(1+\eta^2r_1^4r_2^2)^2}\Big)\psi\cdot\psi \Big] ~,
\\ & Y^{\phi_2} = \eta r_1^2(1-r_2^2)\Big[1+\frac{1}{1+\eta^2r_1^4r_2^2} + \Big(1+\frac{1-\eta^2r_1^4r_2^2}{(1+\eta^2r_1^4r_2^2)^2}\Big)\psi\cdot\psi \Big] ~,
\\ & Y^{\psi^a} = - \frac{\eta}{1+\eta^2}\Big(\eta^2 +\frac{2(1+\eta^2)}{1+\eta^2r_1^2} + \frac{2(1+\eta^2)}{1+\eta^2r_1^4r_2^2}\Big)\psi^a ~,
\end{aligned}
\\ &\begin{aligned}
N=5: \qquad & Z^{r_1} = \frac{2\nu\eta r_1(1-r_1^2)(1-r_1^2r_2^2)}{1+\eta^2r_1^2r_2^4} 
\Big(1 + \frac{1-\eta^2 r_1^4 r_2^2}{1+\eta^2 r_1^4 r_2^2} \psi\cdot\psi \Big) ~,
\\ & Z^{\psi^a} = -2\nu\eta\Big(\frac{2(1-r_1^2)}{1+\eta^2r_1^2} + \frac{1-r_1^4 r_2^2}{1+\eta^2r_1^4r_2^2}\Big)\psi^a ~,
\\ & Y^{\phi_1} = \eta(1-r_1^2)\Big[\frac{1}{1+\eta^2r_1^2}+\frac{2}{1+\eta^2r_1^4r_2^2}
+\Big(\frac{1-\eta^2r_1^2}{(1+\eta^2r_1^2)^2} + \frac{2(1-\eta^2r_1^4r_2^2)}{(1+\eta^2r_1^4r_2^2)^2}\Big)\psi\cdot\psi \Big] ~,
\\ & Y^{\phi_2} = \frac{\eta r_1^2(1-r_2^2)}{1+\eta^2r_1^4r_2^2} \Big(1 + \frac{1-\eta^2r_1^4r_2^2}{1+\eta^2r_1^4r_2^2}\psi\cdot\psi \Big) ~,
\\ & Y^{\psi^a} = \frac{\eta}{1+\eta^2}\Big(1 -\frac{2(1+\eta^2)}{1+\eta^2r_1^2} - \frac{2(1+\eta^2)}{1+\eta^2r_1^4r_2^2}\Big)\psi^a ~,
\end{aligned}
\\ &\begin{aligned}
N=4: \qquad & Z^{r_1} = \nu\eta r_1(1-r_1^2)(1 + \psi\cdot\psi ) ~,
\qquad Z^{\psi^a} = -\nu\eta\Big(1+\frac{2(1-r_1^2)}{1+\eta^2r_1^2} \Big)\psi^a ~,
\\ & Y^{\phi_1} = \eta(1-r_1^2)\Big[1+\frac{1}{1+\eta^2r_1^2}
+\Big(1+\frac{1-\eta^2r_1^2}{(1+\eta^2r_1^2)^2} \Big)\psi\cdot\psi \Big] ~,
\\ & Y^{\psi^a} = - \frac{\eta}{1+\eta^2}\Big(\eta^2 +\frac{2(1+\eta^2)}{1+\eta^2r_1^2} \Big)\psi^a ~,
\end{aligned}
\\ &\begin{aligned}
N=3: \qquad & Z^{\psi^a} = -\frac{2\nu\eta(1-r_1^2)}{1+\eta^2r_1^2} \psi^a ~,
\\ & Y^{\phi_1} = \frac{\eta(1-r_1^2)}{1+\eta^2r_1^2} \Big(1+ \frac{1-\eta^2r_1^2}{1+\eta^2r_1^2}\psi\cdot\psi \Big) ~,
\qquad & Y^{\psi^a} = \frac{\eta}{1+\eta^2}\Big(1 -\frac{2(1+\eta^2)}{1+\eta^2r_1^2} \Big)\psi^a ~,
\end{aligned}
\\ &\begin{aligned}
N=2: \qquad & Z^{\psi^a} = -\nu\eta\psi^a ~, \qquad
Y^{\psi^a} = - \frac{\eta^3}{1+\eta^2}\psi^a ~,
\end{aligned}
\\ &\begin{aligned}
N=1: \qquad & Y^{\psi^a} = \frac{\eta}{1+\eta^2}\psi^a ~.
\end{aligned}
\end{align}
\endgroup

For the two alternative deformations of the $OSP(5|2)$ sigma model, \eqref{eq:deflag52alt1} and \eqref{eq:deflag52alt2}, the non-vanishing components of the vector $Z$ and the one-form $Y$ that solve the Ricci flow equation \eqref{eq:ricci} with the parameters $\nu$ and $\eta$ flowing as in \eqref{eq:rgflow}, i.e. before implementing the analytic continuation in equation \eqref{eq:etakapac}, are\par\vspace{-10pt}
\begin{align}
& \begin{aligned}
& Z^{r_1} = 2\nu\eta r_1^3(1-r_1^2)(1+\eta^2r_1^2)
\Big[\frac{1-r_2^2}{(1+\eta^2r_1^4)(1+\eta^2r_1^4r_2^2)} + 2\Big(\frac{1}{(1+\eta^2r_1^4)^2} - \frac{2x^2}{(1+\eta^2r_1^4r_2^2)^2}\Big) \psi\cdot\psi \Big] ~,
\\ & Z^{\psi^a} = -\frac{2\nu\eta r_1^2(1-r_2^2)}{1+\eta^2r_1^4r_2^2} ~,
\\ & Y^{\phi_1} = \eta(1-r_1^2)\Big[\frac{1}{1+\eta^2r_1^2}-\frac{2}{1+\eta^2r_1^4}+\frac{2}{1+\eta^2r_1^4r_2^2}
+4\eta^2r_1^4\Big(\frac{1}{(1+\eta^2r_1^4)^2} - \frac{2r_2^2}{(1+\eta^2r_1^4r_2^2)^2}\Big)\psi\cdot\psi \Big] ~,
\\ & Y^{\phi_2} = -\frac{1}{\eta r_1^2}\Big(1-\frac{1+\eta^2 r_1^4}{1+\eta^2r_1^4 r_2^2} - \frac{2 \eta^2 r_1^4(1-r_2^2) (1-\eta^2r_1^4r_2^2)}{(1+\eta^2r_1^4r_2^2)^2}\psi\cdot\psi \Big) ~,
\\ & Y^{\psi^a} = \eta r_1^2 \Big(\frac{1}{1+\eta^2 r_1^4} - \frac{2}{1+\eta^2 r_1^4 r_2^2} \Big)\psi^a ~,
\end{aligned}
\end{align}
\allowdisplaybreaks
for case $(1)$ and\par\vspace{-10pt}
\begin{align}
& \begin{aligned}
& Z^{r_1} = -\frac{2\nu\eta r_1^3 r_2^2(1-r_1^2)(1+\eta^2r_1^2)}{1+\eta^2r_1^4 r_2^4}\Big(
\frac{1-r_2^2}{1+\eta^2 r_1^4 r_2^2}-\frac{2r_2^2}{1+\eta^2r_1^4 r_2^4} \psi\cdot\psi \Big) ~,
\\ & Z^{r_2} = -\frac{2\nu\eta r_1^2 r_2(1-r_2^2)}{1+\eta^2r_1^4 r_2^4}\Big(
1-r_2^2 - \frac{2r_2^2(1+\eta^2r_1^4r_2^2)}{1+\eta^2r_1^4r_2^4}\psi\cdot\psi\Big) ~,
\\ & Y^{\phi_1} = \eta(1-r_1^2)\Big(\frac{1}{1+\eta^2r_1^2}+\frac{2}{1+\eta^2r_1^4r_2^2}
-\frac{2}{1+\eta^2r_1^4r_2^4} + \frac{4\eta^2r_1^4r_2^4}{(1+\eta^2r_1^4r_2^4)^2} \psi\cdot\psi \Big) ~,
\\ & Y^{\phi_2} = \eta r_1^2(1-r_2^2)\Big(\frac{1}{1+\eta^2r_1^4r_2^2} -\frac{2}{1+\eta^2r_1^4r_2^4} + \frac{4\eta^2r_1^4r_2^4}{(1+\eta^2r_1^4r_2^4)^2} \psi\cdot\psi \Big) ~,
\quad Y^{\psi^a} = \frac{\eta r_1^2r_2^2}{1+\eta^2r_1^4r_2^4}\psi^a ~,
\end{aligned}
\end{align}
for case $(2)$.

\bibliographystyle{JHEP}
\bibliography{MyBib}

\end{document}